\documentclass{article}
\usepackage[IL2]{fontenc}
\usepackage[letterpaper,margin=1.00in]{geometry}
\usepackage{amsmath, amssymb, amsthm, amsfonts}
\usepackage{bbm}
\usepackage{amscd}
\usepackage{mathrsfs}

\usepackage{comment} % \begin{comment}\end{comment} to comment stuff that may be uncommented later on
\usepackage{ifthen}
\usepackage{tikz}
\usetikzlibrary{positioning,decorations.pathreplacing}
\usepackage{ bbold }
\usepackage{graphicx}
\usepackage{color}
\usepackage{algorithm}
\usepackage[noend]{algpseudocode}
\usepackage{epstopdf}
\usepackage{wrapfig}
\usepackage{paralist}
\usepackage{wasysym}
\usepackage[textsize=tiny]{todonotes}
\usepackage{listings} % listing code
\usepackage{pdflscape} % \clearpage\begin{landscape}...\end{landscape}\clearpage for landscape page

\usepackage{framed}
\usepackage[framemethod=tikz]{mdframed}
\usepackage[bottom]{footmisc}
\usepackage{enumitem}
\setitemize{noitemsep,topsep=3pt,parsep=3pt,partopsep=3pt}
\usepackage[font=small]{caption}
\usepackage{xspace}

\usepackage{thmtools} % not sure what is it good for but probably has to do something with thm-restate
\usepackage{thm-restate} % \begin{restatable}{theorem}{thmmain}..\end{restatable}
\usepackage{hyperref}
\hypersetup{
    unicode=false,          % non-Latin characters in Acrobat’s bookmarks
    colorlinks=true,        % false: boxed links; true: colored links
    linkcolor=red,          % color of internal links (change box color with linkbordercolor)
    citecolor=darkgreen,        % color of links to bibliography
    filecolor=magenta,      % color of file links
    urlcolor=cyan           % color of external links
}

\newtheorem{theorem}{Theorem}[section]
\newtheorem{lemma}[theorem]{Lemma}
\newtheorem{meta-theorem}[theorem]{Meta-Theorem}
\newtheorem{claim}[theorem]{Claim}

\newtheorem{definition}[theorem]{Definition}

\newtheorem{fact}[theorem]{Fact}

\usepackage[capitalize, nameinlink,noabbrev]{cleveref}
\crefname{theorem}{Theorem}{Theorems}
\crefname{proposition}{Proposition}{Propositions}
\crefname{observation}{Observation}{Observations}
\crefname{lemma}{Lemma}{Lemmas}
\crefname{claim}{Claim}{Claims}
\crefname{problem}{Problem}{Problems}
\crefname{conjecture}{Conjecture}{Conjectures}
\crefname{question}{Question}{Questions}
\crefname{example}{Example}{Examples}
\crefname{fact}{Fact}{Facts}
\definecolor{darkgreen}{rgb}{0,0.5,0}

\usepackage{algcompatible}
\algnewcommand\algorithmicswitch{\textbf{switch}}
\algnewcommand\algorithmiccase{\textbf{case}}

\algdef{SE}[SWITCH]{Switch}{EndSwitch}[1]{\algorithmicswitch\ #1\ \algorithmicdo}{\algorithmicend\ \algorithmicswitch}%
\algdef{SE}[CASE]{Case}{EndCase}[1]{\algorithmiccase\ #1}{\algorithmicend\ \algorithmiccase}%
\algtext*{EndSwitch}%
\algtext*{EndCase}%
\newcommand{\fB}{\mathcal{B}}

\newcommand{\eps}{\varepsilon}

\renewcommand{\P}{\textrm{P}}
\newcommand{\Var}{\textrm{Var}}
\newcommand{\e}{\textrm{e}}

\newcommand{\poly}{\operatorname{poly}}

\renewcommand{\phi}{\varphi}

\newcommand{\E}{\mathbb{E}}
\renewcommand{\Pr}{\P}

\newcommand{\R}{\mathbb{R}}
\newcommand{\N}{\mathbb{N}}

\renewcommand{\paragraph}[1]{\vspace{0.15cm}\noindent {\bf #1}:}

\newcommand{\FullOrShort}{full}
\ifthenelse{\equal{\FullOrShort}{full}}{
  
  \newcommand{\fullOnly}[1]{#1}
  \newcommand{\shortOnly}[1]{}
  \renewcommand{\baselinestretch}{1.00}
  
  }{
    \renewcommand{\baselinestretch}{1.00}
    \newcommand{\fullOnly}[1]{}
    \newcommand{\IncludePictures}[1]{}
   
  }

\newcommand{\ctwo}{c^{L\ref{lem:weighted_good_for_large_p}}}
\newcommand{\cthree}{c^{L\ref{lem:hoeffding_chernoff}}}
\newcommand{\failtwo}{prob^{bad,L\ref{lem:weighted_good_for_large_p}}_i}
\newcommand{\ccher}{c^{T\ref{thm:chernoff}}}
\newcommand{\failthree}{prob^{bad,L\ref{lem:hoeffding_chernoff}}_i}
\newcommand{\fail}{prob^{bad}_i}
\newcommand{\failhoef}{prob^{bad,T\ref{thm:hoeffding}}_i}
\newcommand{\failcher}{prob^{bad,T\ref{thm:chernoff}}_i}
\newcommand{\failil}{prob^{bad,T\ref{thm:chernoff}}_{i,\ell}}
\newcommand{\expone}{e^{hoef}_{i}}
\newcommand{\exptwo}{e^{cher}_{i}}
\newcommand{\p}{\mathbf{p}}
\newcommand{\pj}{p_j}
\newcommand{\Deltav}{\mathbf{\Delta}}
\newcommand{\Deltai}{\Delta_i}
\newcommand{\Deltasmall}{\mathbf{\Delta^{small}}}
\newcommand{\Deltasmalli}{\Delta^{small}_i}
\newcommand{\Deltarec}{\mathbf{\Delta^{rec}}}
\newcommand{\Deltareci}{\Delta^{rec}_i}
\newcommand{\q}{\mathbf{q}}
\newcommand{\qj}{q_j}
\newcommand{\precin}{\mathbf{p^{rec,in}}}
\newcommand{\precinj}{p^{rec,in}_j}
\newcommand{\precout}{\mathbf{p^{rec,out}}}
\newcommand{\precoutj}{p^{rec,out}_j}
\newcommand{\psmallin}{\mathbf{p^{small,in}}}
\newcommand{\psmallinj}{p^{small,in}_j}
\newcommand{\psmallout}{\mathbf{p^{small,out}}}
\newcommand{\psmalloutj}{p^{small,out}_j}
\newcommand{\aij}{a_{ij}}
\newcommand{\Arec}{A^{rec}}
\newcommand{\arecij}{a^{rec}_{ij}}
\newcommand{\Asmall}{A^{small}}
\newcommand{\asmallij}{a^{small}_{ij}}
\newcommand{\Ibad}{I^{bad}}
\newcommand{\Ibadsmall}{I^{bad,small}}
\newcommand{\Igoodsmall}{I^{good,small}}
\newcommand{\Ibadrec}{I^{bad,rec}}

\newcommand{\Jsmall}{J^{small}}
\newcommand{\ksmall}{k^{small}}
\newcommand{\pres}{\mathbf{p'}}
\newcommand{\presj}{p'_j}

\newcommand{\kres}{k'}
\newcommand{\Deltares}{\mathbf{\Delta'}}
\newcommand{\Deltaresi}{\Delta'_i}
\newcommand{\Jrec}{J^{rec}}
\newcommand{\Vi}{V_i}
\newcommand{\Vil}{V_{i,\ell}}
\newcommand{\Bil}{B_{i,\ell}}
\newcommand{\Bbigi}{B^{big}_i}
\newcommand{\Bsmalli}{B^{small}_i}
\newcommand{\lmaxi}{\ell^{max}_i}
\newcommand{\alphai}{\alpha_i}
\newcommand{\gammai}{\gamma_i}
\newcommand{\Deltail}{\Delta_{i,\ell}}
\newcommand{\mui}{\mu_i}
\newcommand{\muil}{\mu_{i,\ell}}
\newcommand{\epsone}{\eps^{(1)}_{i,\ell}}
\newcommand{\epstwo}{\eps^{(2)}_{i,\ell}}
\newcommand{\epsthree}{\eps^{(3)}_{i,\ell}}
\newcommand{\epsil}{\eps_{i,\ell}}
\newcommand{\betail}{\beta_{i,\ell}}
\newcommand{\Ibadcher}{I^{bad,bucket}}

\newcommand{\Pot}{\mathrm{Pot}}

\renewcommand{\Pr}{\P}
\usepackage{appendix}
\title{Work-Efficient Parallel Derandomization I: \\ Chernoff-like Concentrations via Pairwise Independence}

\begin{document}
\date{}
\author{Mohsen Ghaffari \\ \small MIT \\ \small ghaffari@mit.edu \and Christoph Grunau \\ \small ETH Zurich \\ \small cgrunau@ethz.ch \and Václav Rozhoň \\ \small ETH Zurich \\ \small vaclav.rozhon@inf.ethz.ch}
\maketitle

    \vspace{-0.5cm}
\begin{abstract}
We present a novel technique for work-efficient parallel derandomization, for algorithms that rely on the concentration of measure bounds such as Chernoff, Hoeffding, and Bernstein inequalities. Our method increases the algorithm's computational work and depth by only polylogarithmic factors. Before our work, the only known method to obtain parallel derandomization with such strong concentrations was by the results of [Motwani, Naor, and Naor FOCS'89; Berger and Rompel FOCS'89], which perform a binary search in a $k$-wise independent space for $k=\poly(\log n)$. However, that method blows up the computational work by a high $\poly(n)$ factor and does not yield work-efficient parallel algorithms. Their method was an extension of the approach of [Luby FOCS'88], which gave a work-efficient derandomization but was limited to algorithms analyzed with only pairwise independence. Pushing the method from pairwise to the higher $k$-wise analysis resulted in the $\poly(n)$ factor computational work blow-up. Our work can be viewed as an alternative extension from the pairwise case, which yields the desired strong concentrations while retaining work efficiency up to logarithmic factors.
    
Our approach works by casting the problem of determining the random variables as an iterative process with $\poly(\log n)$ iterations, where different iterations have independent randomness. This is done so that for the desired concentrations, we need only pairwise independence inside each iteration. In particular, we model each binary random variable as a result of a gradual random walk, and our method shows that the desired Chernoff-like concentrations about the endpoints of these walks can be boiled down to some pairwise analysis on the steps of these random walks in each iteration (while having independence across iterations). Hence, we can fix the randomness of each iteration efficiently before proceeding to the next. 
\end{abstract}

 {   
    % \listoftodos
     \thispagestyle{empty}
 }

{   \newpage
    % \vspace{-0.5cm}
    % \small
    % \smallskip
    \hypersetup{linkcolor=blue}
    \tableofcontents
    \setcounter{page}{0}
    \thispagestyle{empty}
}

\newpage
\setcounter{page}{1}
\section{Introduction}
Probability concentration results such as the Chernoff bound are the bread-and-butter of randomized algorithms~\cite{motwani1995randomized, dubhashi2009concentration}. There are known methods for derandomizing parallel algorithms that rely on these concentrations---see, e.g., \cite{motwani1989probabilistic, berger1989simulating, naor1990small,harris2019deterministic}---but all these methods blow up the algorithm's computational work by a large polynomial. Hence, they fall short of yielding work-efficient deterministic parallel algorithms. In this paper, we present the first generic nearly work-efficient parallel derandomization technique for algorithms that rely on these strong concentrations. This technique increases the algorithm's computational work and depth by only polylogarithmic factors. 

\subsection{Background: parallel work-efficiency, and typical concentration setups}
\label{subsec:background}
\vspace{-5pt}
\noindent\textbf{Work-efficiency in parallel algorithms:} Devising (nearly) work-efficient algorithms with small depth is the ultimate goal in designing parallel algorithms, as that leads to algorithms that can run faster than their sequential counterparts, using a reasonable number of processors. Let us briefly review this. We follow the standard work/depth terminology, see e.g., \cite[Section 1.5]{jaja1992introduction} or \cite{blelloch1996programming}. For any algorithm, its \textit{work} is defined as the total number of operations. Its \textit{depth} is defined as the longest chain of operations with sequential dependencies, in the sense that the $(i+1)^{th}$ operation depends on (and should wait for) the results of operations $i$ in the chain. 
Given $p$ processors, an algorithm with work $W$ and depth $D$ can be run in $W/p+D$ time~\cite{brent1974parallel}. If a parallel algorithm has work polynomially larger than its sequential counterpart, we would need at least polynomially many processors so that the parallel algorithm runs faster than the sequential one. This would severely limit the relevance of such parallel algorithms. As such, the ultimate goal in parallel algorithms is to devise algorithms with work asymptotically equal (or close, e.g., equal up to logarithmic factors) to their sequential counterpart, which also have a small depth. Such algorithms are referred to as \textit{work-efficient} (correspondingly, \textit{nearly work-efficient}) parallel algorithms. See \cite{karp1989survey,jaja1992introduction} for some of the classic examples, and \cite{fineman2018nearly, jambulapati2019parallel, blelloch2020parallelism, li2020faster, andoni2020parallel, cao2020efficient,dhulipala2021theoretically,anderson2021parallel,rozhovn2022undirected,rozhovn2022deterministic} for some of the more recent nearly work-efficient parallel algorithms.

\paragraph{Typical algorithmic setups where concentration bounds are used} Randomized (parallel) algorithms frequently use concentration bounds such as Chernoff, Hoeffding, and Bernstein inequalities. For instance, a simplified form of the Chernoff bound tells us that, for independent random variables $X_1$, $X_2$, $\dots$, $X_\ell \in [0,1]$ and $X:=\sum_{i=1}^{\ell} X_j$, and for any $\eps\in [0,1]$, we have $\P[|X-\E[X]|\geq \eps \E[X]] \leq 2 \exp(- \eps^2 \E[X]/3)$. The prototypical application setup is as follows: we have many independent random variables, and many constraints on these variables (often linear or even polynomial many constraints in terms of the number of variables), each of which requires the summation of a subset of the random variables being in a certain range. Chernoff bound allows us to conclude that each summation will be within a small (constant) factor of its expectation, assuming that the expectations are large enough to allow a union bound over different constraints/summations.  Let us now have a brief look at two of the common scenarios. 
\begin{itemize}
\item \textbf{Sampling.} One routine application is when we use randomness to find a small/sparse \textit{sample} of the data, which provides good representations. Here is a concrete example. Consider an $n$-node undirected graph $G=(V, E)$. Using random sampling, along with Chernoff and union bound, we can sample a subset $T\subseteq V$ of size $(1\pm \eps) pn$ that satisfies the following: for each node $v\in V$, whose degree is denoted by $d_v$, the number of neighbors of $v$ in set $T$ is upper and lower bounded by $(1\pm \eps) d_v p$, assuming $d_v=\Omega((\log n)/p)$, and where $\eps>0$ is a desirably small constant. 
\item \textbf{Rounding.} Another common application is \textit{randomized rounding}, e.g., for linear programs. Roughly speaking, we cast the problem at hand as some integer linear program, solve the fractional relaxation of it efficiently, and then use randomized rounding to turn these fractional solutions into integral solutions, with a small loss in the objectives.  Here is a classic example~\cite{raghavan1987randomized}: Consider a multi-commodity routing instance where for each commodity $c$, we are given a collection $P_c$ of paths from the commodity's source node to its target node, and we should choose exactly one of these (for the integral solution). The congestion of an edge $e$ is the number of chosen paths that include edge $e$, and we want to have small congestion in all edges. A fractional solution provides a probability distribution for each commodity $c$ over $P_c$. Given a fractional solution where each edge has expected congestion $O(\log n)$, via a simple application of Chernoff and union bounds, we get that picking the paths according to their distributions yields an integral solution that has congestion $O(\log n)$.     
\end{itemize}

\subsection{State of the art in parallel derandomization} 
We next overview the known derandomization methods for algorithms that rely on the aforementioned concentrations. We first discuss the standard method of conditional probabilities, frequently used in sequential derandomizations, and then overview the parallel derandomization techniques.  

\paragraph{Method of conditional probabilities} A basic and commonly used method in sequential derandomizations is via conditional probabilities/expectations~\cite{raghavan1986probabilistic}. We use the expected number of violated constraints (e.g., in the sampling example above, the expected number of nodes whose number of neighbors in the sample is outside the allowed range) as a potential function. Initially, this expectation is strictly below $1$. We then iteratively fix the variables one by one (e.g., in the sampling example, determining whether the node is in the sample or not) while ensuring that the potential---i.e., the conditional expectation of the number of violated constraints---does not increase. Once all variables have been fixed, we have an assignment without any violated constraint. The computational efficiency of this approach is heavily dictated by how fast we can compute the conditional expectations in each iteration. Regardless, the method is inherently sequential and relies on fixing the variables one by one. As such, it cannot be used to obtain fast parallel algorithms, i.e., algorithms with depth polylogarithmic or even sublinear in the number of variables.

\paragraph{Using small spaces of randomness} One general theme in derandomizations, especially pioneered in and used for parallel algorithms, is to devise/revise the randomized algorithm such that its analysis relies on weaker properties of the randomness so that we can assume that the randomness comes from a small space. Prominent examples are algorithms analyzed using only pairwise independence~\cite{luby86, alon86} or $k$-wise independence for a constant $k$ ~\cite{alon86}, and algorithms using $\delta$-dependent $k$-wise random variables and small-bias spaces~\cite{naor1990small}. These methods ensure that the randomness of the algorithm comes from a polynomially sized space. Hence, one can enumerate and try all possibilities of randomness (in parallel) and find the one that satisfies all constraints. Of course, that increases the computational work by a factor linear in the space size. Unfortunately, it is well-known that this method requires spaces that are non-trivially large polynomials.\footnote{For instance, assuming $n$ variables, $k$-wise spaces need size $\Theta(n^{k})$, and $\delta$-dependent $k$-wise spaces have size at least $\Omega({1/\delta})$ where we usually need $\delta \leq 1/\poly(n)$ for the desired concentrations to allow union bounds over all constraints~\cite{naor1990small}.} Hence, these methods blow up the algorithm's work by a large polynomial, and cannot yield (nearly) work-efficient parallel algorithms. 

\paragraph{Binary search through the small space of randomness} Luby~\cite{luby1988removing} pointed out that the above derandomization method---in particular the $k$-wise independence used in~\cite{luby86, alon86}---is far from work-efficient. He phrased this as having a prohibitively large penalty on the number of required processors. To fix that, he pioneered a method for parallel derandomization by using a binary search in the small space of randomness. He crafted this for two problems, maximal independent set, and $\Delta+1$ coloring. Concretely, his approach is as follows: both cases can be analyzed using pairwise independence, and hence $\Theta(\log n)$ bits of randomness suffice for them. However, instead of simply brute-forcing through all the $2^{\Theta(\log n)} = \poly(n)$ possibilities of these random strings, Luby effectively performs a binary search in this space, by fixing the bits of randomness one by one. This takes $O(\log n)$ steps, one for each bit. In each step, the method requires computing the conditional expectation of the number of violated constraints (or a suitable pessimistic estimator of it, which is still sufficiently small), for each of the two possible choices of the next bit. Then, we simply choose the bit so that we do not increase the conditional expectation. In a sense, this is an elegant combination of the two methods outlined above, conditional probabilities and small space. The computational efficiency of this method is dictated by how fast we can compute the conditional expectations in each step. 

Luby~\cite{luby1988removing} showed that in the two particular cases of maximal independent set and coloring, the algorithms rely on certain pairwise analyses with extra nice properties.\footnote{See \Cref{subsec:lubyEfficientPairwise} for a precise definition of these nice pairwise analyses.} He then used a well-structured pairwise independent space (where each random variable bit is the dot product of the variable's characteristic vector with the seed of randomness) and showed that the conditional expectations in this space can be computed using near-linear work. We note that this is not always easy, because the $k$-wise independence is lost after some bits are fixed. This led to work-efficient parallel derandomizations for maximal independent set and coloring. 
The method has been extended to some other problems analyzed with pairwise independence, including set cover approximations and minimum dominating set~\cite{berger1989efficient, bezdrighin2022deterministic, ghaffari2023netdecomp}. We remark that these include some instances where the obtained concentration is better than what is implied directly by pairwise independence. However, that is only for a very limited \textit{hitting} type of event\footnote{Hitting events are events of the following form: For a ground set $[n]$, we want to choose a small sample set $T\subset [n]$ such that for every given set $S_i\subset [n]$ for $i\in m$, we have $S_i \cap T\neq \emptyset$. However, these events do not provide good bounds on $|S_i \cap T|$. Moreover, the method used for proving them crucially relies on this simple event, as a single hit suffices and we can use repetitions to amplify the probability of having at least one single hit.} and does not address the wide range of Chernoff applications, including the two prototypical examples mentioned in \Cref{subsec:background}. 

The clear shortcoming of the above is that pairwise independence does not yield strong concentrations. Whether there are (similar) nearly work-efficient parallel derandomizations for stronger concentrations, like the Chernoff bound, has remained open for the past 35 years.

The method of binary search in the space of randomness has been extended to $k$-wise analysis for higher values of $k$, up to $k=\poly(\log n)$~\cite{motwani1989probabilistic, berger1989simulating}. However, this has been at the cost of a polynomial increase in work and does not yield work-efficient derandomizations. Notice that for such high values of $k$, without the binary search, the space size $O(n^{k})$ would be super polynomial, and thus the brute-force approach in the space would require a superpolynomial blowup in computation work. Motwani, Naor, and Naor~\cite{motwani1989probabilistic}, and simultaneously and independently Berger and Rompel~\cite{berger1989simulating}, showed how to perform this binary search to obtain strong concentrations like those implied by the Chernoff bound (this requires choosing $k$ carefully to balance the strength of the obtained concentration with the complexity of computing the related conditional expectations). However, these still increase the computational work by a high polynomial factor (an increase of at least $m^2$ factor where $m$ is the number of constraints). Concretely, when computing the conditional expectations to perform a binary search in a $k$-wise space, once we adjust $k$ to have a failure probability less than $1/m$ for each of $m$ constraints, the conditional expectation can have $\poly(m)$ terms, and computing each of these terms incurs some work. The most recent and the best-known improvement in reducing this polynomial blow-up in the computational work was a paper of Harris~\cite{harris2019deterministic}, but the improvement cuts the polynomial's exponent by only a $2$ factor and remains far away from work-efficiency.\footnote{Concretely, the computational work for simulating $k$-wise independence on $m$ constraints and $n$ variables is reduced from the $mn^k$ bound of \cite{motwani1989probabilistic, berger1989simulating} to $mn^{\lceil k/2\rceil + o(1)}.$ However, we note that this result has the advantage of getting the sharp concentration for all ranges of $\eps$ where Chernoff applies, and this itself was an improvement on an earlier paper of Majan et al.\cite{mahajan2001solving}.} 

Our technique can be viewed as a completely different approach to extending Luby's results~\cite{luby1988removing}, which yields the desired strong concentrations while retaining the work-efficiency up to logarithmic factors. 

\subsection{Our results}
\label{sec:results}

\paragraph{A vanilla version of our derandomization result, as a starter} Our most general and strongest derandomization results are somewhat notation-heavy. Hence, before stating those, we state a weaker but cleaner form that captures some of the basic concentration setups in parallel derandomization.

\begin{theorem}[a vanilla version of our derandomization] \label{thm:vanilla} Suppose we are given sets $S_1, S_2, \dots, S_m \subseteq [1, n]$, where $m\leq \poly(n)$, and two given parameters $p\in[0, 1]$ and $\eps \in [1/\poly(\log n), 1]$ such that for each $i\in [1, m]$, we have $|S_i|\geq \Omega(\frac{\log m}{p\eps^2})$. Randomly including each element of $[1, n]$ in a set $T$ with probability $p$ would provide a set $T$ such that, for every $i\in [1, m]$, we have $|S_i\cap T| \in (1\pm \eps) p |S_i|$.\footnote{We use the notation $z=(1\pm \eps) z'$ as a shorthand for $z\in[(1-\eps)z', (1+\eps)z']$.} There is a deterministic parallel algorithm that computes such a set $T$, using depth $\poly(\log nm)$ and work $\tilde{O}(\sum_{i} |S_i| + n + m)$, which is linear in the input size up to logarithmic factors.
\end{theorem}
For instance, this theorem nicely captures the sampling example in a graph $G=(V, E)$ that we discussed in \Cref{subsec:background}.
%\footnote{With a slight generalization of allowing different probabilities for different variables, the theorem would also capture the rounding for the multicommodity problem discussed in \Cref{subsec:background}. The sets would be all paths that include one edge, and we would get congestion in each edge upper bounded by twice the expectation; depth would be $\poly(\log n)$ and work would be linear in the input size for the rounding (the summation of the lengths of the paths). We do not present a theorem for this exact case. Instead, we will try to present our theorems in their strongest forms in few statements.} 
There, we would have $m=n=|V|$ sets, one for each node and simply equal to its set of neighbors. Then we get a sample set $T$ such that each node $v$ has $(1\pm \eps) p d_v$ neighbors in $T$. The algorithm would have depth $\poly(\log n)$ and work $\tilde{O}(\sum_{j} |S_j| + m + n) = \tilde{O}(|E|+ |V|).$  

Another remark is that the astute reader may have noticed that the theorem statement has a lower bound on $\eps$ by assuming $\eps \in [1/\poly(\log n), 1]$. This is unlike the Chernoff bound where we could allow $\eps$ to be an arbitrarily small value in $[0, 1]$. We could phrase the theorem in a more general manner without a lower bound assumption on $\eps$, but then the algorithm's work and depth would increase by a $\poly(\log n/\eps)$ factor. This is indeed a drawback of our current method, though we note that many applications in parallel algorithms are fine with a constant or polylogarithmically sub-constant deviation factor $\eps$. See \Cref{subsec:followup} for a follow-up work that removes this assumption and obtains an efficient parallel algorithm with the optimal concentration for all ranges of $\eps$.

\paragraph{Generalizations, and the relevant notations} There are several dimensions to consider in generalizations. We next discuss these and gradually introduce the notations that we use in our statements. 
\begin{itemize}
\item[(1)] Different random variables might have different probabilities. We consider $n$ binary random variables and use the notation $p_j$ to denote the probability of the $j^{th}$ variable being equal to $1$. Hence, a part of the input to the problem is a vector $\p \in [0,1]^n$. The output is to compute a $0/1$ setting of variables, i.e., a vector $\q \in \{0,1\}^n$, such that $\q$ is close to $\p$ in the way that we discuss next.  
\item[(2)] We consider $m$ different constraints on the variables, and each constraint might have different coefficients for different random variables that it involves. To encode these constraints, we use a matrix $A \in \mathbb{R}_{\geq 0}^{m \times n}$, where $a_{ij}$ denotes the entry in the $i^{th}$ row and $j^{th}$ colum. Given an $m$-dimensional vector $\Deltav \in \mathbb{R}_{>0}^m$ of the allowed deviations in different constraints, our ideal objective is to compute an output vector $\q \in \{0,1\}^n$ such that for every $i \in [m]$ we have $|\sum_{j \in [n]} \aij(\pj - \qj)| \leq \Deltai.$ For instance, a common useful setting of $\Deltav \in \mathbb{R}_{>0}^n$ would be to set $\Deltai=\eps (\sum_{j \in [n]} \aij\pj),$ for a desirably small deviation factor $\eps\in [0, 1]$. In this case, our constraints would boil down to $A\q \in (1\pm \eps) A\p$, in a coordinate-wise sense. Notice that the input size is  $\tilde{O}(\max(nnz(A),n,m))$, where $nnz(A)$ denotes the number of non-zero entries in $A$. Our derandomizations provide deterministic algorithms that have computational work nearly linear in this input size, and depth polylogarithmic in it.

\item[(3)] We might want something weaker than satisfying all the constraints simultaneously. This is to widen the applicability range and it often brings useful flexibility to the algorithms. For instance, we could wish to have only most of the constraints satisfied. Even stronger, we can have the number of violated constraints upper bounded by the bound that we would have on the expected number of violations if we apply the Chernoff bound to each constraint.\footnote{Instead of just the number of violated constraints, we could generalize by assuming an importance score for each constraint and get a bound on the importance score of the violated constraints. We omit that generalization, as its proof is the same.} Concretely, for each constraint $i\in [m]$, we can provide an upper bound $\fail$ on $\Pr[|\sum_{j \in [n]} \aij(\pj - \qj)| > \Deltai]$, ideally similar to the upper bound provided by Chernoff or Hoeffding bounds, and we would ensure that for the set $\Ibad \subseteq [m]$ of violated constraints, we have $\Ibad \subseteq [m]$ with $|\Ibad| \leq \sum_{i \in [m]} \fail$.   
\item[(4)] Finally, the concentration exponents that we obtain depend on a granularity parameter $k$ to be chosen. We defer the reader to \Cref{sec:overview} for more on this parameter. Our concentration exponents often have a term that grows linearly with $k$. We cannot set $k$ arbitrarily large at no cost, because $k$ appears also in the computational work and depth of our deterministic algorithms as a polynomial factor. For most applications, the reader can view $k$ as a large $\poly(\log n)$. That usually gives strong enough concentrations that can be union bounded over polynomial many constraints, for allowed deviations that are an $\eps$ factor of the expectation for $\eps\in [1/\poly(\log n), 1]$. Furthermore, it keeps our algorithm's computational work nearly linear in the input size and depth polylogarithmic.
\end{itemize}
Having introduced these notations, and the general setup for our derandomization results, we are now ready to state our main derandomization theorems.

\bigskip
\paragraph{Main results---Hoeffding-like concentrations} Our first result provides concentrations that are comparable to Hoeffding's inequality~\cite[Theorem 2]{hoeffding1994probability}. Recall that for independent random variables $X_1$, $X_2$, $\dots$, $X_\ell \in [0,a_j]$ and $X:=\sum_{j=1}^{\ell} X_{j}$, and given a $\Delta\geq 0$, Hoeffding shows that $\P[|X-\mathbb{E}[X]|\geq \Delta] \leq 2 \exp(- \Delta^2/(\sum_{j} a^2_j)).$ The following theorem presents our parallel derandomization analogue. 

\begin{theorem}[Hoeffding-Like Concentration]
\label{thm:hoeffding}
There exists a constant $c > 0$ such that the following holds. Let $n,m,k \in \mathbb{N}$ be arbitrary and $A \in \mathbb{R}_{\geq 0}^{m \times n}$. Also, let $\p \in [0,1]^n$ and $\Deltav \in \mathbb{R}_{>0}^m$. For each $i \in [m]$, let
\[\fail := c \exp \left(-(1/c) \min\left(\frac{\Deltai^2}{\sum_{j\in [n]} \aij^2}, \frac{\Deltai k}{\sum_{j\in [n]} \pj \aij}\right)\right).\]
There exists a deterministic parallel algorithm with $\tilde{O}(\max(nnz(A),n,m)\poly(k))$ work and $\poly(\log(nm),k)$ depth that outputs a vector $\q \in \{0,1\}^n$ and a set $\Ibad \subseteq [m]$ with $|\Ibad| \leq \sum_{i \in [m]} \fail$ such that for every $i \in [m] \setminus \Ibad$ it holds that
\[|\sum_{j \in [n]} \aij(\pj - \qj)| \leq \Deltai.\]
\end{theorem}
\smallskip

Let us provide some intuition for the provided guarantee and compare it to Hoeffding's inequality. Notice that in the definition of $\fail$, the exponent of the exponential is equal to the minimum of two terms, up to a constant. The first term exactly matches the exponent provided by Hoeffdings inequality, as reviewed above. The second term is impacted by our granularity parameter $k$. For instance, for the intuitive and frequently used parameter range where we allow deviation $\Deltav=\eps A\p$, and thus want $A\q \in (1\pm \eps) A\p$, we can simplify the second exponent term as $\frac{\Deltai k}{\sum_{j\in [n]} \pj \aij} = \eps k$. Hence, our concentration exponent is the minimum of that of Hoeffding and $\eps k$. As stated before, the main target regime for us is deviation factors $\eps \in [1/\poly(\log n), 1]$. Therefore, by setting $k = \Theta(\frac{1}{\eps} \log n) \in \poly(\log n)$, we can easily ensure that the second exponent term is $\Omega(\log m)$, which means it would be good enough to union bound over all constraints and thus it is not a limiting factor. As such, we would recover the guarantees provided by Hoeffdings inequality while keeping the algorithm's work bound $\tilde{O}(\max(nnz(A),n,m))$ and its depth  $\poly(\log(nm))$.
\bigskip

\paragraph{Main results---Chernoff-like concentrations} Our second result provides concentrations that are comparable with Chernoff's bound. See \Cref{subsec:background} for a reminder on the statement. The following theorem has concentrations that match Chernoff's in the case where the constraint coefficients are uniform. Thus, it proves the Vanilla version stated in \Cref{thm:vanilla}. This statement is presented more generally, and itself will be used for proving our Bernstein-like concentration statement (\cref{thm:variance}).  
\begin{theorem}[Chernoff-Like Concentration]
\label{thm:chernoff}
There exists a constant $c > 0$ such that the following holds. Let $n,m,k \in \mathbb{N}$ be arbitrary and $A \in \mathbb{R}_{\geq 0}^{m \times n}$. Also, let $\p \in [0,1]^n$ and $\Deltav \in \mathbb{R}_{>0}^m$. For each $i \in [m]$, let
\[\fail := c \exp \left(-(1/c) \min \left(\frac{\Deltai^2}{\sum_{j\in [n]} \pj \aij (\max_{j \in [n]} \aij)}, \frac{\Deltai}{\max_{j\in [n]} \aij}, \frac{\Deltai k}{\sum_{j\in [n]} \pj \aij} \right)\right).\]
There exists a deterministic parallel algorithm with $\tilde{O}(\max(nnz(A),n,m)\poly(k))$ work and $\poly(\log(nm),k)$ depth that outputs a vector $\q \in \{0,1\}^n$ and a set $\Ibad \subseteq [m]$ with $|\Ibad| \leq \sum_{i \in [m]} \fail$ such that for every $i \in [m] \setminus \Ibad$ it holds that
\[|\sum_{j \in [n]} \aij(\pj - \qj)| \leq \Deltai.\]
\end{theorem}
\smallskip
Again, let us discuss the provided guarantee intuitively. An interesting and simple regime to consider, which corresponds to \Cref{thm:vanilla}, is when each nonzero $a_{ij}$ is equal to $1$, and where we define $\Deltav=\eps A\p$ for $\eps\in [0,1]$. Let us focus on one constraint $i\in [m]$ and define $\mu=\sum_{j} a_{ij} p_{j}$, that is, $\mu$ is the expectation of the value of the $i^{th}$ constraint under the input random assignment. Then, in $\fail$, up to a constant factor, the concentration exponent is equal to $\min(\eps^2\mu, \eps \mu, \eps k) = \min(\eps^2\mu, \eps k)$. Notice that $\eps^2\mu$ is the familiar exponent of the Chernoff bound. As before, the main target regime for us is deviation factors $\eps \in [1/\poly(\log n), 1]$. Therefore, by setting $k = \Theta(\frac{1}{\eps} \log n) \in \poly(\log n)$, we can easily ensure that the second exponent term is $\Omega(\log m)$, which means it would be good enough to union bound over all constraints and thus it is not a limiting factor. Hence, \Cref{thm:vanilla} is simply a special case of \Cref{thm:chernoff}.
\bigskip

\paragraph{Main results---Bernstein-like concentrations} Our third result provides concentrations that are comparable with Bernstein's bound~\cite[Theorem 8.1]{dubhashi2009concentration} (also \cref{lem:concentration_inequality}), modulo a small loss. Recall that for independent random variables $X_1$, $X_2$, $\dots$, $X_\ell$ such that $X_j \in \{0,a_j\}$ and $X:=\sum_{j=1}^{\ell} X_{j}$, and given a $\Delta\geq 0$, Bernstein shows that $\P[|X-\mathbb{E}[X]|\geq \Delta] \leq 2 \exp\left(- \Delta^2/\left(4\max\left(\sigma^2, b\Delta\right)\right)\right).$ Here, $\sigma^2$ is the variance of $X$, i.e., $\sigma^2 = \sum_{j\in [\ell]} (a^2_j \cdot p_j (1-p_j))$, and $b=\max_{j\in [\ell]} a_j$. By assuming $p_j\leq 1/2$ for all $j$, we get $\sigma^2 = \Theta(\sum_{j} p_j a^2_j)$.\footnote{If $p_j>1/2$, we could do a similar simplification by looking at the complement of this binary random variable; either way we can get a result comparable to Bernstein's. Proceeding with this simplification makes our bounds slightly more readable.} Therefore, in this case, we can restate Bernstein's inequality as 
$$\P\left[|X-\mathbb{E}[X]|\geq \Delta\right] \leq 2 \exp\left(- \Theta\left(\min\left(\frac{\Delta^2}{\sum_{j\in [\ell]} p_j a^2_j}, \frac{\Delta}{\max_{j\in [\ell]} a_j}\right)\right)\right).$$
\begin{restatable}{theorem}{variance}[Bernstein-Like Concentration]
\label{thm:variance}
There exists a constant $c > 0$ such that the following holds. Let $n,m,k \in \mathbb{N}$ be arbitrary and $A \in \mathbb{R}_{\geq 0}^{m \times n}$. Also, let $\p \in [0,1]^n$ and $\Deltav \in \mathbb{R}_{>0}^m$. For each $i \in [m]$, let $\alpha_i := \left(c \cdot \log \left(\frac{\sum_{j\in [n]} \pj\aij}{\Deltai} + 2\right)\right)^{-1}$ and 

\[\fail := (1/\alpha_i) \cdot \exp\bigg( - \alpha_i \cdot \min \left( \left( \frac{\Delta^2_i}{\sum_{j \in [m]} \pj \aij^2}, \frac{\Deltai}{\max_{j \in [m]} \aij},\frac{\Deltai k}{\sum_{j \in [m]} \pj \aij} \right)\right) .\]
There exists a deterministic parallel algorithm with $\tilde{O}(\max(nnz(A),n,m)\poly(k))$ work and $\poly(\log(nm),k)$ depth that outputs a vector $\q \in \{0,1\}^n$ and a set $\Ibad \subseteq [m]$ with $|\Ibad| \leq \sum_{i \in [m]} \fail$ such that for every $i \in [m] \setminus \Ibad$ it holds that
\[|\sum_{j \in [n]} \aij(\pj - \qj)| \leq \Deltai.\]
\end{restatable}
\smallskip

The reader will notice that the concentration provided by the above theorem asymptotically matches that of Bernstein's inequality, as reviewed before, except for two differences: (1) Instead of a $c$ factor in front and a $1/c$ factor in the exponent, for constants $c$, we have a $1/\alpha_i$ factor in front and more importantly an $\alpha_i$ factor in the exponent. Here,  $\alpha_i<1$ is a loss factor, and it slightly weakens the concentration. Let us take a closer look at the extent of this weakening, focusing on the exponent in particular. A helpful setup to consider is again where we set $\Deltav=\eps A\p$ for $\eps\in [0,1]$ and allow a relative $\eps$ deviation factor in each constraint, compared to the expectation. In this case, we have $\alpha_i = \left(c \cdot \log \left(\frac{\sum_{j\in [m]} \aij\pj}{\Deltai} + 2\right)\right)^{-1} = \Theta(1/\log(1/\eps))$. Hence, the loss is only logarithmic in the inverse of the deviation factor $\eps$. For the main target regimes where $\eps$ is a positive constant or at least $1/\poly(\log n)$, this is respectively a loss of only a small constant or in the worst case an $\Theta(\log\log n)$ factor in the concentration exponent. (2) The second difference in comparison to Bernstein's inequality is the appearance of the last term in the concentration exponent. This is similar to the two previous theorems. Again, for the commonly used setting of $\Deltav=\eps A\p$ for $\eps\in [0,1]$, by setting $k=\Omega(\frac{1}{\eps} \log m)$, we can remove the effect of this term, while keeping the computational work and depth near-linear and polylogarithmic in the input size, respectively.

\subsection{Follow up work}
\label{subsec:followup}
In a follow-up work, Ghaffari and Grunau~\cite{GG2023Chernoff} show a work-efficient deterministic parallel algorithm, with polylogarithmic depth, that achieves the tight concentration for all ranges of $\eps$ where the Chernoff bound is applicable. That is, for instance, in the context of the vanilla theorem stated here (\Cref{thm:vanilla}), their algorithm will ensure that each set $S_i$ for $i\in [m]$ will have size $p|S_i| \pm O(\sqrt{p|S_i|\log m } + \log m).$ Their method is different than the one here. They present a bootstrapping method that makes black-box usage of the rougher partitioning results presented in this paper and recursively sharpens the concentration to the optimal bound, while keeping the overall algorithm nearly work-efficient.

\section{An overview of our method}
\label{sec:overview}
Our approach can be viewed as going opposite to the conventional derandomization wisdom of directly reducing the number of random bits in the randomized algorithm (and thus the space of the required randomness). Instead, we cast the task of fixing the random values as a gradual process with $\poly(\log n)$ iterations. We use only pairwise independence inside each iteration, but full independence across different iterations. Hence, the total number of bits in the random process is a higher $\poly(\log n)$, in contrast with the $O(\log^2 n)$ bits of $O(\log n)$-wise independence, or the $O(\log n)$ bits of $(1/\poly(n))$-dependent $O(\log n)$-wise independent space, which are often sufficient for the desired concentrations in randomized algorithms~\cite{motwani1989probabilistic, naor1993small}. In a sense, we will view each binary random variable as a gradual random walk, taking only a small step in each iteration in a random direction (we will use pairwise independence between the directions of different walks). Our slowed-down process gives us a lot of useful structure within each iteration: we can boil down the desired concentrations of the final random variables to some nice pairwise analyses inside each iteration. Utilizing this, we use known methods to fix the randomness of each iteration in a work-efficient manner, before proceeding to the next iteration.

Next, we provide a more concrete overview of our method, in the context of a simple concentration problem. This will be even weaker than our Vanilla result stated in \cref{thm:vanilla}, but it already delivers most of the key ideas. We then mention some of the ideas one can use to generalize from this and get \cref{thm:vanilla}. We will provide the actual proofs only for our main results, which are more general; this generality is needed in connecting different results. We do not attempt to summarize their ideas here.

\subsection{A randomized algorithm, with only pairwise analysis inside each step}
Our aim is to prove the following theorem. 

\begin{theorem}[A simple discrepancy problem that further simplifies \cref{thm:vanilla}]
\label{thm:simple}
We assume that we are given sets $S_1, \dots, S_m \subseteq [n]$ such that for each $i \in [m]$ we have $|S_i| = \Omega(\log n)$. We show how one can split the elements into three classes \texttt{YES}, \texttt{NO}, and \texttt{MAYBE} so that, for every $i \in [m]$,  set $S_i$ contains at least $|S_i|/5$ many \texttt{YES} elements and at least $|S_i|/5$ many \texttt{NO} elements, using a deterministic algorithm with work $\tilde{O}(\sum_{i \in [m]} |S_i|)$ and depth $\tilde{O}(1)$. 
\end{theorem}
In this subsection, we will show how to achieve this result with a randomized algorithm that uses $n$ pairwise independent random variables $\poly\log(n)$ times. 

\paragraph{Trivial randomized solution with full independence}
Notice that one can solve the problem from \cref{thm:simple} trivially by assigning each element a label \texttt{YES} or \texttt{NO} with probability $1/2$, using standard Chernoff bounds. However, Chernoff bounds require $\Omega(\log n)$-wise independence. Our first step is to design a randomized algorithm with $\poly\log(n)$ steps such that in every step we need only $n$ pairwise independent random bits (where we have full independence across different steps). 

A key intuition is to notice that the issue with the trivial randomized solution is that we are deciding on the value of every random bit too abruptly. For every element $i \in [n]$, we want to substitute the process of sampling the $i$-th random  \texttt{YES} / \texttt{NO} bit by a more gradual process in which, in every step, we reveal only a small amount of information about the final $i$-th random bit. The hope is that such a slower process can be analyzed only using the first and second moments inside each step, i.e., it can be made work only with pairwise independent random variables in each step. 

% \todo{maybe these things deserve to be subsections}
\paragraph{Sampling with random walks}
One way to sample one random bit gradually is via a random walk. So, let us set some granularity parameter $k$ -- later, we will see that setting $k = \Theta(\log n)$ is the right choice -- and consider the following simple randomized process of \cref{alg:randomized}. In this algorithm, we are working with values $p_{j,t}$ in every step $t$ that one should think of as our current estimate of the probability that the element $j$ gets \texttt{YES} or \texttt{NO}. In the beginning, we set these values to be $1/2$. In every subsequent step, we decide randomly in which direction we nudge each probability: in every step, we either increase it by $1/k$, or decrease it by $1/k$, where each of these two choices happens with probability $1/2$. Once the value $p_{j,t}$ becomes equal to $0$ or $1$, we stop the walk of this variable, because at this point we have managed to sample our random \texttt{YES/NO} bit. 

\begin{algorithm}[ht]
\caption{Simple randomized algorithm}
\label{alg:randomized}
{\bf Input:}  Set family $S_1, \dots, S_m$ on elements $[n]$, $k > 0$ and $k$ is even\\
{\bf Output:} Partial \texttt{YES/NO} assignment of elements $[n]$
\begin{algorithmic}
\STATE Define $T = 100k^2$. 
\STATE Define $p_{j, 0} = 1/2$ for every $j \in [m]$.    
\FOR{$t = 1, 2, \dots, T$}
    \STATE For every element $u$ define $p_{j,t} = p_{j,t-1} + X_{j,t}$ where $(X_{j, t})_{1 \le j \le n}$ are defined as follows. 
    \IF{$0 < p_{j,t-1} < 1$}
        \STATE $\P(X_{j,t} = \pm 1/k) = 1/2$, all these variables are pairwise independent. 
    \ELSE
        \STATE $X_{j,t} = 0$ deterministically.     
    \ENDIF
\ENDFOR
\STATE For every $j \in [n]$, whenever $p_{j, T} = 0$, output \texttt{NO}, whenever $p_{j,T} = 1$, output \texttt{YES}, otherwise \texttt{MAYBE}
\end{algorithmic}
\end{algorithm}

The value $T = \Theta(k^2)$ is chosen intentionally so that we can use the well-known fact that a $T$ step balanced random walk is expected to be $\Theta(\sqrt{T})$ steps away from the origin. Put differently, every element $i \in [n]$ is assigned a label \texttt{YES} or \texttt{NO} by our algorithm with probability at least $3/4 $, if $T = \Omega(k^2)$ with a sufficiently large constant in the asymptotic notation. Moreover, by the symmetrical nature of our process, we conclude that every element gets labeled \texttt{YES} with probability at least $3/8$ and the same bound holds also for the probability of \texttt{NO}. 

We can already now see that \cref{alg:randomized} solves our problem, provided that all randomness of that algorithm is fully independent instead of just pairwise independent: Since every set $S_i$ expects at least $3|S_i|/8$ \texttt{YES}es and at least $3|S_i|/8$ \texttt{NO}s, standard Chernoff bound allows us to conclude that every set gets at least $|S_i|/5$ \texttt{YES}es and at least $|S_i|/5$ \texttt{NO}s with probability $1 - 1/\poly(n)$. 

Let us now sketch how we can arrive at the same conclusion even if random variables $X_{j,t}, 1 \le j \le m$ are only pairwise independent for every fixed $t$, but we still have full independence between different steps $t$.

\paragraph{Pairwise analysis of the deviation}
To analyze the random process using pairwise analysis inside one step, we will need to analyze two random variables for each set $S_i$ that we call $\phi_{i,t}$ and $\psi_{i,t}$. First, the variable $\phi_{i,t}$ is measuring the current deviation from the initial perfectly balanced setup in every step $t$. Formally, 
\begin{align*}
    \phi_{i,t} = \sum_{j \in S_i} p_{j, t}. 
\end{align*}
Our aim is to show that $|S_i|/3 \le \phi_{i,T} \le  2|S_i|/3$ with probability $1 - 1/\poly(n)$. This can be seen by the martingale version of Bernstein's inequality. More concretely, we start by noting that $\phi_{i,t}$ is a martingale, i.e., $\E_{\text{step $t$}}[\phi_{i, t}] = \phi_{i,t-1}$. Moreover, even with pairwise independence, we can compute the variance $\Var\left[\phi_{i,t} - \phi_{i,t-1} | \phi_{i,t-1} \right]$. Since $\phi_{i,t} - \phi_{i,t-1} = \sum_{j \in S_i} X_{j, t}$, we can use our definition of random variables $X_{j,t}$, together with pairwise independence, to conclude that
\begin{align*}
    \Var\left[ \sum_{j \in S_i} X_{j, t}\right] \le |S_i|/k^2. 
\end{align*}
We note that this bound is much better than the worst case possible change of $\phi_{i,t} - \phi_{i,t-1} $ which can be up to $|S_i|/k$. We now have enough information to plug in inside Bernstein's inequality from \cref{lem:concentration_inequality}; we get:
\begin{align}
\label{eq:rand1}
    \P\left( |\phi_{i,T} - |S_i|/2| > |S_i|/6\right)
    &\le 2\exp\left( - \frac{(|S_i|/6)^2}{4\max \left( 100k^2 \cdot |S_i|/k^2 ,  |S_i|/6 \cdot |S_i|/k \right) } \right) \\ \nonumber
    &= \exp\left(-\Omega(\min(|S_i|,k)) \right)
\end{align}
Choosing $k,S = \Omega(\log n)$ with a large enough constant gives us $1-1/\poly(n)$ probability that the final deviation is small. The reason for Bernstein's inequality is that applying Chebyshev would result only in a much weaker concentration of $1 - 1/\poly(k, |S|)$ that we cannot union bound over. Bernstein's inequality very fittingly relies only on a second moment bound in every step but leverages the independence of different steps to improve exponentially upon Chebyshev's inequality. 

\paragraph{Pairwise analysis of the number of fixed variables}
Understanding the variable $\phi_{i, T}$ would not be useful if none of the variables $p_{j, T}$ are $0$ or $1$ at the end of the algorithm. Thus, in the second part of our analysis, we need to argue that a constant proportion of random variables is going to be set to \texttt{YES} or \texttt{NO}. Somehow, we need an argument generalizing the fact that the random walk is spreading to distance $\sqrt{T}$ after $T$ steps. The simplest proofs of such results analyze the squared distance from the origin, and we will similarly look at the following, quadratic, expression that measures the ``spread'' of random variables in one set $S_i$. We define:
\begin{align}
\label{def:smallpsi}
    \psi_{i, t} = \sum_{j_1, j_2 \in S_i^2} (p_{j_1, t} - p_{j_2, t})^2
\end{align}
This random variable is equal to $0$ at the beginning and we expect it to grow over time. More precisely, a quick computation relying on pairwise independence reveals that
\begin{align*}
    \E[\psi_{i,t} -  \psi_{i, t-1} | \psi_{i, t-1}] \ge |S_i||S_i^t|/(4k^2)
\end{align*}
Here, $S_i^t \subseteq S_i$ is the subset of elements of $S_i$ that are still moving (are not set to $0$ or $1$) in step $t$. In particular, once $S_i^t = \emptyset$, the random variable $\psi$ stops growing. However, while at least, say, $|S_i|/10$ variables are still moving,  we expect $\psi$ to increase by $|S_i|^2/(40k^2)$ in every step. Multiply this by $T = 100k^2$ steps and we would expect $\psi$ to be larger than $2|S_i|^2$ at the end of the algorithm, which is not possible since the definition of $\psi_{i,t}$ implies it is always at most $|S_i|^2$.  

We thus need to prove that the value of $\psi_{i,T}$ is sufficiently concentrated around its mean, which we will again do using Bernstein's inequality. To do so, we need to compute an upper bound on the variance of $\psi_{i,t} - \psi_{i, t-1}$: using only pairwise independence (we note that this time we need to consider expressions that are fourth order terms in $X_{j,t}$), one can compute that $\Var[\psi_{i,t} - \psi_{i, t-1} | \psi_{i, t-1}] \le 100(|S_i|^3/k^2 + |S_i|^4/k^3)$ which is for large $k$ substantially better than the worst case possible fluctuation $\psi_{i,t} - \psi_{i,t-1}$ which can be up to $4|S_i|^2/k$. We plug in to Bernstein's inequality \cref{lem:concentration_inequality} again:
\begin{align}
\label{eq:rand_second}
    \P(|S_i^T| > |S_i|/10) 
    &\le \P(\psi_{i, T} < \E[\psi_{i, T}] - |S_i|^2)\\ 
    &\le \exp\left( -\frac{\left( |S_i|^2 \right)^2}{4 \max \left( 100k^2 \cdot  100(|S_i|^3/k^2 + |S_i|^4/k^3), |S_i|^2\cdot 4|S_i|^2/k \right) } \right) \\
    &= \exp \left( -\Omega(\min(|S_i|, k)) \right)
\end{align}
Choosing $k,S = \Omega(\log n)$ with large enough constant gives us $1 - 1/\poly(n)$ probability that only at most a $1/10$ fraction of elements of $S_i$ end up as \texttt{MAYBE}. Putting this together with the bounds on $\phi_{i,T}$, we conclude that at least $|S_i|/3 - |S_i|/10 > |S_i|/5$ elements are \texttt{YES}es and the same holds for \texttt{NO}s, which is what we wanted to prove. 

\subsection{A sequential derandomization}
Before explaining how we derandomize \cref{alg:randomized} in the parallel setting, let us make a detour and explain how one can solve the problem from \cref{thm:simple} with a simple sequential deterministic algorithm. A parallel variant of this approach will be a part of our technique, and it is easier to understand the approach by first getting familiar with the methodology and the notations in this much simpler sequential setup.

\paragraph{Sequential derandomization of the concentration bounds}
 This sequential algorithm directly derandomizes the trivial randomized solution of sampling the value of each element randomly. This is using a certain method of conditional expectations~\cite{raghavan1986probabilistic}, with pessimistic estimators coming from the proof of the Chernoff bound. In particular, we determine the values of random variables $X_1, X_2, \dots, X_n \in \{-1, +1\}$, where $X_j = 1$ iff the element $j$ is in class \texttt{YES}, one by one, while maintaining a certain potential function that we discuss next. 

At the time step $t$ where we have already fixed $X_1=x_1, X_2=x_2, \dots, X_t=x_t$, for every set $S_i$, we keep track of the following two potentials, $\Phi^{(1)}_{i, t}$ and $\Phi^{(2)}_{i, t}$. 

\begin{align}
\label{eq:introphi}
\Phi^{(1)}_{i, t} = \prod_{\substack{1 \le t' \le t \\ t' \in S_i}} \left( 1 + \lambda x_{t'} + \lambda^2 x_{t'}^2 \right) \cdot \left( 1 +  \lambda^2\right)^{|S_i \cap [t+1, n]|}
\end{align}
\begin{align}
\label{eq:introphi2}
\Phi^{(2)}_{i, t} = \prod_{\substack{1 \le t' \le t \\ t' \in S_i}} \left( 1 - \lambda x_{t'} + \lambda^2 x_{t'}^2 \right) \cdot \left( 1 +  \lambda^2 \right)^{|S_i \cap [t+1, n]|}
\end{align}
for
\begin{align*}
    \lambda = 1/10. 
\end{align*}
These potentials can be seen as corresponding to the ingredients of proof of a simple version of the standard Chernoff bound~\cite{dubhashi2009concentration}: To understand the connection, note that one way to derive a version of Chernoff bound for a sum of independent random variables $X_1, \dots, X_n$, where $\P(X_j = \pm 1) = 1/2$, is to consider the expression $\prod_{j \in [n]} \e^{\lambda X_j}$ and apply Markov inequality to its expectation. Importantly, to (approximately) compute the value of $\E\left[\prod_{j \in [n]} \e^{\lambda X_j}\right] = \prod_{j \in [n]} \E\left[ \e^{\lambda X_j}\right]$, it is enough to approximate each partial term in the multiplication up to second order terms, i.e., the proof uses $\e^{\lambda X_j} \le 1 + \lambda X_j + \lambda^2 X_j^2$. This quadratic approximation is what we see in \cref{eq:introphi,eq:introphi2}. The right terms of \cref{eq:introphi,eq:introphi2} are simply the values of $\E[1 \pm \lambda X_j +\lambda X_j^2]$ for variables $X_{t+1}, X_{t+2}, \dots, X_n$ which are still uniformly random in $\{-1, +1\}$ and remain to be decided. 

We also define an overall potential $\Pot_t$ that corresponds to the fact that in the end, we want to union bound over all bad events that can happen (each set contributes two bad events: too many \texttt{YES}es or too many \texttt{NO}s). At time step $t$, the overall potential is a normalized summation of the individual potentials, concretely defined as follows:
\begin{align*}
    \Pot_t = \frac{1}{2n} \sum_{i \in [m]}  \left( \frac{\Phi^{(1)}_{i, t}}{\Phi^{(1)}_{i, 0}} + \frac{\Phi^{(2)}_{i, t}}{\Phi^{(2)}_{i, 0}}\right)
\end{align*}

Our deterministic algorithm will simply iterate over the elements of $[n]$ and for each one of them it sets the value of $X_t$ to $-1$ or $1$ so that $\Pot_t \le \Pot_{t-1}$. Let us now argue that it can always be done and that at the end, no bad event occurs.  
\begin{enumerate}
    \item \textbf{The overall potential $\Pot_t$ is monotonically non-increasing:} We start by noting that $\E[1 \pm \lambda x_{t'} + \lambda^2 x_{t'}^2 ] = 1 \pm 0 + \lambda^2$. This implies that $\E_{X_t}[\Pot_t] = \Pot_{t-1}$, hence one of the two values of $X_j$ leads to $\Pot_t \le \Pot_{t-1}$. 
    \item \textbf{No bad event happens:}
    Note that we set up our potential in such a way that $\Pot_0 = 1$. Our task is to prove that a bad event would lead to $\Pot_T > 1$. To see this, assume that the set $S_j$ has less than $|S_j|/3$ \texttt{NO}s at the end of the algorithm (the \texttt{YES}es case is analogous). Recall the convention that \texttt{NO} corresponds to $-1$ and \texttt{YES} corresponds to $+1$. We then consider the corresponding potential $\Phi^{(1)}_{i, t}$. Using the fact that $1 + z + z^2 \ge \e^z$ for $z=\lambda x_{t'} \le 1$, for every $t'$, we can compute that 
    \begin{align*}
    \Phi^{(1)}_{i, n}
        \ge \prod_{1 \le t' \le n} \left(1 + \lambda x_{t'} + \lambda^2 x_{t'}^2\right)
        \ge \prod_{1 \le t' \le n} \e^{\lambda x_{t'}}
        = \e^{\lambda \sum_{j \in S_i} x_j } 
        \ge \e^{ |S_i|/30}
    \end{align*}
    On the other hand, we compute that
    \begin{align*}
        \Phi^{(1)}_{i, 0} = \left( 1 + \lambda^2\right)^{|S_i|} \le \e^{|S_i|/100}
    \end{align*}
    We conclude that $\Phi^{(1)}_{i, T} / \Phi^{(1)}_{i, 0} > \e^{|S_i|/50} > 3n$, by using the assumption that $S_i = \Omega(\log n)$ with large enough constant, which implies $\Pot_{T}>3/2$. Hence, if we had less than $|S_j|/3$ \texttt{NO}s in $S_j$, we would have $\Pot_{T}>3/2$, but we explicitly maintained that $\Pot_T\leq \Pot_{T-1}\leq \dots \leq \Pot_0=1$.
\end{enumerate}

\subsection{Parallel derandomization, using pairwise independence per step}
Let us now go back to our task -- work-efficient parallel derandomization of the randomized process in \cref{alg:randomized} that proves \cref{thm:simple}. 
We do this using two ingredients. The first one is the method discussed above to derandomize Chernoff bounds; we apply it to derandomize Bernstein's inequality used in the analysis of \cref{alg:randomized}. This argument handles the part of the randomized argument that depends on the independence of random variables across different steps of the process. Crucially, the randomized process has only $T = O(k^2) = O(\log^2 n)$ steps, so using the standard derandomization idea results in only $\poly\log(n)$ loss in depth. In contrast, the setup of the above with the naive sequential Chernoff bound derandomization would require $\Omega(n)$ depth and would be mostly uninteresting from a parallel derandomization viewpoint. 

One notable difference with the sequential Chernoff bound derandomization is this: in the sequential derandomization of Chernoff, in every step, we only needed to decide on two possible values of a binary random variable. In our parallel derandomization, one step of our process consists of assigning values to $n$ pairwise independent random variables. To handle this part of the argument for randomness inside one step of the algorithm, we make use of Luby's work-efficient derandomization technique for nice pairwise independent analyses~\cite{luby1988removing}, as we state in \cref{lem:pairwise}. Roughly speaking, this result says that whenever we work with quadratic expression $Q(X_1, \dots, X_n)$ over binary pairwise independent random variables $X_1, \dots, X_n$, which satisfy certain niceness conditions as formalized in \Cref{def:quadratic_term}, we can find an instantiation of the variables $x_1, \dots, x_n$ such that $Q(x_1, \dots, x_n) \le \E[Q(X_1, \dots, X_n)]$ in small work and depth. In fact, the result of Luby is why we work with pairwise independent random variables in the first place. 

Let us now write down the potential function that our deterministic algorithm keeps monotonically non-increasing. This is similar to the Chernoff bound analysis we have already seen and mimics the standard proofs of Bernstein's inequality (see \cite[Section 1.7]{dubhashi2009concentration}).  
%, but we note that there is an additional multiplier $\lambda$ in all expressions. This is the same multiplier used in most proofs of nontrivial concentration inequalities. In the proof of Bernstein's inequality for a martingale $Z_1, \dots, Z_T$, we set $\lambda = \eps \cdot \min(1/W, \Delta/V)$, where $\eps$ is some sufficiently small constant, $W$ is the maximum possible change $Z_t - Z_{t-1}$ for any $t$, and $V$ is the maximum possible value of $\Var\left[Z_t - Z_{t-1} | Z_{t-1}\right]$ for any $t$. 
We define
\begin{align}
\label{detphi}
\Phi^{(1)}_{i, t} = \prod_{1 \le t' \le t} \left( 1 + \lambda_i \sum_{j \in S_i} x_{j, t'} + \left( \lambda_i \sum_{j \in S_i} x_{j, t'} \right)^2 \right) \cdot \left( 1 + \lambda_i^2 |S_i|/k^2 \right)^{T - t}
\end{align}
\begin{align}
\label{detphi2}
\Phi^{(2)}_{i, t} = \prod_{1 \le t' \le t} \left( 1 - \lambda_i \sum_{j \in S_i} x_{j, t'} + \left( \lambda_i \sum_{j \in S_i} x_{j, t'} \right)^2 \right) \cdot \left( 1 + \lambda_i^2 |S_i|/k^2 \right)^{T - t} 
\end{align}
with
\begin{align*}
    \lambda_i = \frac{1}{10^{10}} \min\left( 1, \frac{k}{|S_i|} \right)
\end{align*}
%\todo{Mohsen: it would be good to relate these to the pairwise analysis, especially saying something about where $\Psi$ came from.}
Next, we define
\begin{align}
\label{eq:detpsi}
\Psi_{i, t} = 
&\prod_{1 \le t' \le t} \left( 1 - \lambda'_i \left( \sum_{j_1, j_2 \in S_i^2} (p_{j_1, t} - p_{j_2, t})^2 \right) + \lambda^{\prime 2}_i \left( \sum_{j_1, j_2 \in S_i^2} (p_{j_1, t} - p_{j_2, t})^2 \right)^2 \right) \\ \nonumber
&\cdot \left( 1 - \lambda'_i \cdot \frac{|S_i|^2}{100k^2} + \lambda^{\prime 2}_i \cdot 100\left( \frac{|S_i|^3}{k^2} + \frac{|S_i|^4}{k^3} \right)^2\right)^{T - t}     
\end{align}
with 
\begin{align*}
    \psi_{i, t} = \sum_{j_1, j_2 \in S_i^2} (p_{j_1, t} - p_{j_2, t})^2
\end{align*}
analogously to \cref{def:smallpsi}
and
\begin{align*}
    \lambda_i' = \frac{1}{10^{10}} \min\left( \frac{1}{|S_i|}, \frac{k}{|S|^2} \right).
\end{align*}
Finally, we again define the overall potential to be the normalized sum of all partial potentials:
\begin{align}
\label{detfin}
    \Pot_t = \frac{1}{3m} \sum_{i \in [m]} \left( \frac{\Phi^{(1)}_{i, t}}{\Phi^{(1)}_{i, 0}} + \frac{\Phi^{(2)}_{i, t}}{\Phi^{(2)}_{i, 0}} + \frac{\Psi_{i, t}}{\Psi_{i, 0}} \right). 
\end{align}

We can now analyze the deterministic algorithm that applies Luby's pairwise derandomization in every step $t$ to find values $x_{j, t}$ for every $j \in [n]$ such that $\Pot_t(x_{1, t}, \dots, x_{n, t}) \le \E[\Pot_t(X_1, \dots, X_n)]$. This turns out to be a little bit more technical because \cref{eq:detpsi} includes the term $\left( \sum_{j_1, j_2 \in S_i^2} (p_{j_1, t} - p_{j_2, t})^2 \right)^2$, which involves the fourth powers in the variables $x_{1, t}, \dots, x_{n, t}$. This would seem to make the pairwise derandomization inapplicable. However, it is possible to upper bound this term with a certain expression quadratic in $x_{1, n}, \dots, x_{n, t}$, so the actual definition of $\Psi$ in \cref{def:Psi} in the formal proof uses this quadratic expression instead. 

\begin{enumerate}
    \item First, we need to prove that $\Pot_t$ is monotonically non-increasing, i.e., we always have $\Pot_t \le \Pot_{t-1}$. This boils down to computing that the first and the second moments of $\sum_{j \in S_i} x_{j, t}$ are $0$ and $|S_i|/k^2$, respectively in the case of $\Phi^{(1)}$ and $\Phi^{(2)}$, and that the first and the second moment of $\sum_{j_1, j_2 \in S_i^2} (p_{j_1, t} - p_{j_2, t})^2 $ are at most $\frac{|S_i|^2}{100k^2}$ and $100\left( \frac{|S_i|^3}{k^2} + \frac{|S_i|^4}{k^3} \right)$, respectively. This computation was already discussed in the analysis of the randomized algorithm. 
    \item Next, we prove that whenever a bad event happens, we have $\Pot_t > 1$ which contradicts $\Pot_0 = 1$.  Here, we discuss this only for the case of the bad event of having one set $S_i$ with $\phi_{i,T} - |S_i|/2 > |S_i|/6$ (cf. \cref{eq:rand1}). Such an event implies
    \begin{align*}
        \Phi_{i,T}^{(1)}
        &=\prod_{1 \le t' \le T}\left( 1 + \lambda_i \sum_{j \in S_i} x_{j, t'} + \left( \lambda_i \sum_{j \in S_i} x_{j, t'} \right)^2 \right)\\
        &\ge \prod_{1 \le t' \le T} \e^{\lambda_i \sum_{j \in S_i} x_{j, t'}} && \text{\footnotesize{\% As $z=\lambda_i \sum_{j \in S_i} x_{j, t'} \leq 1$ and thus $1 + z + z^2 \ge \e^z$.}} \\
        &= \exp\left\{\lambda_i \sum_{1 \le t' \le T} \sum_{j \in S_i} x_{j, t'}\right\}
        = \exp\left\{ \lambda_i \left( \phi_{i,t} - |S_i|/2 \right)\right\}\\
        &\ge \exp\left\{\lambda_i |S_i|/6\right\}
        = \exp\left\{\Omega(\min(|S_i|, k))\right\}
    \end{align*}
    On the other hand, we have
    \begin{align*}
        \Phi_{i, 0}^{(1)}
        = \left( 1 + \lambda_i^2|S_i|/k^2\right)^{T}
        \le \exp\left\{T \cdot \lambda_i^2|S_i|/k^2\right\}
        = \exp\left\{O(\min(|S_i|, k^2/|S_i|)\right\}
    \end{align*}
    Using the fact that the constant in the definition of $\lambda_i$ is chosen to be small enough, together with the fact that $\min(|S_i|, k^2/|S_i|) \le \min(|S_i|, k)$, we conclude that $\Phi_{i,T}^{(1)} / \Phi_{i, 0}^{(1)} = \exp\left\{\Omega(\min(|S_i|, k))\right\} > 3m$ provided that we choose $k, |S_i| = \Omega(\log n)$ with large enough constant. The computation for the potentials $\Phi^{(2)}$ is the same and for $\Psi$ it is similar. 
\end{enumerate}
This finishes the proof sketch of \cref{thm:simple}. 

\subsection{Generalizing to prove the vanilla derandomization} We next discuss some of the ideas that would help to finish a proof of \cref{thm:vanilla} (beyond the sketch provided above for \cref{thm:simple}). %

We review the additional ideas one can use to prove \cref{thm:vanilla} in three steps. First, we need to look back at the above proof sketch of \cref{thm:simple} and note that, with some work, it can prove a more general statement. Second, we sketch how to handle the problem that our argument keeps some values nonintegral (labeled as \texttt{MAYBE} for the purposes of \cref{thm:simple}). Third, we show how one can deal with sampling probabilities $p \ll 1/2$ by repeatedly ``subsampling'' with probability $1/2$.

As the first step, we note that the argument for the above proof of \cref{thm:simple} can be extended. In particular, the argument allows for sublinear deviations between the \texttt{YES}es and \texttt{NO}s. Also, we do not need to require that the starting probability of every element is $1/2$, but an arbitrary value. With these extensions, one can get the following more general statement. 
\begin{lemma}[Simplified and unweighted version of the Partial Fixing Lemma (\cref{lem:main})]
\label{lem:easy_partial_fixing}
Let $n,m, \in \mathbb{N}$ with $m = \poly(n)$. Also, let $S_1,S_2,\ldots,S_m \subseteq [n]$, $\p \in \left[0,1\right]^n$ and $\Deltav \in \mathbb{R}^m_{>0}$ such that for every $i \in [m]$, 
\[\Deltai \geq \Omega(\max(\sqrt{|S_i| \log(n)}, \frac{1}{\poly(\log n)}|S_i|)).\]

There exists a deterministic parallel algorithm with $\tilde{O}(n + \sum_{i \in [n]} |S_i|)$ work and $\poly(\log n)$ depth that outputs a vector $\q \in [0,1]^n$ such that for every $i \in [m]$,
\[|\sum_{j \in {S_i}}(\pj - \qj)| \leq \Deltai\]
and
\[|\{j \in S_i \colon \qj \notin \{0,1\}\}| \leq 0.9 |S_i|.\]
\end{lemma}
%Formally, we can prove the following lemma, which is a simplified version of our main technical result \cref{lem:main}.

As the second step, we now show how we can repeatedly apply \cref{lem:easy_partial_fixing} to get the same statement as in \cref{lem:easy_partial_fixing} but where $\qj \in \{0,1\}$ for every $j \in [n]$ instead of the weaker condition $|\{j \in S_i \colon \qj \notin \{0,1\}\}| \leq 0.9 |S_i|$. 
To do this, we use the fact that we can repeatedly apply \cref{lem:easy_partial_fixing} to compute a sequence of vectors $\p^{(0)} := \p, \p^{(1)},\p^{(2)},\ldots$ where $\p^{(\ell + 1)}$ is computed by using the algorithm of \cref{lem:easy_partial_fixing} with $\p^{(\ell)}$ as input, ignoring all 0/1 entries in $\p^{(\ell)}$. Moreover, we set $\Delta_i^{(\ell )} = 0.99^{\ell} \cdot \Delta_i$. 
This way, we inductively get
\begin{align*}
|\{j \in S_i \colon \pj^{(\ell)}\notin \{0,1\}\}| \leq 0.9^\ell |S_i|    
\end{align*}
which implies we are done after $O(\log n)$ invocations of the lemma. Here, we are relying on the fact that the rate in which the number of nonintegral elements of $S_i$ shrinks is much more rapid than the rate in which $\Delta_i$ shrinks, which implies that for every invocation $\ell$ of \cref{lem:easy_partial_fixing} we have
\begin{align*}
    \Deltai^{\ell} \geq \Omega\left(\max(\sqrt{|S_i^{\ell}| \log(n)}, \frac{1}{\poly(\log n)}|S_i^\ell|)\right), 
\end{align*}
and thus the lemma is always applicable. The total deviation we incur by all the calls to the lemma can be computed as the summation
\begin{align*}
    \Delta_i + 0.99 \Delta_i + 0.99^2\Delta_i + \dots = O(\Delta_i). 
\end{align*}
Hence, the final result of our new algorithm indeed satisfies all requirements of \cref{lem:easy_partial_fixing}, up to constant factors; with the additional improvement that now all values of $\q$ are integral. 

%The only remaining step that needs to be sketched is 
We now discuss the third step. We informally sketched that by iteratively applying \cref{lem:easy_partial_fixing}, we can obtain the same guarantees as in \cref{lem:easy_partial_fixing} and where additionally $\q$ is integral. 
However, even this stronger version is not sufficient to prove our vanilla derandomization result \cref{thm:vanilla} - this would require us to replace the condition $\Deltai = \Omega(\max(\sqrt{|S_i|\log(n)},\frac{1}{\poly(\log n)}|S_i|)$ with the more relaxed condition $\Deltai = \Omega(\max(\sqrt{\mui\log(n)},\frac{1}{\poly(\log n)}\mui)$, where $\mui := \sum_{j \in S_i} \pj$. Note that the conditions are equivalent as long as $\pj \geq 1/2$ for every $j \in [n]$ but for small sampling probabilities, the former condition is much stricter. To fix this, we consider repeatedly using \cref{lem:easy_partial_fixing} to ``subsample'' with probability $1/2$. To illustrate the idea, we now give a proof sketch of \cref{thm:vanilla} using the integral version of \cref{lem:easy_partial_fixing}. Recall that the input to \cref{thm:vanilla} is a uniform sampling probability $p \in [0,1]$ and a precision parameter $\eps \in [1/\poly\log n,1]$. Given that each set satisfies $|S_i| = \Omega\left(\frac{\log n}{p \eps^2}\right)$, the goal is to compute a set $T \subseteq [n]$ such that $|S_i \cap T| \in (1\pm \eps)p|S_i|$ for every $i \in [m]$.

For simplicity, we assume that $p = \frac{1}{2^R} \geq \frac{1}{n}$ for some $R$. 
We compute a sequence of sets $T_0 := [n] \supseteq T_1 \supseteq \ldots \supseteq T_R := T$.
Let $\eps_\ell = \frac{\max(0.9^{R - \ell},1/\poly(\log n)}{100}\eps$.
We compute $T_{\ell + 1}$ from $T_\ell$ by using the integral version of \cref{lem:easy_partial_fixing} as follows:
We choose $T_\ell$ as our ground set and consider the sets $S_{i,\ell} := S_i \cap T_\ell$ for every $i \in [m]$. We set $p_j = 1/2$ for every $j \in T_\ell$ and allow a deviation of $\Deltail = \epsil|S_{i,\ell}|$. 
%%%%%%%%%%%%%%%
% In order to apply \cref{lem:easy_partial_fixing}, we have to verify that
%
% \[\Deltail = \Omega \left(\max(\sqrt{|S_{i,\ell}\log n},\frac{1}{poly(\log n)}|S_i,\ell|) \right).\]
%
% The first condition follows from \todo{Should we spell out the calculation or just says it follows from this and that?}
% \[\Delta_{i,\ell} = \Omega \left( 0.9^{R - \ell} \eps |S_{i,\ell}|\right) = \Omega \left( 0.9^{R - \ell} \eps \sqrt{S_{i,\ell}} \sqrt{\frac{|S_i|}{2^\ell}}\right) = \Omega(),\]
% where the last inequality follows from 
%
%
% \todo{verify conditions of \cref{lem:easy_partial_fixing} are satisfied} 
% %%%%%%%%%%%%%%%
Therefore, we can compute a set $T_{\ell + 1}$ satisfying
\[|S_i \cap T_{\ell + 1}| \in (1\pm \eps_\ell)\frac{1}{2}|S_i \cap T_\ell|.\]
Setting $T= T_R$, a simple induction then gives
\[|S_i \cap T| \in \prod_{\ell \in [1,T-1]} (1\pm \eps_\ell)\frac{1}{2^T}|S_i| \in (1+\eps)p|S_i|,\]
and thus $T$ satisfies the condition of \cref{thm:vanilla}.

% \newpage

\section{Preliminaries}
\paragraph{Notation}
We use the following notation to sum over set products: We use $\sum_{(j_1,j_2) \in B^2}$ to denote summation over all $|B|^2$ pairs $j_1,j_2$ with $j_1 \in B, j_2 \in B$. We also sometimes index matrices and vectors by sets. For example, we write $A \in \mathbb{R}^{I \times J}$ if $A$ has one row corresponding to each $i \in I$ and one column corresponding to each $j \in J$.

\begin{fact}
\label{fact:exponential}
$1 + x + x^2 \ge \e^{x}$ for all $x \le 1$. 
\end{fact}

\paragraph{Recalling Bernstein's inequality} Next, we state a version of Bernstein's inequality. This inequality is a crucial tool to analyze \cref{alg:randomized}. We then ``white box'' the proof of this inequality inside our main technical lemma, \cref{lem:main}. After various reductions, we then get its parallel derandomization in \cref{thm:variance}. 

\begin{lemma}[Bernstein's inequality for martingales, a variant of Theorem 8.1 from \cite{dubhashi2009concentration}]
\label{lem:concentration_inequality}
Let $Z_0, Z_1, \dots, Z_T$ be a martingale satisfying the conditions
\[
\max_{1 \le t \le T}|Z_t - Z_{t-1}| \le c
\]
and
\[
\max_{1 \le t \le T} \sup \Var[ Z_t - Z_{t-1} | Z_{t-1}] \le v. 
\]
for some $c,v > 0$.
For any $\Delta > 0$ we have
\[
\P(Z_T \ge Z_0 + \Delta) \le \exp\left( - \frac{\Delta^2}{4\max(T \cdot v , \Delta \cdot c)}\right) 
\]
and
\[
\P(Z_T \le Z_0 - \Delta) \le \exp\left( - \frac{\Delta^2}{4\max(T \cdot v , \Delta \cdot c)}\right) 
\]
Moreover, the first inequality also holds if $(Z_t)$ is a submartingale and the second inequality holds also if $(Z_t)$ is a supermartingale. 
\end{lemma}

\paragraph{Work-efficient derandomization for nice quadratic functions}
Here, we recall Luby's method~\cite{luby1988removing} for work-efficient derandomization of nice pairwise analysis. We comment that our terminology is slightly different.
\label{subsec:lubyEfficientPairwise}
\begin{definition}
\label{def:quadratic_term}
We say that a function $Q(x_1, \dots, x_n)$ is a \emph{nice quadratic term} if it can be written as
\begin{align}
    Q(x_1, \dots, x_n) = \left( \sum_{i \in A} \alpha_i x_i \right) \left( \sum_{j \in B} \beta_j x_j \right) + \sum_{i' \in C} \gamma_{i'} x_{i'} + \delta
\end{align}
for some sets $A, B, C \subseteq [n]$ and $\alpha_i,\beta_j, \gamma_{i'}, \delta \in \R$ for every $i \in A, j\in B, i'\in C$. Moreover, the \emph{complexity} of a nice quadratic term $||Q(x_1, \dots, x_n)||$ is defined as 
\begin{align}
    ||Q(x_1, \dots, x_n)|| = |A| + |B| + |C| + 1. 
\end{align}
%and we use $W(Q(x_1, \dots, x_n))$ for the amount of work such that one can compute  all constants $(\alpha_i)_{i \in A}, (\beta_j)_{ j \in B}, (\gamma_{i'})_{i'\in C}, \delta$ with $W(Q_k)$ work and $\poly\log(\max(n, W(Q_k)))$ depth. 
Finally, whenever 
\begin{align}
    f(x_1, \dots, x_n) = \sum_{k = 1}^{K} Q_k(x_1, \dots, x_n)
\end{align}
we write
\begin{align}
    ||f(x_1, \dots, x_n)|| = \sum_{k = 1}^{K} || Q_k(x_1, \dots, x_n)||%\; , && 
%    W(f(x_1, \dots, x_n)) = \sum_{k = 1}^{K} W( Q_k(x_1, \dots, x_n)). 
\end{align}
\end{definition}

\begin{lemma} [\textbf{Luby's derandomization for nice quadratic functions}]
\label{lem:pairwise}
Let $f(x_1, \dots, x_n)$ be a function that can be written as 
\begin{align}
    f(x_1, \dots, x_n) = \sum_{k = 1}^{K} Q_k(x_1, \dots, x_n)
\end{align}
where each $Q_k(x_1, \dots, x_n)$ is a nice quadratic term from \cref{def:quadratic_term}. 
%Moreover, for each $1 \le k \le K$, let $W(Q_k)$ be such that all constants $(\alpha_i)_{i \in A}, (\beta_j)_{ j \in B}, (\gamma_{i'})_{i'\in C}, \delta$ can be computed in $W(Q_k)$ work and $\poly\log(n)$ depth.  
Then, there exist a probability space $\Omega$ with pairwise independent binary random variables $X_1, \dots, X_n$ on it, and an algorithm with work 
\begin{align}
 ||f(x_1, \dots, x_n)||  \cdot \poly\log(n)    
\end{align}
and depth $\poly( ||f(x_1, \dots, x_n)||,\log n)$ that produces values $\bar{x}_1, \dots, \bar{x}_n \in \{-1,1\}$ such that
\begin{align}
    f(\bar{x}_1, \dots, \bar{x}_n) \le \E\left[ f(X_1, \dots, X_n) \right]. 
\end{align}
Here, we assume that we get all constants of all nice quadratic terms $Q_k$ as part of the input. 
\end{lemma}
The result is implicit in the work of ~\cite{luby1988removing} and has been used under different phrasings. See \cite[Section 3.2]{berger1994efficient} or Harris~\cite[Section 2]{harris2019deterministic}. We provide a proof sketch in \Cref{app:pairwise}.

\section{Proof of the key lemma}
\label{sec:partialfixing}

This section is devoted to the proof of \cref{lem:main} which is the basis for our results discussed in \cref{sec:results}. This lemma generalizes \cref{thm:simple} and \cref{thm:vanilla} whose proof was discussed in \cref{sec:overview}. 

\begin{restatable}{lemma}{partialfixing}[\textbf{The Partial Fixing Lemma}]
\label{lem:main}
There exists a constant $c > 0$ such that the following holds. Let $n,m,k \in \mathbb{N}$ be arbitrary and $A \in \mathbb{R}_{\geq 0}^{m \times n}$. Also, let $\p \in \{0,1/k,2/k,\ldots,1\}^n$ and $\Deltav \in \mathbb{R}_{>0}^m$. For each $i \in [m]$, let

\begin{align}
\label{eq:wind-up}
prob^{bad,L\ref{lem:main}}_i := c\exp\left( - (1/c)\min \left\{  \frac{\Delta^2_i}{\sum_{j \in [n]} a^2_{ij}},\frac{\Delta_i k}{\sum_{j \in [n]} \aij}  \right\} \right).    
\end{align}

There exists a deterministic PRAM algorithm with $\tilde{O}(\max(nnz(A),n,m)\poly(k))$ work and $\poly(\log(nm),k)$ depth that outputs for each $i \in [m]$ a set of indices $I^{ignore}_i \subseteq [n]$ with $\sum_{j \in I^{ignore}_i} \aij \leq \frac{\Delta_i}{1000}$, a set $I^{bad} \subseteq [m]$ with $|I^{bad}| \leq \sum_{i \in [m]} prob_i^{bad,L\ref{lem:main}}$ and a vector $\q \in [0,1]^n$ such that for $J^{non-int} := \{j \in [n] \colon \qj \notin \{0,1\}\}$, it holds for every $i \in [m] \setminus I^{bad}$ that

\begin{enumerate}
\item $\sum_{j \in J^{non-int} \setminus I^{ignore}_i} \aij \leq 0.99 \sum_{j \in [n]} \aij$, \label{op1}
\item $\sum_{j \in J^{non-int} \setminus I^{ignore}_i} a^2_{ij} \leq 0.99\sum_{j \in [n]} \aij^2$ and\label{op2}
\item $|\sum_{j \in [n]} \aij(\pj - \qj)| \leq \frac{\Delta_i}{1000}$.\label{op3}
\end{enumerate}

\end{restatable}

\paragraph{Changes between \cref{lem:main} and the proof sketch from \cref{sec:overview}}
\cref{lem:main} generalizes the analysis sketch of \cref{thm:simple} from \cref{sec:overview}. In particular, in \cref{def:Phi1,def:Phi2,def:Psi,eq:pot_definition} we are going to define a potential that generalizes the definitions of potentials \cref{detphi,detphi2,eq:detpsi,detfin} from \cref{sec:overview}. There are two main additional difficulties. First, as we already mentioned in \cref{sec:overview}, the expression $\left( \sum_{j_1, j_2 \in S_i^2} (p_{j_1, t} - p_{j_2, t})^2 \right)^2 $ contains fourth order terms in variables $x_{1, t}, \dots, x_{n,t}$ (recall that $p_{j, t} = \sum_{1 \le t' \le t} x_{j,t}$). This is problematic since we can apply Luby's derandomization only for quadratic terms. This is only a technicality, however, since later in \cref{cl:aqualung} we prove that there is a worst-case upper bound of this expression:
\begin{align}
\label{eq:saturday}
&\left( \sum_{j_1, j_2 \in S_i^2} (p_{j_1, t} - p_{j_2, t})^2 \right)^2\\
&\le
4\sum_{(j_1,j_2,j_1',j_2') \in B^4}            \left(x_{j_1, t} - x_{j_2, t}\right)\left(x_{j_1', t} - x_{j_2', t}\right)(p_{j_1,t-1} - p_{j_2, t-1})(p_{j'_1,t-1} - p_{j'_2, t-1}) 
        + \frac{100|B|^4}{k^3} \label{eq:sunday}
\end{align}
where we can see that the right-hand side of \cref{eq:sunday} is only a quadratic term in $x_{1, t}, \dots, x_{n, t}$. Moreover, the value of this quadratic term is always at most $4|B|^4/k^2$, which implies the $\frac{100 |B|^3}{k^2}\left( 1 + \frac{|B|}{k}\right)$ upper bound on $\left( \sum_{j_1, j_2 \in S_i^2} (p_{j_1, t} - p_{j_2, t})^2 \right)^2 $ used in \cref{sec:overview}. Thus, in our definition of the potential $\Psi$ in \cref{def:Psi}, we already plug in the right-hand side of \cref{eq:sunday} to the place where we want the left-hand side, \cref{eq:saturday}, to be.

The second major problem is that \cref{lem:main} is a weighted generalization of the result proven in \cref{sec:overview}. This is problematic for the analysis of the potential $\Psi_{i,t}$. We solve this problem by splitting every set $S_i$ into buckets of elements whose weights $a_{i,j}$ are the same up to $2$-factor. For  a carefully selected subset of buckets $B \subseteq S_i$, we then define their potential $\Psi_{B, t}$; there is no single potential $\Psi_{i,t}$. We split the buckets into several types, where the most important distinction is a split of buckets into \emph{small} and \emph{large} ones by \cref{eq:def_small_bucket}. In \cref{cl:danil_genius}, we prove that for every set $S_i$, the sum of the values $a_{i,j}$ over its all small buckets is at most $O(\Delta_i)$, so we can ignore these buckets using the definition of the set $I_i^{ignore}$ from the statement of \cref{lem:main}. On the other hand, large buckets happen to be large enough so that in case of a bad event, even just one large bucket $B$ gives us a sufficiently large value of $\Psi_{B, t} / \Psi_{B, 0}$.

\cref{lem:main} is proven in the rest of this section. In the proof, we first describe the potentials that we need to take care of and \cref{alg:deterministic}, based on a pairwise derandomization of \cref{alg:randomized}, that keeps the main potential monotone. Then, we analyze that algorithm.

\paragraph{Flexibility in $\Delta_i$}
Notice that to prove \cref{lem:main}, in its statement, we can replace for every set $i$ the requirements 
\begin{align}
 \sum_{j \in I_i^{ignore}} \aij \le \Delta_i/1000, &&    | \sum_{j \in [n]} \aij (\pj - \qj)| \le \Delta_i / 1000
\end{align}
by
\begin{align}
\label{eq:stinky}
 \sum_{j \in I_i^{ignore}} \aij = O(\Delta_i), &&    | \sum_{j \in [n]} \aij (\pj - \qj)| = O(\Delta_i)
\end{align}
This is because the only other place where $\Delta_i$ is used in the statement is \cref{eq:wind-up} where we require only an asymptotic guarantee. Thus, we will only prove how to achieve the weaker requirements in \cref{eq:stinky}. 

\paragraph{Boring sets}
We identify two noninteresting cases. 
We say that a set $i \in [m]$ has \emph{boringly large $\Delta_i$} if  
\begin{align}
    \label{eq:boring1}
\Delta_i \ge \sum_{j \in [n]} a_{i, j}
\end{align}
Similarly, we say that $i$ has \emph{boringly small $\Delta_i$} if
\begin{align}
    \label{eq:boring2}
\Delta_i^2 < \sum_{j \in [n]} a^2_{i, j}
\end{align}
Whenever $i$ has boringly large or small $\Delta_i$, we also say that $i$ is boring. 
Intuitively, in the first case, the set $i$  will satisfy the requirement \cref{op3} for any choice of the output vector $\q$, thus we will never include it to the bad set $I^{bad}$. On the other hand, in the second case, we have $prob_i^{bad, L \text{\ref{lem:main}}} \ge 1$ for $c$ large enough. 

\paragraph{Setting up notation}
We now define the potential function that our algorithm, \cref{alg:deterministic}, keeps monotone. 
The potential function for the $t$-th step of the algorithm, $1 \le t \le T$, is formed by adding partial potentials functions $\Phi^{(1)}_{i,t}, \Phi^{(2)}_{i,t}, \Psi_{i,t}$ that are defined next for every nonboring set $i \in [m]$. These potentials are functions of variables $x_{j, t} \in \{-1/k, 0, 1/k\}$, $j \in [n], 1 \le t \le T$; in every step $t$, our algorithm fixes the values of variables $x_{j,t}$ for every $j \in [n]$. We also use the following notation for every element $j\in [n]$ and $0 \le t \le T$:
\begin{align}
    p_{j, t} = p_{j} + \sum_{t' = 1}^{t} x_{j, t'}, 
\end{align}
i.e., $p_{j,t}$ is the ``current'' value of the vector $p$ for element $j$ in time $t$. 

\paragraph{Potential functions $\Phi^{(1)}_{i,t}, \Phi^{(2)}_{i,t}$}
In the following definition of the partial potentials $\Phi^{(1)}, \Phi^{(2)}$, we choose 
\begin{align}
\label{def:lambda1}
\lambda_i =  \frac{1}{10^8 } \cdot \min\left(\frac{\Delta_i}{\sum_{j \in [n]} a^2_{i, j}}, \frac{k}{\sum_{j \in [n]} a_{i, j}} \right)
\end{align}
We define for every nonboring set $i$:
\begin{align}
\label{def:Phi1}
\Phi^{(1)}_{i, t} = \prod_{1 \le t' \le t} \left( 1 + \lambda_i \sum_{j \in [n]} a_{i, j} \cdot x_{j, t'} + (\lambda_i \sum_{j \in [n]} a_{i, j} \cdot x_{j, t'})^2 \right) \cdot \left( 1 + \lambda_i^2 \sum_{j \in [n]} a^2_{i, j}/k^2 \right)^{T-t }
\end{align}
and
\begin{align}
\label{def:Phi2}
\Phi^{(2)}_{i, t} = \prod_{1 \le t' \le t} \left( 1 - \lambda_i \sum_{j \in [n]} a_{i, j} \cdot x_{j, t'} + (\lambda_i \sum_{j \in [n]} a_{i, j} \cdot x_{j, t'})^2 \right) \cdot \left( 1 + \lambda_i^2 \sum_{j \in [n]} a^2_{i, j}/k^2 \right)^{T-t }
\end{align}

\paragraph{Potential function $\Psi_{B,t}$}
Every nonboring set $i \in [m]$ defines several partial potential functions $\Psi_{B, t}$ as follows. 
%We note that in the uniform case, we would simply write down an analogue of \cref{def:Psi}, but due to the complecations of weights, we need to split the set into buckets of similar-weight items in the general case. 
We split $[n]$ into \emph{buckets} $[n] = \dots \sqcup B_{i, -1} \sqcup B_{i, 0} \sqcup B_{i, 1} \sqcup \dots$ where $B_{i, \iota}$ is the set of elements $j \in [n]$ such that \begin{align}
    \label{eq:def_bucket}
    2^\iota \le a_{i, j} < 2^{\iota+1}. 
\end{align}
Let $\fB_i$ be the set of all buckets $B_{i, \iota}$. 
We split these buckets into \emph{size-one}, \emph{small} and \emph{large} buckets, i.e., $\fB_i = \fB_i^{size-one} \sqcup \fB_i^{small} \sqcup \fB_i^{large}$. 
Namely, a bucket $B_\iota$ is a size-one bucket if $|B_\iota| = 1$. Next, we say that a bucket $B_\iota \in \fB_i \setminus \fB_i^{size-one}$ is small if
\begin{align}
    \label{eq:def_small_bucket}
    |B_\iota| < \frac{\Delta_i^2}{\sum_{j \in [n]} a_{i,j}^2}
\end{align}
and the remaining buckets are large. Moreover, for large buckets, we pick a set of \emph{representative buckets} $\fB_i^{represent} \subseteq \fB_i^{large}$ as follows. For every maximal group of large buckets $B_{\iota_1}, B_{\iota_2}, \dots, B_{\iota_\ell}$ of the same size, i.e., $|B_{\iota_1}| = |B_{\iota_2}| = \dots = |B_{\iota_\ell}| $ with $\iota_1 < \iota_2 < \dots < \iota_\ell$, only the bucket $B_{\iota_\ell}$ is the representative bucket. 

For every representative bucket $B$, we define a corresponding potential $\Psi_{B, t}$. 
%These potentials will later define $\Psi_{i,t}$ in \cref{def:Psi_outer}. 
First, we define the value $y_{B, t}$ as follows. We say that an index $i$ is \emph{moving}  in $t$-th iteration whenever $p_{j,t-1} \not\in \{0,1\}$. Whenever 
\begin{align}
\label{eq:cookies}
    \left| \{ j \in B, p_{i, t-1} \not\in \{0,1\} \}\right| 
    > |B|/10
\end{align}
that is, too many indices for the set $i$ are still moving, we define 
\begin{align}
\label{eq:cookies2}
y_{B, t} = \sum_{(j_1, j_2) \in B^2} (p_{j_1, t} - p_{j_2, t})^2. 
\end{align}

Otherwise, we define  $y_{B, t}$ inductively from $y_{B, t-1}$ as 
\begin{align}
\label{eq:cookies3}
 y_{B, t} = y_{B, t-1} + \frac{|B|^2}{100k^2}   .
\end{align}
Now, using the following definition of a multiplier $\lambda'_{\beta}$ parametrized by some $\beta \in \R$
\begin{align}
\label{def:lambda2}
    \lambda'_{\beta} = \frac{k}{10^6 \beta (\beta + k)}.
\end{align}
we write
\begin{align}
\label{def:Psi}
\Psi_{B, t} 
&= \prod_{1 \le t' \le t} \Big( 1 - \lambda'_{|B|} (y_{B, t'} - y_{B, t'-1}) \\
&+ \lambda_{|B|}^{\prime 2}
\Big(
 4\sum_{(j_1,j_2,j_1',j_2') \in B^4}
        \left(x_{j_1, t'} - x_{j_2, t'}\right)\left(x_{j_1', t'} - x_{j_2', t'}\right)(p_{j_1,t'-1} - p_{j_2, t'-1})(p_{j'_1,t'-1} - p_{j'_2, t'-1}) 
        + \frac{100|B|^4}{k^3} 
\Big)
\Big)\\
&\cdot  \left(1 - \lambda'_{|B|} \cdot \frac{|B|^2}{100k^2} + \lambda_{|B|}^{\prime 2} \cdot \frac{100 |B|^3}{k^2}\left( 1 + \frac{|B|}{k}\right)\right)^{T - t}
\end{align}

% We are now ready to define the potential $\Psi_{i, t}$. We set
% \begin{align}
% \label{def:Psi_outer}
% \Psi_{i, t} = \sum_{B \in \fB_i^{represent}} \frac{\Psi_{B, t}}{\Psi_{B, 0}} \cdot \exp\left( -  \min(|B|, k)/10^{30} \right)
% \end{align}\todo{perhaps not even define psi(S,t)}

\paragraph{The Overall Potential}
The potential $\Pot_t$ for the iteration $t$ is defined as follows.  
\begin{align}
\label{eq:pot_definition}
\Pot_t 
&= \sum_{\substack{i \in [m] \\ \text{$i$ not boring}}} prob_{i}^{bad, L\text{\ref{lem:main}}} \cdot
\left( \frac{\Phi^{(1)}_{i, t}}{\Phi^{(1)}_{i, 0}} +  \frac{\Phi^{(2)}_{i, t}}{\Phi^{(2)}_{i, 0}} + \sum_{B \in \fB_i^{represent}} \frac{\Psi_{B, t}}{\Psi_{B, 0}} \cdot \exp\left( - \min\left(|B|, k\right)/10^{30} \right) \right)
\end{align}
where $prob_i^{bad, L \text{\ref{lem:main}}}$ is defined by \cref{eq:wind-up}. The values in the denominators are constants whose responsibility is the correct normalization of terms. 

\paragraph{Algorithm and its Time Complexity}
We will use \cref{alg:deterministic} to construct the output vector $\mathbf{q}$. In this algorithm, we keep track of the value of $\Pot_t$ across the steps $t = 1, 2, \dots, T$ of the algorithm and make sure it only decreases, i.e., $\Pot_{t} \le \Pot_{t-1}$.

\begin{algorithm}[ht]
\caption{Derandomization of \cref{alg:randomized}}
\label{alg:deterministic}
{\bf Input:}  $n,m,k \in \N, A \in \R^{m \times n}_{\ge 0}, \p \in (\{0, 1/k, \dots, 1\})^n, \Deltav \in \R^n_{> 0}$ \\
{\bf Output:} Values $\q \in (\{0, 1/k, \dots, 1\})^n$
\begin{algorithmic}
\STATE Define $T = 100k^2$
\STATE Define $p_{j, 0} = \pj$ for every $j \in [n]$. 
\FOR{$t = 1, 2, \dots, T$}
    \STATE \label{line:pairwise} Use \cref{lem:pairwise} to find values $x_{1, t}, \dots, x_{n, t}$ such that 
    $$\Pot_{t}(x_{1, t}, \dots, x_{n,t}) \le \E[\Pot_{t}(X_1, \dots, X_n)]  \le \Pot_{t-1}.$$
\ENDFOR
\STATE Set $\qj = p_{j, T}$ for every $j\in [n]$.
\STATE For every set $i$ we set $I_i^{ignore} = [n]$ whenever $i$  has boringly large $\Delta_i$, we set $I_i^{ignore} = \emptyset$ whenever $i$ has boringly small $\Delta_i$, and otherwise \begin{align}
    \label{eq:ignore}
    I_i^{ignore} = \bigcup_{B \in \fB_i^{small} \cup \fB_i^{size-one}}B. 
\end{align}
\STATE $I^{bad}$ contains every set $i$ that fails at least one of the three conditions \cref{op1,op2,op3}. 
\STATE Return $\q, I^{bad}, (I_i^{ignore})_{i\in[m]}$
\end{algorithmic}
\end{algorithm}

In particular, in step $t$ we have already fixed the values of $x_{j, t'}$ for all $j \in [m], t' < t$. This enables us to think of $\Pot_t$ as a function of variables $x_{j, t}$ for all $j \in [n]$, i.e.,
\begin{align}
\Pot_t = \Pot_t(x_{1, t}, x_{2, t}, \dots, x_{n, t}).     
\end{align}
We will also later sometimes use the analogous notation for partial potentials $\Phi^{(1)}_{i,t}, \Phi^{(2)}_{i,t}, \Psi_{B,t}$ whenever we want to emphasize that we think of these potentials of functions of variables $x_{1, t}, \dots, x_{n,t}$ that we are fixing in the $t$-th step. 

To understand \cref{line:pairwise} of \cref{alg:deterministic}, we note that all expectations are over an arbitrary family of pairwise independent random variables $(X_j)_{j \in [n]}$ where the marginal for every $j \in [m]$ is $\P(X_j = \pm 1/k) = 1/2$ if $p_{i, t-1} \not\in \{0,1\}$ and $\P(X_j = 0) = 1$ otherwise. 
Later, we will prove that $ \E\left[ 
    \Pot_t \left( X_1, X_2, \dots, X_n \right)
 \right]
 \le \Pot_{t-1}$ which means that in \cref{line:pairwise} we are, in fact, making sure that $\Pot_t \le \Pot_{t-1}$. However, first, we discuss why \cref{lem:pairwise} can be applied and what the overall work and depth of our algorithm are. 

\paragraph{Roadmap for the rest of the proof}
Now that the algorithm to prove \cref{lem:main} is fixed, we give a short roadmap for the rest of the proof. First, in \cref{cl:time}, we analyze the work and depth of the algorithm, by casting our potential as the sum of nice quadratic terms from \cref{def:quadratic_term}. Next, we continue with the analysis of the potentials $\Phi_{i,t}^{(1)}, \Phi_{i,t}^{(2)}, \Psi_{B, t}$. For each potential, we have one claim where we analyze the relevant first and second moments as in the analysis of the randomized \cref{alg:randomized} in \cref{sec:overview}; these are \cref{cl:phi_monotone,cl:phi2_monotone,cl:psi_monotone} that together imply the monotonicity of $\Pot_t$ in \cref{cl:monotone}. Next, every potential corresponds to one item from \cref{op1,op2,op3} in the statement of \cref{lem:main}: $\Phi_{i,t}^{(1)}$ corresponds to one inequality of \cref{op3}, $\Phi_{i,t}^{(2)}$ corresponds to the other inequality of \cref{op3}, and $\Psi_{B,t}$ captures conditions \cref{op1,op2}. We prove in \cref{cl:Phi_ratio,cl:Phi2_ratio,cl:psi_concentration} that whenever one of the conditions \cref{op1,op2,op3} fails, the corresponding potential is large. These claims then together give the required upper bound on $|I^{bad}|$ in  \cref{cl:at_last}. Finally, in \cref{cl:danil_genius} we prove that the sets $I_i^{ignore}$ have size $O(\Delta_i)$.

\paragraph{Time Complexity}
We start by computing the work and depth of \cref{alg:deterministic}

\begin{claim}
\label{cl:time}
\cref{alg:deterministic} has $\tilde{O}(\max(nnz(A),n,m)\poly(k))$ work and $\poly(\log(nm),k)$ depth.   
\end{claim}
We need to make sure that $\Pot_t(x_{1, t}, \dots, x_{n, t})$ can be written as a summation of nice quadratic terms from \cref{def:quadratic_term}, with $x_{1, t}, \dots, x_{n, t}$ being the variables. Moreover, we need to inspect the complexity $||Q||$ of each such nice quadratic term $Q$ and the work and depth necessary to compute the constants in those terms. This will be done by inspecting  \cref{def:Phi1,def:Phi2,def:Psi,eq:pot_definition}. 

For $\Phi_{i,t}^{(1)}, \Phi_{i,t}^{(2)}$ we have that the only part of the defining expression \cref{def:Phi1,def:Phi2} that is not constant is the nice quadratic term
\begin{align}
     1 \pm \lambda_i \sum_{j \in [n]} a_{i, j} \cdot x_{j, t'} + (\lambda_i \sum_{j \in [n]} a_{i, j} \cdot x_{j, t'})^2. 
\end{align}
that occurs in the product $\prod_{1 \le t' \le t}(\dots)$ for $t' = t$. 
Thus
\begin{align}
    ||\Phi_{i, t}^{(1)}|| = ||\Phi_{i, t}^{(2)}|| = O(1 + nnz(A_{i*})) 
\end{align}
where $A_{i*}$ is the number of nonzero entries in the $i$-th column of $A$. 
Moreover, by inspecting the definition of $\lambda_i$ and the constant factors in the definition of $\Phi_{i,t}^{(1)}, \Phi_{i,t}^{(2)}$, we conclude that we can compute the constant factors of the nice quadratic expressions in $\poly\log(nm)$ depth and work $O\left(k^2 \left( 1 + nnz(A_{i*})\right)\right)  $. 

Similarly, we now inspect the definition of $\Psi_{B, t}$ for some bucket $B$ of some set $i$. The only parts of the definition in \cref{def:Psi} that are not constant are the two expressions
\begin{align}
\label{t1}
    \lambda'_{|B|} (y_{B, t} - y_{B, t-1}) 
\end{align}
and
\begin{align}
\label{t2}
\sum_{(j_1,j_2,j_1',j_2') \in B^4}\left(x_{j_1, t} - x_{j_2, t}\right)\left(x_{j_1', t} - x_{j_2', t}\right)(p_{j_1,t-1} - p_{j_2, t-1})(p_{j'_1,t-1} - p_{j'_2, t-1}) 
\end{align}
that occur in the product $\prod_{1 \le t' \le t}(\dots)$ for $t' = t$. 
Regarding \cref{t1}, the potentially nonconstant term is $y_{B, t}$, which is either constant or it is defined via \cref{eq:cookies2} which can be rewritten as
\begin{align}
y_{B, t} 
&= \sum_{(j_1, j_2) \in B^2} (p_{j_1, t-1} + x_{j_1, t} - p_{j_2, t-1} - x_{j_2, t})^2. \\
&= 2|B|\sum_{j \in B}  x_{j,t}^2  
- 2\left( \sum_{j_1 \in B} x_{j_1, t}\right)\left( \sum_{j_2 \in B} x_{j_2, t}\right)\\
&+2|B|\sum_{j \in B}  p_{j, t-1}^2  
- 2\left( \sum_{j_1 \in B} p_{j_1, t-1}\right)\left( \sum_{j_2 \in B} p_{j_2, t-1}\right) \label{uf}
\end{align}
Note that the expression $\sum_{j \in B} x_{j,t}^2$ can be seen as a sum of $|B|$ nice quadratic terms $Q_j$, each with $||Q_j|| = O(1)$ and the rest of \cref{uf} is one nice quadratic expression. 
We conclude that 
\begin{align}
    ||y_{B,t}|| = O(|B|). 
\end{align}
Similarly, the work necessary to compute constants in these nice quadratic expressions $y_{B, t}$ is $O(|B|)$. 

We continue by inspecting \cref{t2}. We see that the expression is a summation of four analogous expressions, one of them being 
\begin{align}
    &\sum_{(j_1,j_2,j_1',j_2') \in B^4}
    x_{j_1, t} x_{j_1', t} (p_{j_1,t-1} - p_{j_2, t-1})(p_{j'_1,t-1} - p_{j'_2, t-1})\label{purgatory}\\
    &= |B|^2 \left( \sum_{(j_1, j_2) \in B^2}x_{j_1, t} x_{j_1', t}p_{j_1, t-1} p_{j'_1, t-1} \right)\label{hell1}
    - 2|B|\left( \sum_{(j_1, j_1', j_2') \in B^3}x_{j_1, t}x_{j_1', t} p_{j_1, t-1} p_{j_2', t-1}\right)\\
    &+ \sum_{(j_1,j_2,j_1',j_2') \in B^{4}} x_{j_1, t} x_{j_1', t} p_{j_2,t-1}p_{j_2',t-1}\label{hell3}\\
    &= |B|^2\left( \sum_{j_1 \in B} x_{j_1,t} p_{j_1, t-1} \right) 
    \left( \sum_{j_1' \in B} x_{j_1', t} p_{j_1', t-1} \right)
    -2|B|\left( \sum_{j_1 \in B} x_{j_1, t} p_{j_1, t-1} \right)\left(\sum_{j_1' \in B} x_{j_1', t}\right)
    \left(\sum_{j_2' \in B} p_{j_2', t-1}\right)\\    
    &+ \left( \sum_{j_1 \in B} x_{j_1, t} \right)
     \left( \sum_{j_1' \in B} x_{j_1', t} \right)
     \left( \sum_{j_2 \in B} p_{j_2, t-1} \right)
     \left( \sum_{j_2' \in B} p_{j_2', t-1} \right). 
\end{align}
We conclude that
\begin{align}
    ||\text{\cref{t2}}|| = O(|B|). 
\end{align}
Similarly, the work to compute all constants in these nice quadratic expressions is also $O(|B|)$. 

Finally, we inspect \cref{eq:pot_definition} and notice that 
\begin{align}
    \label{lleh3}
    ||\Pot_t|| = O(\max(nnz(A), n, m)). 
\end{align}
and the complexity of constructing the constants in the nice quadratic expressions that $\Pot_t$ consists of is the same. 
The work and depth bounds from the statement follow.

\paragraph{Overview of the analysis of potentials}
%To simplify notation, from now on, $X_1, \dots, X_n$ always denotes an arbitrary family of pairwise independent random variables, with $\P(X_1 = 1/k) = \P(X_1 = -1/k) = 1/2$. All expectations are over the randomness of such a family. 
%Whenever we write a potential indexed by $t$, e.g., $\Phi^{(1)}_{i, t}$, we think of this potential as a function $\Phi^{(1)}_{i, t}(X_1, X_2, \dots X_n)$. 
%Similarly, note that the value of $y_{B, t}$ as defined in \cref{eq:cookies2} is a function of $p_{j,t}$ and thus function of random variables $X_j$. To emphasize that $y_{B, t}$ is itself a random variable, we capitalize it and write $Y_{B,t}$. 
For every potential $\Phi^{(1)}, \Phi^{(2)}, \Psi$, we first prove its monotonicity. Namely, this is done in \cref{cl:phi_monotone,cl:psi_monotone} that together imply \cref{cl:monotone}. Next, for every potential, we prove that whenever a certain bad event happens, it implies that the overall potential $\Pot_T$ attains a much larger value than at the beginning of the algorithm. This type of argument corresponds to a concentration argument in probabilistic analysis. 
%In the following computations, we always work with a fixed set $i$ to analyze $\Phi^{(1)}_{i, t}$ or $\Psi_{i, t}$. To avoid clutter in the notation, we write $a_{i,j}$ instead of $a_{i, j}$ except for claim statements. 

\paragraph{Analysis of $\Phi^{(1)}_{i, t}$}
We will now analyze the potential $\Phi^{(1)}_{i, t}$. We first prove that it is monotone in \cref{cl:phi_monotone}. Next, we prove that the value of $\Phi_{i,t}^{(1)} / \Phi_{i,0}^{(1)}$ is large in case that $\sum_{j \in [n]} p_{j, T}a_{i,j} - \sum_{j \in [n]} a_{i,j}/2 \ge \Delta_i$. 

\begin{claim}
\label{cl:phi_monotone}
For every nonboring set $i$ we have $\E \left[  \Phi^{(1)}_{i, t}\left(X_1, \dots, X_n \right) \right] \le \Phi^{(1)}_{i, t-1}$. 
\end{claim}
\begin{proof}
By definition of $\Phi^{(1)}_{i, t}$ and $\Phi^{(1)}_{i, t-1}$ from \cref{def:Phi1}, it suffices to prove that
\begin{align}
    \label{eq:lunch}
    \E\left[ 1 + \lambda_i \sum_{j \in [n]} a_{i,j} X_j + \left(\lambda_i \sum_{j \in [n]} a_{i,j} X_j\right)^2 \right] \le \left(1 + \lambda_i^2 \sum_{j \in [n]} a_{i,j}^2/k^2\right). 
\end{align}
That is, the first and the second moment of each $X_j$ needs to be computed. 
We first note that 
\begin{align}
 \E[\sum_{j \in [n]} a_{i,j} X_j] = \sum_{j \in [n]} a_{i,j} \E[X_j] = 0.   
\end{align}
by the linearity of expectation. 
We continue with the second moment. We have 
\begin{align}
\E\left[ \left( \sum_{j \in [n]} a_{i,j} X_j \right)^2 \right] = \sum_{j \in [n]} a_{i,j}^2 \E[X_j^2]
\end{align}
by the pairwise independence of $(X_j)_{j \in [m]}$ and $\E[X_j] = 0$. Next, we note that 
\begin{align}
    \E[ X_j^2] \le 1 / k^2
\end{align}
since either $\E[X_j^2] = 1/k^2$ if $p_{i, t-1} \not\in \{0,1\}$ and $\E[X_j^2] = 0$ otherwise. The proof follows. 
\end{proof}

We continue by showing that the ratio of $\Phi_{i, t}^{(1)} / \Phi_{i, 0}^{(1)}$ is large whenever $\sum_{j \in [n]} p_{j, t} a_{i,j} > \sum_{j \in [n]} p_{i, 0} a_{i,j} + \Delta_i$, i.e., we deviate by more than $\Delta_i$ from the starting state.  
\begin{claim}
\label{cl:Phi_ratio}
For any nonboring set $i$ and iteration $t$, whenever 
\begin{align}
\label{eq:fun}
\sum_{j \in [n]} p_{j, t} a_{i, j} > \sum_{j \in [n]} p_{i, 0} a_{i, j} + \Delta_i, 
\end{align}
we have 
\begin{align}
\label{eq:conclusion}
\Phi_{i, t}^{(1)} / \Phi_{i, 0}^{(1)} 
\ge  \exp\left(\frac{1}{10^9 } \min\left( \frac{ \Delta_i^2  }{\sum_{j \in [n]} a^2_{i, j}}, \frac{k \Delta_i}{\sum_{j \in [n]} a_{i, j}}\right)\right).
\end{align}
 
\end{claim}
\begin{proof}
On one hand, we have 
\begin{align}
\Phi_{i, 0}^{(1)} 
&= \prod_{1 \le t \le T} \left( 1 + \lambda_i^2 \sum_{j \in [n]} a_{i,j}^2 /k^2 \right) \\
&\le \exp\left(\sum_{1 \le t \le T} \lambda_i^2 \sum_{j \in [n]} a_{i,j}^2 / k^2\right) && 1+x \le \e^x\\
&= \exp\left(100\lambda_i^2 \sum_{j \in [n]} a_{i,j}^2\right) && T = 100k^2\\
&= \exp\left(\frac{1}{10^{12}} \min\left( \frac{\Delta_i^2}{\sum_{j \in [n]} a_{i,j}^2}, \frac{k^2\sum_{j \in [n]} a_{i,j}^2}{(\sum_{j \in [n]} a_{i,j})^2} \right)\right). && \text{\cref{def:lambda1}} \label{eq:happy}
\end{align}

Note that 
\begin{align}
    \frac{k \Delta_i}{\sum_{j \in [n]} a_{i,j}} = \sqrt{\frac{ \Delta_i^2  }{\sum_{j \in [n]} a_{i,j}^2} \cdot \frac{k^2\sum_{j \in [n]} a_{i,j}^2}{(\sum_{j \in [n]} a_{i,j})^2}}
\end{align}
i.e., $\frac{k \Delta_i}{\sum_{j \in [n]} a_{i,j}}$ is a geometric mean of the two terms in \cref{eq:happy}. Replacing one of those two terms by their geometric mean can only increase their minimum, hence we conclude that
\begin{align}
\label{eq:blabla}
\Phi_{i, 0}^{(1)} 
&\le \exp\left(\frac{1}{10^{12} } \min\left( \frac{ \Delta_i^2  }{\sum_{j \in [n]} a_{i,j}^2}, \frac{k \Delta_i}{\sum_{j \in [n]} a_{i,j}}\right)\right) 
\end{align}

On the other hand, we note that for any $t'$ we have
\begin{align}
\lambda_i \sum_{j \in [n]} a_{i,j} \cdot x_{j, t'} 
\le \left( \frac{1}{10^{8}} \frac{k}{\sum_{j \in [n]} a_{i,j}} \right) \cdot \frac{\sum_{j \in [n]} a_{i,j}}{k} < 1.    
\end{align}
Thus, we can apply   \cref{fact:exponential} to the left-hand side and bound
\begin{align}
\label{eq:blablah}
    \Phi_{i, t}^{(1)} 
    &\ge \prod_{1 \le t' \le t} \left( 1 + \lambda_i \sum_{j \in [n]} a_{i,j} \cdot x_{j, t'} + (\lambda_i \sum_{j \in [n]} a_{i,j} \cdot x_{j, t'})^2 \right) \\
    &\ge \prod_{1 \le t' \le t} \exp\left(\lambda_i \sum_{j \in [n]} a_{i,j} \cdot x_{j, t'}\right)&& \text{\cref{fact:exponential}} \\
    &= \exp\left(\lambda_i \sum_{1 \le t' \le t}\sum_{j \in [n]} a_{i,j} x_{j, t'}\right)\\ 
    &= \exp\left( \lambda_i \left( \sum_{j \in [n]} p_{j, T}a_{i,j} - \sum_{j \in [n]}p_{i, 0} a_{i,j}\right)\right)\\ 
    &\ge \exp\left(\lambda_i \Delta_i\right)  && \text{\cref{eq:fun}} \\
    &= \exp\left(\frac{1}{10^8 } \min\left( \frac{ \Delta_i^2  }{\sum_{j \in [n]} a_{i,j}^2}, \frac{k \Delta_i}{\sum_{j \in [n]} a_{i,j}}\right)\right). && \text{\cref{def:lambda1}} 
\end{align}

Putting \cref{eq:blabla,eq:blablah} together finishes the proof. 
\end{proof}

\paragraph{Analysis of $\Phi_{i, t}^{(2)}$}
The analysis is done analogously to $\Phi_{i, t}^{(1)}$. 
\begin{claim}
\label{cl:phi2_monotone}
For every nonboring set $i$ we have $\E \left[  \Phi^{(2)}_{i, t}\left(X_1, \dots, X_n \right) \right] \le \Phi^{(2)}_{i, t-1}$. 
\end{claim}
\begin{proof}
The proof is the same as in \cref{cl:phi_monotone}. 
\end{proof}
\begin{claim}
\label{cl:Phi2_ratio}
For any nonboring set $i$ and iteration $t$, whenever 
\begin{align}
\label{eq:fun2}
\sum_{j \in [n]} p_{j, T} a_{i, j} < \sum_{j \in [n]} p_{i, 0} a_{i, j} - \Delta_i, 
\end{align}
we have 
\begin{align}
\label{eq:conclusion2}
\Phi_{i, t}^{(2)} / \Phi_{i, 0}^{(2)} 
\ge  \exp\left(\frac{1}{10^9 } \min\left( \frac{ \Delta_i^2  }{\sum_{j \in [n]} a^2_{i, j}}, \frac{k \Delta_i}{\sum_{j \in [n]} a_{i, j}}\right)\right).
\end{align}
 
\end{claim}
\begin{proof}
We use the same proof as in \cref{cl:Phi_ratio}. Note that we start with the assumption $\sum_{j \in [n]} p_{j, T} a_{i,j} < \sum_{j \in [n]} p_{i, 0} a_{i,j} - \Delta_i$ where $\Delta_i$ has opposite sign than the original assumption \cref{eq:fun}. However, the definition of $\Phi_{i, t}^{(2)}$ leads to an opposite sign in \cref{eq:blablah} and the two opposite signs cancel; we thus arrive at the same conclusion \cref{eq:conclusion} as in the analysis of $\Phi_{i, t}^{(1)}$.     
\end{proof}

\paragraph{Analysis of $\Psi_{B, t}$}
We analyze separately each term $\Psi_{B, t}$ for a representative bucket $B \in \fB_i$ of some nonboring set $i$. In the following analysis, we use crucially that all the elements of the same bucket have roughly the same size, i.e., there exists some $\ell_B$ such that \begin{align}
\label{eq:bucket}
2^{\ell_B} \le a_{i,j} < 2^{\ell_{B} + 1}.    
\end{align}

Recall \cref{eq:cookies2,eq:cookies3} that define the value of $y_{B,t}$. At least in the case of \cref{eq:cookies2}, $y_{B, t}$ is a function of variables $x_{j,t}$ and we will write
\begin{align}
    y_{B, t} = y_{B, t}(x_{1, t}, \dots, x_{n, t})
\end{align}
to emphasize that $y_{B,t}$ depends on variables that are being fixed in the $t$-th step of our algorithm. 

\begin{claim}
\label{cl:aqualung}
    Let $B$ be any bucket of some nonboring set $i$ such that the condition \cref{eq:cookies} is satisfied for $i$ in step $t$. Then, for any choice of values $X_j$, we have 
    \begin{align}
        &(y_{B, t}(X_1, \dots, X_n) - y_{B, t-1})^2 \\
        &\le4\sum_{(j_1,j_2,j_1',j_2') \in B^4}
        \left(X_{j_1} - X_{j_2}\right)\left(X_{j_1'} - X_{j_2'}\right)(p_{j_1,t-1} - p_{j_2, t-1})(p_{j'_1,t-1} - p_{j'_2, t-1})   
        + \frac{100|B|^4}{k^3}
    \end{align}
\end{claim}
\begin{proof}
We start by writing down the value of $ y_{B, t}(X_1, \dots, X_n) - y_{B, t-1} $ using \cref{eq:cookies2}:
    \begin{align}
 y_{B, t}(X_1, \dots, X_n) - y_{B, t-1} 
 &= \sum_{(j_1, j_2) \in B^2} \left( (p_{j_1, t-1} + X_j) - (p_{j_2, t-1} + X_j)  \right)^2
 - \sum_{(j_1, j_2) \in B^2} \left( p_{j_1, t-1}  - p_{j_2, t-1} \right)^2\\
 &= \sum_{(j_1, j_2) \in B^2} \left( \left(X_{j_1} - X_{j_2}\right)^2 + 2(X_{j_1} - X_{j_2})(p_{j_1,t-1} - p_{j_2, t-1})\right)  . \label{eq:boring}
\end{align}

Thus, we can write
\begin{align}
    \left( y_{B, t}(X_1, \dots, X_n) - y_{B, t-1} \right)^2  
    &=
    \sum_{(j_1,j_2,j_1',j_2') \in B^4}
    \Big(
        \left(X_{j_1} - X_{j_2}\right)^2 \cdot  \left(X_{j_1'} - X_{j_2'}\right)^2 
    \Big)
    \label{eq:first_term} \\
    &+ 
    4\sum_{(j_1,j_2,j_1',j_2') \in B^4}
    \Big(
        \left(X_{j_1} - X_{j_2}\right)\left(X_{j_1'} - X_{j_2'}\right)(p_{j_1,t-1} - p_{j_2, t-1})(p_{j'_1,t-1} - p_{j'_2, t-1}) 
    \Big) \label{eq:second_term}\\
    &+  
    4\sum_{(j_1,j_2,j_1',j_2') \in B^4}
    \Big(
        \left(X_{j_1} - X_{j_2}\right)^2 \left(X_{j_1'} - X_{j_2'}\right)(p_{j'_1,t-1} - p_{j'_2, t-1}) 
    \Big)
    \label{eq:third_term}
\end{align}
The first term, \cref{eq:first_term}, is always upper bounded by $4|B|^4/k^4$. The third term, \cref{eq:third_term}, is always upper bounded by $|B|^4/k^3$. 
The claim follows. 
\end{proof}

\begin{claim}
\label{cl:psi_monotone}
For every bucket $B$ of some  nonboring set we have $\E[\Psi_{B, t}(X_1, \dots, X_n)] \le \Psi_{B, t-1}$. 
\end{claim}
\begin{proof}
In view of the definition of $\Psi_{B,t}$ in \cref{def:Psi}, we start with the computation of the first moment of $y_{B, t}(X_1, \dots, X_n)$, using \cref{eq:boring} to that end. 

\newcommand{\Bmov}{B_\mathrm{moving}}
Recall that an index $j$ is \emph{moving} whenever $p_{j,t-1} \not\in \{0,1\}$. We define $\Bmov \subseteq B$ to be the subset of moving indices. Assuming the condition of \cref{eq:cookies}, we have 
\begin{align}
\label{eq:part1}
\E[y_{B, t}(X_1, \dots, X_n) - y_{B, t-1}] 
&= \sum_{(j_1, j_2) \in B \times B} \E\left[ \left( X_{j_1} - X_{j_2} \right)^2\right] && \text{\cref{eq:boring}}\\
&\ge \sum_{j_1 \in \Bmov}  \sum_{j_2 \in B, j_2 \not=j_1} \E\left[ X_{j_1}^2 + X_{j_2}^2 \right] && j_1\not=j_2 \Rightarrow \E[X_{j_1} X_{j_2}] = 0\\
&\ge \sum_{j_1 \in \Bmov}  \sum_{j_2 \in B, j_2 \not=j_1} \E\left[ X_{j_1}^2 \right] \\
&\ge \frac{|\Bmov|(|B|-1)}{k^2} && \text{$j$ moving} \Rightarrow \E[X_j^2] = 1/k^2\\
&\ge \frac{|B|^2}{100k^2} && \text{\cref{eq:cookies}}, \text{$B$ is not size-one}
\end{align}
If \cref{eq:cookies} is not true, note that we arrive at the same conclusion by definition from \cref{eq:cookies3}. 

Next, we observe that 
\begin{align}
     &\E\left[ 4\sum_{(j_1,j_2,j_1',j_2') \in B^4}
        \left(X_{j_1} - X_{j_2}\right)\left(X_{j_1'} - X_{j_2'}\right)(p_{j_1,t-1} - p_{j_2, t-1})(p_{j'_1,t-1} - p_{j'_2, t-1}) \right] \label{eq:cross_eyed_mary} + \frac{100|B|^4}{k^3}\\
        &\le
        4\max_{X_1, \dots, X_n}\left( 
        \sum_{\substack{(j_1,j_2,j_1',j_2') \in B^4\\|\{i,j,i',j'\}| < 4}} \left(X_{j_1} - X_{j_2}\right)\left(X_{j_1'} - X_{j_2'}\right)(p_{j_1,t-1} - p_{j_2, t-1})(p_{j'_1,t-1} - p_{j'_2, t-1})\right)
        +\frac{100|B|^4}{k^3}\\
        &\le\frac{100|B|^3}{k^2}\left( 1 + |B|/k\right) \label{eq:part2}
\end{align}
since the value inside the expectation, \cref{eq:cross_eyed_mary}, is equal to zero for terms where all the indices $j_1, j_2, j_1', j_2'$ are different, by pairwise independence. 

Finally, going back to the statement of the claim, we need to prove that 
\begin{align}
    &\E \big[
    \big( 1 - \lambda'_{|B|} (y_{B, t}(X_1, \dots, X_n) - y_{B, t-1}) + \\
    &\lambda_{|B|}^{\prime 2}\big(  4\sum_{(j_1,j_2,j_1',j_2') \in B^4}
        \left(X_{j_1} - X_{j_2}\right)\left(X_{j_1'} - X_{j_2'}\right)(p_{j_1,t-1} - p_{j_2, t-1})(p_{j'_1,t-1} - p_{j'_2, t-1}) \big) 
        \big)  
    \big]\\
    &\le 1 - \lambda'_{|B|} \cdot \frac{|B|^2}{100k^2} + \lambda_{|B|}^{\prime 2} \cdot \frac{100 |B|^3}{k^2}\left( 1 + \frac{|B|}{k}\right)
\end{align}
and this inequality follows from \cref{eq:part1,eq:part2}. 
\end{proof}

We can now derive the monotonicity claim that is necessary for \cref{line:pairwise} of \cref{alg:deterministic} to ensure that the potentials $\Pot_t$ are monotone. 
\begin{claim}
\label{cl:monotone}
For any step $t$, we have
\begin{align}
    \Pot_t \le \Pot_{t-1}. 
\end{align}
\end{claim}
\begin{proof}
First, we observe that \cref{cl:phi_monotone,cl:phi2_monotone,cl:psi_monotone}, together with the fact that the definition of overall potentials $\Pot_{t-1}, \Pot_t$ in \cref{eq:pot_definition} is linear in $\Phi^{(1)}, \Phi^{(2)}$, and $\Psi$, imply that
\begin{align}
    \E[\Pot_t(X_1, \dots, X_n)] \le \Pot_{t-1}
\end{align}
for any pairwise independent family $X_1, \dots, X_n$. 
Then, we note that on \cref{line:pairwise} of \cref{alg:deterministic} we use \cref{lem:pairwise} to get values $x_{1, t}, \dots, x_{n, t}$ with 
\begin{align}
    \Pot_t(x_{1, t}, \dots, x_{n, t}) \le \E[\Pot_t(X_1, \dots, X_n)]. 
\end{align}
Note that we can use \cref{lem:pairwise}, this is a special case of \cref{cl:time}. The claim follows. 
\end{proof}

Let $B_{\text{leftover}}$  be the number of moving indices $j \in B$ after step $T$. Next, we prove that whenever $|B_{\text{leftover}}| $ is at least a small constant fraction of $|B|$, the potential ratio $\frac{\Psi_{B, T}}{\Psi_{B, 0}}$ is very large. 
\begin{claim}
\label{cl:Psi_ratio}
Let $B$ be a bucket of some nonboring set. 
Whenever $|B_{\text{leftover}}| > |B|/10$, we have 
\begin{align}
 \frac{\Psi_{B, T}}{\Psi_{B, 0}} \ge \exp\left(\min(|B|, k) / 10^7\right).   
\end{align}
\end{claim}
\begin{proof}
Note that while it is the case that $|\{ j \in B, p_{j,t} \not\in \{0,1\}\}| \ge |B|/10$ for some step $t$, we have $y_{B, t} = \sum_{(j_1, j_2)\in B^2} (p_{j_1,t} - p_{j_2,t})^2 \le |B|^2$ by \cref{eq:cookies,eq:cookies2}. Thus, whenever $|B_{\text{leftover}}| > |B|/10$, \cref{eq:cookies} was satisfied for all values $1 \le t \le T$ and, in particular,
\begin{align}
\label{eq:fin_ass}
    y_{B, T} \le |B|^2
\end{align}
We will now use this inequality and the assumption that \cref{eq:cookies} was satisfied for all steps to bound $\Psi_{B, T} / \Psi_{B, 0}$. We start with $\Psi_{B, 0}$. First, note that by definition of $\lambda'_\beta$ in \cref{def:lambda2}, we have
\[
\lambda'_{|B|} \cdot \frac{|B|^2}{100k^2} 
= \frac{|B|}{10^8 k(|B| + k)}  
\]
and
\[
\lambda_{|B|}^{\prime 2} \cdot \frac{10^3 |B|^3}{k^2} \left( 1 + \frac{|B|}{k}\right)
= \frac{k^2}{10^{12} |B|^2 (|B|+k)^2} \frac{10^3|B|^3}{k^2}\left( 1 + \frac{|B|}{k}\right)
= \frac{|B|}{10^9 k (|B| + k)}
\]
Thus, we have
\begin{align}
\label{eq:pat}
\Psi_{B, 0}
&= \prod_{1 \le t \le T} \left(
1 - \lambda'_{|B|} \frac{|B|^2}{100k^2} + \lambda_{|B|}^{\prime 2} \frac{10^3|B|^3}{k^2}\left( 1 + \frac{|B|}{k}\right) \right)\\
&\le \prod_{1 \le t \le T} \left( 1 -\frac{|B|}{5\cdot 10^8 k (|B| + k)} \right)\\ 
&\le \exp\left( - \sum_{1 \le t \le T} \frac{|B|}{5\cdot 10^8 k (|B| + k)} \right) && 1 - x \le \e^{-x}\\ 
&= \exp\left( -\frac{|B|k}{5\cdot 10^6(|B|+k)}\right) && T = 100k^2\\
&\le \exp\left( -\min(|B|, k) / 10^7 \right)
\end{align}

Next, we bound $\Psi_{B, T}$. 
Note that for every choice of values of $x_{1, t}, \dots, x_{n, t}$, we can use \cref{eq:boring} to conclude that
\begin{align}
&-\lambda'_{|B|} (y_{B, t} - y_{B, t-1}) 
= -\lambda'_{|B|} \sum_{(j_1, j_2) \in B^2} \left( \left(x_{j_1, t} - x_{j_2, t}\right)^2 + 2(x_{j_1, t} - x_{j_2, t})(p_{j_1,t-1} - p_{j_2, t-1})\right) \\
&\leq \frac{k}{10^5|B|(|B|+k)} \cdot \frac{4|B|^2}{k} 
\leq 1   
\end{align}
Therefore, we may apply \cref{fact:exponential} for $x = - \lambda'_{|B|} (y_{B, t} - y_{B, t-1})$  which we use in the following computation, alongside with our assumption $y_{B, T} \le |B|^2$ from \cref{eq:fin_ass}. 
Also, note that we can use \cref{cl:aqualung} since the condition \cref{eq:cookies} was satisfied for all steps $t$. 
\begin{align}
\label{eq:mat}
\Psi_{B, T}
&= \prod_{1  \le t \le T} \Big(1 - \lambda'_{|B|} (y_{B, t} - y_{B, t-1}) \\
&+ \lambda_{|B|}^{\prime 2}
\Big(  4\sum_{(j_1,j_2,j_1',j_2') \in B^4}
        \left(x_{j_1, t} - x_{j_2, t}\right)\left(x_{j_1', t} - x_{j_2', t}\right)(p_{j_1,t-1} - p_{j_2, t-1})(p_{j'_1,t-1} - p_{j'_2, t-1}) \Big)\\
&        + \frac{100|B|^3}{k^2}\left( 1 + |B|/k\right) 
\Big) \\
&\ge \prod_{1  \le t \le T} \left(1 - \lambda'_{|B|} (y_{B, t} - y_{B, t-1}) + \lambda_{|B|}^{\prime 2}(y_{B, t} - y_{B, t-1})^2 \right) && \text{\cref{cl:aqualung}}\\
&\geq \exp\left(-\sum_{1 \le t \le T} \lambda'_{|B|} (y_{B, t} - y_{B, t-1}) \right) && \text{\cref{fact:exponential}}\\
&= \exp\left(- \lambda'_{|B|} (y_{B, T} - y_{B, 0}) \right) \\
&\geq \exp\left(- \lambda'_{|B|} |B|^2\right) && \text{\cref{eq:fin_ass}}\\
&= \exp\left(- \frac{k|B|}{10^6(|B|+k)}\right)\\
&\ge \exp\left(- \min(|B|, k)/10^6\right)
\end{align}

Putting \cref{eq:pat,eq:mat} together, we conclude that $\Psi_{B, T}  / \Psi_{B, 0} \ge \exp\left( \min(|B|, k)/10^7 \right)$ as needed. 

\end{proof}

\paragraph{Small and Large Buckets}
Now we continue with the analysis of potentials $\Psi_{B, t}$ for buckets $B \in \fB_i$. 
Remember that in \cref{eq:def_small_bucket} we classified buckets as small and large, and chose a subset of large ones that we called representative. We will prove that the sum of weights of elements in small buckets is $O(\Delta_i)$ in \cref{cl:danil_genius} which means that the set $I_i^{ignore}$ defined in \cref{alg:deterministic} satisfies the required properties. 

\begin{claim}
\label{cl:danil_genius}
For every set $i$, we have
\begin{align}
        \sum_{j \in I_i^{ignore} } a_{i, j} 
        \le 10\Delta_i. 
    \end{align}
\end{claim}
\begin{proof}
Notice that the inequality holds by definition for sets with boringly large $\Delta_i$, and it also holds for sets with boringly small $\Delta$ as for those we set $I_i^{ignore} = \emptyset$. Next, assume that the set $i$ is not boring. Recall that in \cref{alg:deterministic}, we define for such sets
\begin{align}
    I_i^{ignore} = \bigcup_{B \in \left( \fB_i^{small} \cup \fB_i^{size-one}\right)} B.  
\end{align}
First, we analyze the size-one buckets. Let us call them  $B^{one}_1, \dots, B^{one}_\tau$ and let $\ell_1 < \dots < \ell_\tau$ be such that for the unique element $a_{i,j}$ with $B_i^{one} = \{j\}$ we have $2^{\ell_i} \le a_{i,j} < 2^{\ell_i + 1}$. We note that 
\begin{align}
\sum_{j \in B \in \fB_i^{size-one}} a_{i,j} \le 2\cdot 2^{\ell_\tau + 1} \le 4 \cdot 2^{\ell_\tau}. 
\end{align}
We consider two possibilities. Either $\Delta_i \ge 2^{\ell_\tau}$, in which case we conclude
\begin{align}
\label{eq:spongebog1}
\sum_{j \in B \in \fB_i^{size-one}} a_{i,j} \le 4\Delta_i.    
\end{align}
In the other case, we have $\Delta_i < 2^{\ell_\tau}$. However, this implies $\Delta_i^2 < \sum_{j \in [n]} a_{i,j}^2$, hence $i$ is boring by \cref{eq:boring2}, contradicting the assumption that $i$ is not boring.  

We next argue about small buckets. 
For every $B \in \fB_i^{small}$ we let $\ell_B$ be such that all elements $j \in B$ satisfy
\begin{align}
\label{eq:float}
2^{\ell_B} \le a_{i,j} \le 2^{\ell_B + 1}
\end{align}
Recall that by definition of small buckets in \cref{eq:def_small_bucket} we have that 
\begin{align}
\label{eq:wing}
|B| \le \frac{\Delta_i^2}{\sum_{j \in [n]} a_{i,j}^2}.     
\end{align}
Let us write $\tau = |\fB_i^{small}|$ and sort the small buckets according to their size, i.e., we write 
\begin{align}
    |B_1| \le |B_2| \le \dots \le |B_{\tau}|
\end{align}
and use $\ell_i = \ell_{B_i}$ to simplify notation. Next, define numbers $\beta_1, \beta_2, \dots, \beta_\tau$ such that $\beta_1 = |B_1|$ and for every $1 \le \tau' \le \tau$ we have $\beta_{\tau'+1} = |B_{\tau'+1}| - |B_{\tau'}|$.  In other words, we have for each $B_{\tau'}$ that
\begin{align}
\label{eq:section}
    |B_{\tau'}| = \beta_1 + \beta_2 + \dots + \beta_{\tau'}
\end{align}
Note that all variables $\beta_i$ are nonnegative. 
We will now write
\begin{align}
    \sum_{j \in B \in \fB_i^{small}} a_{i, j} 
    &\le 2\sum_{B \in \fB_i^{small}} |B| \cdot 2^{\ell_B} && \text{\cref{eq:float}} \\
    &= 2 \sum_{1 \le \tau' \le \tau} \left( \beta_1 + \beta_2 + \dots + \beta_{\tau'}\right) \cdot 2^{\ell_{\tau'}}\\
    &= 2 \sum_{1 \le \tau' \le \tau} \beta_{\tau'} \cdot \left( 2^{\ell_{\tau'}} + 2^{\ell_{\tau'+1}} + \dots + 2^{\ell_{\tau}} \right) \label{eq:apathy}
\end{align}

For each $\tau'$, we define 
\begin{align}
L_{\tau'} = \max\left(\ell_{\tau'}, \ell_{\tau'+1}, \dots, \ell_{\tau}\right).     
\end{align}
We can then upper bound \cref{eq:apathy} and get
\begin{align}
    \label{eq:disk}
    \sum_{j \in B \in \fB_i^{small}} a_{i, j} 
    \le 4 \sum_{1 \le \tau' \le \tau} \beta_{\tau'} \cdot 2^{L_{\tau'}}
\end{align}
Next, we will use the Cauchy-Schwarz inequality for the vectors $(\sqrt{\beta_{\tau'}})_{1 \le \tau' \le \tau}$ and $(\sqrt{\beta_{\tau'}}2^{L_{\tau'}})_{1 \le \tau' \le \tau}$ that says that
\begin{align}
\label{eq:cs}
    \sum_{1 \le {\tau'} \le \tau} \beta_{\tau'} \cdot 2^{L_{\tau'}}
    \le \sqrt{ \left( 
    \sum_{1\le {\tau'} \le \tau} \beta_{\tau'}
    \right) \cdot
    \left(
    \sum_{1 \le {\tau'} \le \tau} \beta_{\tau'} 2^{2L_{\tau'}}
    \right)}
\end{align}
We will now rewrite the two terms on the right-hand side. First, we note that
\begin{align}
\label{eq:us}
    \sum_{1\le {\tau'} \le \tau} \beta_{\tau'} = |B_\tau|
\end{align}
by \cref{eq:section}. 
Next, we write
\begin{align}
    \sum_{1 \le {\tau'} \le \tau} \beta_{\tau'} 2^{2L_{\tau'}}
    &\le \sum_{1 \le {\tau'} \le \tau} \beta_{\tau'} \left(2^{2\ell_{\tau'}} + 2^{2\ell_{{\tau'}+1}} + \dots + 2^{2\ell_{\tau}} \right)\\
    &= \sum_{1 \le {\tau'} \le \tau} \left( \beta_1 + \dots + \beta_{\tau'} \right) 2^{2\ell_{\tau'}} \\
    &= \sum_{1 \le {\tau'} \le \tau} |B_{\tau'}| 2^{2\ell_{\tau'}}\\
    &\le \sum_{j \in B \in \fB_i^{small}} a_{i,j}^2 
    \le \sum_{j \in [n]} a_{i,j}^2 \label{eq:evoke}
\end{align}

Putting \cref{eq:disk,eq:cs,eq:us,eq:evoke} together, we conclude that
\begin{align}
    \label{eq:confront}
    \sum_{j \in B \in \fB_i^{small}} a_{i, j} 
    \le 4 \sqrt{|B_\tau| \cdot \left( \sum_{j \in [n]} a_{i,j}^2\right)}
\end{align}
Finally, we use the assumption \cref{eq:wing} on $B_\tau$ and conclude that
\begin{align}
    \label{eq:jungle}
    \sum_{j \in B \in \fB_i^{small}} a_{i, j} 
    \le 4 \sqrt{\frac{\Delta_i^2}{\sum_{j \in [n]}a_{i,j}^2} \cdot \left( \sum_{j \in [n]} a_{i,j}^2\right)} = 4\Delta_i. 
\end{align}
The proof follows from \cref{eq:spongebog1} together with \cref{eq:jungle}. 
\end{proof}

On the other hand, for the large buckets we prove in \cref{cl:psi_concentration} that  whenever the sum of terms $a_{i,j}$ or $a_{i,j}^2$ over indices still moving in the end does not drop by a constant factor, the potential increase in $\Pot_t$ is very large. 

\begin{claim}
\label{cl:psi_concentration}
For any nonboring set $i$, the following holds. Whenever
    \begin{align}
        \label{eq:cable}
        \sum_{\substack{j \in B \in \fB_i^{large} \\ p_{j, T} \not\in \{0,1\}}} a_{i, j} 
        > 0.99 \sum_{j \in [n]} a_{i, j},
    \end{align}
    or 
    \begin{align}
        \label{eq:cable2}
        \sum_{\substack{j \in B \in \fB_i^{large} \\ p_{j, T} \not\in \{0,1\}}} a^2_{i, j} 
        > 0.99 \sum_{j \in [n]} a_{i, j}^2,
    \end{align}
    we have 
    \begin{align}
        \label{eq:ceremony}
\sum_{B \in \fB_i^{represent}} \frac{\Psi_{B, t}}{\Psi_{B, 0}} \cdot \exp\left( - \min(|B|, k)/10^{30} \right)
\ge  \exp\left(\frac{1}{10^{12} } \min\left( \frac{ \Delta_i^2  }{\sum_{j \in [n]} a^2_{i, j}}, \frac{k \Delta_i}{\sum_{j \in [n]} a_{i, j}}\right)\right).
    \end{align}
\end{claim}

The proof has two parts; first, we prove that whenever \cref{eq:cable} or \cref{eq:cable2} holds, then there is necessarily a representative bucket $B \in \fB_i^{represent}$ such that 
$    |B_{leftover}| \ge |B|/10$. Next, we use \cref{cl:psi_concentration} to derive \cref{eq:ceremony}. 
\begin{proof}

First, recall the definition of $\fB_i^{represent}$: we group the buckets of $\fB_i^{large}$ into maximal \emph{groups} of buckets of the same size. For any such group $\fB_i^\beta$ of buckets of size $|B| = \beta$, we order the buckets as $B_1, B_2, \dots, B_\tau$ such that $\ell_1 < \ell_2 < \dots < \ell_\tau$ where we use $\ell_\iota = \ell_{B_\iota}$ to simplify notation. The bucket $B_\tau$ is then chosen as the representative bucket for that group. 

First, assume \cref{eq:cable2} holds (we later deal with \cref{eq:cable}). 
We can then write
\begin{align}
        \sum_{\substack{j \in B \in \fB_i^{large} \\ p_{j, T} \in \{0,1\}}} a_{i, j}^2 
&=         \sum_{\substack{j \in B \in \fB_i^{large} }} a_{i, j}^2 
        - \sum_{\substack{j \in B \in \fB_i^{large} \\ p_{j, T} \not\in \{0,1\}}} a_{i, j}^2\\
        &< \sum_{\substack{j \in B \in \fB_i^{large} }} a_{i, j}^2 
        - 0.99\sum_{\substack{j \in [n] }} a_{i, j}^2 && \text{\cref{eq:cable2}}\\
        &< 0.01 \sum_{\substack{j \in B \in \fB_i^{large} }} a_{i, j}^2 
\end{align}
The inequality  implies the existence of at least one group $\fB_i^\beta$ such that
\begin{align}
\label{eq:pokemon}
        \sum_{\substack{j \in B \in \fB_i^\beta \\ p_{j,T} \in \{0, 1\}}} a_{i,j}^2 
        < 0.01 \sum_{\substack{j \in B \in \fB_i^\beta}} a_{i,j}^2,
    \end{align}
Let us now fix this group $\beta$ and let the respective buckets of the group be $B_1 < \dots < B_\tau$. We next focus on the bucket $B_\tau$, which is the representative bucket of the group $\beta$. 
We start by writing
\begin{align}
    \sum_{j \in B \in \fB_i^{\beta}} a_{i,j}^2 
    =     \sum_{j \in B_\tau} a_{i,j}^2  + \sum_{ \in B_1 \cup \dots \cup B_{\tau-1} } a_{i,j}^2
\end{align}
On one hand,  we have
\begin{align}
     \sum_{j \in B_\tau} a_{i,j}^2 \ge \beta \cdot 2^{2\ell_\tau}. 
\end{align}
On the other hand, we have
\begin{align}
     \sum_{j \in B_1 \cup \dots \cup B_{\tau-1} } a_{i,j}^2 
     \le\left( \beta 2^{2\ell_\tau} + 2^{2(\ell_\tau - 1)}+ 2^{2(\ell_\tau - 2)} + \dots\right) 
     \le 2\beta\cdot 2^{2\ell_\tau}
\end{align}
where we used that the size of elements of the largest non-representative bucket is at most $2^{\ell_t}$. 
We conclude that
\begin{align}
\label{eq:veryfun}
    \sum_{j \in B_\tau} a_{i,j}^2
    \ge
    \frac13 \sum_{j \in B \in \fB_i^{\beta}} a_{i,j}^2 .
\end{align}
Putting \cref{eq:pokemon,eq:veryfun} together, we get
\begin{align}
        |\{ j \in B_\tau, p_{j, T} \in \{0,1\} \}| \cdot 2^{2\ell_\tau} 
        &\le \sum_{\substack{j \in B_\tau \\ p_{j,T} \in \{0, 1\}}} a_{i,j}^2 && a_{i,j}^2  \ge 2^{2\ell_\tau} \\
        &\le \sum_{\substack{j \in B \in \fB_i^\beta\\ p_{j,T} \in \{0, 1\}}} a_{i,j}^2\\
        &< 0.01 \sum_{\substack{j \in B \in \fB_i^\beta}} a_{i,j}^2 && \text{\cref{eq:pokemon}}\\
        &\le  0.03 \sum_{j \in B_\tau} a_{i,j}^2 && \text{\cref{eq:veryfun}}     \\ 
        &\le 0.12 |B_\tau| 2^{2\ell_\tau} && a_{i,j}^2 \le 2^{2\ell_\tau + 2}
    \end{align}
We conclude that
\begin{align}
    \label{eq:onestep2}
    |\{ j \in B_\tau, p_{j, T} \in \{0,1\} \}| 
    \le 0.12 |B_\tau|
\end{align}
Next, we notice that if instead of assuming \cref{eq:cable2} we started with \cref{eq:cable} as our assumption, we could do the same argument and still arrive at \cref{eq:onestep2} (in fact, even with a slightly better constant factor loss. In particular, both possible assumptions \cref{eq:cable,eq:cable2} allow us to apply \cref{cl:Psi_ratio} to the bucket $B_\tau$. We conclude that
\begin{align}
\sum_{B \in \fB_i^{represent}} \frac{\Psi_{B, t}}{\Psi_{B, 0}} \cdot \exp\left( - |B|/10^{30} \right)
&\ge \frac{\Psi_{B_\tau, t}}{\Psi_{B_\tau, 0}} \cdot \exp\left( - \min(|B_\tau|,k)/10^{30} \right)\\
&\ge\exp\left( \min(|B_\tau|,k)/10^{10} \right) && \text{\cref{cl:Psi_ratio}}
\end{align}
Note that by definition, any representative bucket $B_\tau$ is large and hence, by \cref{eq:def_small_bucket} it satisfies
\begin{align}
    |B_\tau| \ge \frac{\Delta_i^2}{\sum_{j \in [n]} a_{i,j}^2}
\end{align}
and thus we conclude that 
\begin{align}
    \text{LHS of \cref{eq:ceremony}} \ge\exp\left( \min\left(\frac{\Delta_i^2}{\sum_{j \in [n]} a_{i,j}^2},k\right)/10^{10} \right)
\end{align}
Finally, we use the assumption from the claim statement that the set $i$ is not boring. In particular, we can assume that $100\Delta_i < \sum_{j \in [n]} a_{i,j}$ 
which implies \cref{eq:ceremony} that was to be proven.

\begin{comment}
\begin{align}
    \sum_{\substack{j \in B \in \fB_i^{represent} \\ p_{j, T} \not\in \{0,1\}}} a_{i,j}^2 
    \ge 
    \frac{1}{10^{20}} \cdot \sum_{\substack{j \in B \in \fB_i^{large} \\ p_{j, T} \not\in \{0,1\}}} a_{i,j}^2 
\end{align}
To see this, note that we grouped all buckets $B \in \fB_i^{large}$ with the same value of $|B|$ together and assigned the bucket with largest value of $\ell_B$ to the set $\fB_i^{represent}$. Consider any such group $B_1, B_2, \dots, B_\tau$ with $B_\tau$ the representing bucket and for easy of notation use $\ell_\iota = \ell_{B_\iota}$ and $b = |B_1| = \dots = |B_\tau|$. 
We have
\begin{align}
    \sum_{1 \le j \le t} \sum_{i \in B_\iota} a_{i, j}^2
    &\le 4\sum_{1 \le j \le t} \sum_{i \in B_\iota} 2^{2\ell_\iota} && \text{\cref{eq:def_bucket}}\\
    &= 4b \sum_{1 \le j \le t} 2^{2\ell_\iota}
    \le 8b 2^{2\ell_\tau}
\end{align}
\end{comment}
\end{proof}

\paragraph{Bounding the number of bad sets}
We can now finish the proof of \cref{lem:main} by proving that the set $I^{bad}$ of bad sets is sufficiently small. 
\begin{claim}
    \label{cl:at_last}
    For the set $I^{bad}$ we have
  \begin{align}
    |I^{bad}| \le \sum_{i \in [m]} prob_i^{bad, L\text{\ref{lem:main}}}        
    \end{align}
\end{claim}
\begin{proof}
Note that sets with boringly large $\Delta_i$ have $I_i^{ignore} = [n]$ and thus they satisfy both \cref{op1} and \cref{op2}. Moreover, by definition, they satisfy \cref{op3} weakened in \cref{eq:stinky}. 
The sets with boringly small $\Delta$ on the other hand satisfy $prob_i^{bad, L\text{\ref{lem:main}}} > 1$ for $c$ in \cref{eq:wind-up} large enough. 
Thus, it suffices to prove
  \begin{align}
  \label{omg}
    |\{ i \in I^{bad}, \text{$i$ not boring}\} | \le \sum_{\substack{i \in [m] \\ \text{$i$ not boring}}} prob_i^{bad, L\text{\ref{lem:main}}}  
    \end{align}

Recall that our algorithm makes sure that in every step, the potential $\Pot$ from \cref{eq:pot_definition} is monotone, as proven in \cref{cl:monotone}. In particular, we conclude that $\Pot_T \le \Pot_0$, i.e., 
\begin{align}
\label{lhs}
&\sum_{\substack{i \in [m]\\ \text{$i$ is not boring}}}prob_i^{bad, L \text{\ref{lem:main}}}\left( \frac{\Phi^{(1)}_{i, T}}{\Phi^{(1)}_{i, 0}} +  \frac{\Phi^{(2)}_{i, T}}{\Phi^{(2)}_{i, 0}} + \sum_{B \in \fB_i^{represent}} \frac{\Psi_{B, T}}{\Psi_{B, 0}} \cdot \exp\left( - \min\left(|B|, k\right)/10^{30} \right) \right)\\
&\le \sum_{\substack{i \in [m]\\ \text{$i$ is not boring}}} prob_i^{bad, L \text{\ref{lem:main}}}\left( 1 + 1 + \sum_{B \in \fB_i^{represent}} 1 \cdot \exp\left( - \min\left(|B|, k\right)/10^{30}\right) \right)    \label{rhs}
\end{align}

Recall that all representative buckets have different sizes. For those representative buckets $B$ for which $|B| \le k$, we can upper bound
\begin{align}
    \sum_{\substack{B \in \fB_i^{represent}\\ |B| \le k}} \exp\left( - |B|/10^{30}\right)
    \le \sum_{x = 1}^{\infty} \exp( -x / 10^{30})
    \le 10^{40}
\end{align}
Next, we note that we can assume that $k \ge 10^{40} \log n$ since all bounds on work and depth are up to $\poly\log(n)$ factors. This enables us to conclude that for representative buckets $B$ for which $|B| > k$ we have
\begin{align}
    \sum_{\substack{B \in \fB_i^{represent} \\ |B| > k}} \exp\left( - k/10^{30}\right)
    \le n \cdot \exp\left( - k/10^{30}\right)
    \le 1
\end{align}
We conclude that
\begin{align}
\label{eq:ch}
&\text{RHS of \cref{rhs}}
\le \sum_{\substack{i \in [m]\\ \text{$i$ is not boring}}} prob_i^{bad, L \text{\ref{lem:main}}}
\end{align}
if the constant $c$ in the definition of $prob_i^{bad, L \text{\ref{lem:main}}}$ in \cref{eq:wind-up} is chosen large enough. 

Next, assume that a set $i \in [m]$ is bad. Then, if \cref{op3} fails, we use \cref{cl:Phi_ratio,cl:Phi2_ratio} to conclude that 
\begin{align}
\label{eq:11}
\frac{\Phi^{(1)}_{i, T}}{\Phi^{(1)}_{i, 0}} +  \frac{\Phi^{(2)}_{i, T}}{\Phi^{(2)}_{i, 0}}
\ge   
\exp\left(
    \frac{1}{10^{9} } \min\left( 
        \frac{ \Delta_i^2  }{\sum_{j \in [n]} a^2_{i, j}}, \frac{k \Delta_i}{\sum_{j \in [n]} a_{i, j}}
    \right)
\right)
\end{align}
Otherwise, if \cref{op1} or \cref{op2} fails, we use \cref{cl:psi_concentration} to conclude that
\begin{align}
\label{eq:22}
\sum_{B \in \fB_i^{represent}} \frac{\Psi_{B, t}}{\Psi_{B, 0}} \cdot \exp\left( - \min(|B|, k)/10^{30} \right)
\ge  \exp\left(\frac{1}{10^{10} } \min\left( \frac{ \Delta_i^2  }{\sum_{j \in [n]} a^2_{i, j}}, \frac{k \Delta_i}{\sum_{j \in [n]} a_{i, j}}\right)\right).
\end{align}
Plugging in \cref{eq:11,eq:22} back to \cref{lhs}, we conclude that
\begin{align}
 \label{eq:doma}
    &\text{LHS of \cref{lhs}}\\
    &\ge
    \sum_{\substack{i \in I^{bad}\\ \text{$i$ is not boring}}} prob_i^{bad, L \text{\ref{lem:main}}} \cdot \exp\left(\frac{1}{10^{10} } \min\left( \frac{ \Delta_i^2  }{\sum_{j \in [n]} a^2_{i, j}}, \frac{k \Delta_i}{\sum_{j \in [n]} a_{i, j}}\right)\right)\\
&\ge    \sum_{\substack{i \in I^{bad}\\ \text{$i$ is not boring}}}  \exp\left(\frac{1}{10^{11} } \min\left( \frac{ \Delta_i^2  }{\sum_{j \in [n]} a^2_{i, j}}, \frac{k \Delta_i}{\sum_{j \in [n]} a_{i, j}}\right)\right) && \text{definition of $prob_i^{bad, L \text{\ref{lem:main}}}$}\\
&\ge |\{i \in I_{bad}, \text{$i$ is not boring}\}|
\end{align}
Putting \cref{eq:ch,eq:doma} together finishes the proof of \cref{omg}, as needed. 
\end{proof}

\section{Applications of the partial fixing lemma}

In this section, we present the proofs of our main theorems. The following result can be proven by repeatedly applying \cref{lem:main}.

\begin{lemma}
\label{lem:weighted_good_for_large_p}
There exists a constant $c > 0$ such that the following holds. Let $n,m,k \in \mathbb{N}$ be arbitrary and $A \in \mathbb{R}_{\geq 0}^{m \times n}$. Also, let $\p \in [0,1]^n$ and $\Deltav \in \mathbb{R}_{>0}^m$. For each $i \in [m]$, let

\[\failtwo := c\exp\bigg( - (1/c)\min \{  \frac{\Delta^2_i}{\sum_{j \in [n]} a^2_{ij}},\frac{\Delta_i k}{\sum_{j \in [n]} \aij}  \} \bigg).\]

There exists a deterministic PRAM algorithm with $\tilde{O}(\max(nnz(A),n,m)\poly(k))$ work and $\poly(\log(nm),k)$ depth that outputs a set $I^{bad} \subseteq [m]$ with $|I^{bad}| \leq \sum_{i \in [m]} prob^{bad}_i$ and a vector $\q \in \{0,1\}^n$ such that for every $i \in [m] \setminus I^{bad}$ it holds that

\[|\sum_{j \in [n]} \aij(\pj - \qj)| \leq \Delta_i.\]
\end{lemma}

We note that if we replace $\frac{\Deltai k}{\sum_{j \in [n]} \aij}$ in the exponent with $\frac{\Deltai k}{\sum_{j \in [n]} \pj\aij}$, then we would obtain the Hoeffding-like derandomization result \cref{thm:hoeffding}. In particular, \cref{lem:weighted_good_for_large_p} matches the guarantees of \cref{thm:hoeffding} if $\pj \geq \frac{1}{2}$ for every $j \in [n]$ but is in general much weaker. Nevertheless, we can use repeated applications of \cref{lem:weighted_good_for_large_p} to prove both \cref{thm:hoeffding} and \cref{thm:chernoff}. We remark that the proof below is conceptually easy but tedious.

\begin{proof}
First, observe that we can assume without loss of generality that for every $i \in [m]$, $\failtwo < 1$ and therefore also that $\Delta_i \geq \frac{1000\sum_{j \in [n]} \aij}{k}$ for every $i \in [m]$. Let

\[ratio(A,\Deltav) := \max_{i \in [m]} \frac{\sum_{j \in [n]} \aij}{\Delta_i}.\]

By our assumption, we have that $ratio(A,\Deltav) \leq \frac{k}{1000}$. If $ratio(A,\Deltav) \leq 1$, then setting $\q$ as the all-zero vector fulfills the conditions of \cref{lem:weighted_good_for_large_p}.
Now, assume that $ratio(A,\Deltav) > 1$.

We compute the output vector $\q$ in the following way: First, let $\mathbf{p^{partial,in}} \in \mathbb{R}^n$ be the vector one obtains from $\p$ by rounding each entry to the closest multiple of $1/k$. Note that

\[|\sum_{j \in [n]} \aij(\pj - p^{partial,in}_j)| \leq \frac{1}{k} \sum_{j \in [n]} \aij \leq \frac{\Delta_i}{1000},\]

where the last inequality follows from our assumption $\Delta_i \geq \frac{1000\sum_{j \in [n]} \aij}{k}$. 

By applying \cref{lem:main} with input $A, \Deltav$ and $\mathbf{p^{partial,in}}$, we can compute in $\tilde{O}(\max(nnz(A),n,m)\poly(k))$ work and $\poly(\log(nm),k)$ depth sets  $(I^{ignore}_i)_{i \in [m]}$ and $ I^{bad,partial} \subseteq [m]$ and a vector $\mathbf{p^{partial,out}} \in [0,1]^n$ such that $\sum_{j \in I^{ignore}_i} \aij \leq \frac{\Delta_i}{1000}$ for every $i \in [m]$, $|I^{bad,partial}| \leq \sum_{i \in [m]} prob_i^{bad,L\ref{lem:main}}(A,\Deltav,k)$ and such that for $J^{non-int} := \{j \in [n] \colon p^{partial,out}_j \notin \{0,1\}\}$, it holds for every $i \in I^{good,partial} := [m] \setminus I^{bad,partial}$ that

\begin{enumerate}
\item $\sum_{j \in J^{non-int} \setminus I^{ignore}_i} \aij \leq 0.99 \sum_{j \in [n]} \aij$,
\item $\sum_{j \in J^{non-int}\setminus I^{ignore}_i} a^2_{ij} \leq 0.99\sum_{j \in [n]} \aij^2$ and
\item $|\sum_{j \in [n]} \aij(p^{partial,in}_j - p^{partial,out}_j)| \leq \frac{\Delta_i}{1000}$.
\end{enumerate}

Let $A^{rec} \in \mathbb{R}^{I^{good,partial} \times J^{non-int}}_{\geq 0}, \precin \in [0,1]^{J^{non-int}}$ and $\mathbf{\Delta^{rec}} \in \mathbb{R}_{>0}^{I^{good,partial}}$ be defined as 

\begin{enumerate}
\item $
\arecij =
\begin{cases}
0, \text{if $j \in I^{ignore}_i$}\\
\aij ,\text{otherwise}
\end{cases} 
$ for every $i \in I^{good,partial}, j \in J^{non-int}$,
\item $\precinj = p^{partial,out}_j$ for every $j \in J^{non-int}$ and
\item $\Delta^{rec}_i = 0.997\Delta_i$ for every $i \in I^{good,partial}$.
\end{enumerate}

Let $I^{bad,rec} \subseteq I^{good,partial}$ with $|I^{bad,rec}| \leq \sum_{i \in I^{good,partial}} \failtwo(A^{rec},\Deltav^{rec},k)$ 
and $\precout \in \{0,1\}^{J^{non-int}}$ such that for every $i \in I^{good,partial} \setminus I^{bad,rec}$, it holds that

\[|\sum_{j \in J^{non-int}} \arecij(\pj^{rec,in} - \pj^{rec,out})| \leq \Delta^{rec}_i.\]

Note that these are exactly the output guarantees of \cref{lem:weighted_good_for_large_p} with input $A^{rec}, \precin$ and $\Deltav^{rec}$, the lemma we want to prove. Also, note that

\[ratio(A^{rec},\mathbf{\Delta^{rec}}) := \max_{i \in I^{good,partial}} \frac{\sum_{j \in J^{non-int}} \arecij}{\Delta^{rec}_i} \leq \max_{i \in I^{good,partial}} \frac{0.99}{0.997}\frac{\sum_{j\in [n]} \aij}{\Delta_i} \leq 0.999ratio(A,\Deltav).\]
Thus, we can inductively assume that $I^{bad,rec}$ and $\precout$ can be efficiently computed in parallel. We provide more details of this induction at the end of the proof.

Finally, we output $I^{bad} := I^{bad,partial} \cup I^{bad,rec}$ and $\q \in \{0,1\}^n$ with

\[\qj =
\begin{cases}
\precoutj &, \text{if $j \in J^{non-int}$}\\
p^{partial,out}_j &,\text{otherwise}
\end{cases} \text{ for } j \in [n].\]

We start by verifying that $I^{bad}$ and $\q$ satisfy the output guarantees of \cref{lem:weighted_good_for_large_p}.
To that end, consider first an arbitrary $j \in [m] \setminus I^{bad}$. First, note that

\begin{align*}
|\sum_{j \in [n]} \aij(p^{partial,out}_j - \qj)| &= |\sum_{j \in J^{non-int}} \aij(p^{partial,out}_j - \qj)| \\
                                                   &  \leq |\sum_{j \in J^{non-int}} \arecij(p^{partial,out}_j - \qj)| + \sum_{j \in J^{non-int}} |\aij - \arecij| \\
                                                   &\leq |\sum_{j \in J^{non-int}} \arecij(\precinj - \precoutj)| + \sum_{j \in I^{ignore}_i} \aij \\
&\leq \Delta^{rec}_i + \frac{\Delta_i}{1000} = 0.997\Delta_i + \frac{\Delta_i}{1000} = 0.998\Delta_i.
\end{align*}

Therefore,

\begin{align*}
|\sum_{j \in [n]} \aij(\pj - \qj)| &\leq |\sum_{j \in [n]} \aij(\pj - p^{partial,in}_j)| + |\sum_{j \in [n]} \aij(p^{partial,in}_j - p^{partial,out}_j)| + |\sum_{j \in [n]} \aij(p^{partial,out}_j - \qj)| \\
&\leq \frac{\Delta_i}{1000} + \frac{\Delta_i}{1000} + 0.998\Delta_i = \Delta_i,
\end{align*}

which is one of the output conditions of \cref{lem:weighted_good_for_large_p}.

Consider now an arbitrary $i \in [m]$. Let $c^{L\ref{lem:main}} \geq 1$ be the constant used in \cref{lem:main} and recall our assumption that $prob^{bad, L\ref{lem:weighted_good_for_large_p}}_i < 1$. Let 

\[\alpha_i := \min \left( \frac{\Delta^2_i}{\sum_{j \in [n]} a^2_{ij}},\frac{\Delta_i k}{\sum_{j \in [n]} \aij} \right).\]

For $c$ being sufficiently large, we have

\begin{align*}
prob^{bad,L\ref{lem:main}}_i(A,\Deltav,k) &:= c^{L\ref{lem:main}} \exp(-(1/c^{L\ref{lem:main}})\alpha_i) \\
&\leq c^{L\ref{lem:main}} \exp(-(1/c)\alpha_i)^{c/c^{L\ref{lem:main}}} \\
&\leq c\exp(-(1/c)\alpha_i)^{10} \\
&= \failtwo(A,\Deltav,k) \cdot \exp(-(1/c)\alpha_i)^9 \\
&\leq \frac{1}{2}\failtwo(A,\Deltav,k),
\end{align*}
where the last inequality follows from $\exp(-(1/c)\alpha_i)) < 1/c \leq \frac{1}{2}$ as otherwise $prob^{bad, L\ref{lem:weighted_good_for_large_p}}_i \geq 1$.

Also, for every $i \in I^{good,partial}$ and $c$ being sufficiently large, we have

\begin{align*}
    \failtwo(A^{rec},\mathbf{\Delta^{rec}},k) &:= c\exp \left( -(1/c)\min \left(\frac{(\Delta^{rec}_i)^2}{\sum_{j \in J^{non-int}} (\arecij)^2}, \frac{\Delta^{rec}_i k}{\sum_{j \in J^{non-int}} \arecij}\right)\right) \\
    &\leq c\exp \left( -(1/c)\min \left(\frac{(\Delta^{rec}_i)^2}{\sum_{j \in J^{non-int}} (\arecij)^2}, \frac{\Delta^{rec}_i k}{\sum_{j \in J^{non-int}} \arecij}\right) \right) \\
    &\leq c\exp \left( -(1/c)\min \left(\frac{0.997^2}{0.99^2}\frac{\Delta_i^2}{\sum_{j \in [n]} \aij^2}, \frac{0.997}{0.99}\frac{\Delta_i k}{\sum_{j \in [n]} \aij} \right) \right) \\
    &\leq c\exp \left( -(1/c)\min \left(\frac{\Delta_i^2}{\sum_{j \in [n]} \aij^2}, \frac{\Delta_i k}{\sum_{j \in [n]} \aij}\right) \right)^{1.001} \\
    &\leq \failtwo(A,\Delta,k) \cdot (1/c)^{0.001} \\
    &\leq \frac{1}{2} \failtwo(A,\Delta,k).
\end{align*}

Therefore,
\begin{align*}
|I^{bad}| \leq |I^{bad,partial}| + |I^{bad,rec}| &\leq \sum_{i \in [m]} prob_i^{bad,L\ref{lem:main}}(A,\Deltav,k) + \sum_{i \in I^{good,partial}} \failtwo(A^{rec},\Deltav^{rec},k) \\
&\leq \sum_{i \in [m]} \frac{1}{2}prob_i^{bad,L\ref{lem:main}}(A,\Deltav,k) + \sum_{i \in [m]} \frac{1}{2}\failtwo(A,\Deltav,k) \\
&\leq \sum_{i \in [m]} \failtwo(A,\Deltav,k),
\end{align*}
as needed. Thus, the output indeed satisfies the guarantees of \cref{lem:weighted_good_for_large_p}. In particular, we devised a correct recursive algorithm. It remains to argue about the work and depth of the recursive algorithm.
Note that we argued above that we can assume that $ratio(A,\Deltav) \leq \frac{k}{1000}$.
The base case $ratio(A,\Deltav) \leq 1$ trivially takes $\tilde{O}(\max(nnz(A),n,m)\poly(k))$ work and $\poly(\log(nm),k)$ depth. On the other hand, if $ratio(A,\Deltav) > 1$,  we need $\tilde{O}(\max(nnz(A),n,m)\poly(k))$ work and $\poly(\log(nm),k)$ depth for running the algorithm of \cref{lem:main} and then perform one recursive call with $ratio(A^{rec},\mathbf{\Delta^{rec}}) \leq 0.999 ratio(A,\Deltav)$. Thus, the recursion depth of the algorithm is upper bounded by $O(\log k)$ and thus the overall work and depth is $\tilde{O}(\max(nnz(A),n,m)\poly(k))$ and $\poly(\log(nm),k)$, respectively. This finishes the proof of \cref{lem:weighted_good_for_large_p}.
\end{proof}

\subsection{Chernoff and Hoeffding like concentration}

We now use repeated applications of \cref{lem:weighted_good_for_large_p} to prove \cref{lem:hoeffding_chernoff}. Informally speaking, we use \cref{lem:weighted_good_for_large_p} to repeatedly "sample" the elements which are within a $2$-factor of the minimum sampling probability with probability $1/2$. The elements that are sampled increase their remaining sampling probability by a $ 2$ factor. For the elements that are not sampled, we set $\qj = 0$. Thus, in each iteration, we essentially increase the minimum sampling probability by a factor of $2$. As we assume $\pj \geq \frac{1}{2^k}$, only $k$ such subsamplings are needed. We note that dropping the assumption $\pj \geq \frac{1}{2^k}$ for every $j \in [n]$ and dropping the $k$ in the exponent of $\failthree$ would directly imply \cref{thm:hoeffding,thm:chernoff}. Note that we can assume without loss of generality that $k \gg \log(nm)$. Hence, the first condition only requires $\pj \geq \frac{1}{n}$ for every $j \in [n]$ and thus can be easily removed by setting $\qj = 0$ for every $j$ with $\pj < \frac{1}{n}$. Also, because $k \gg \log(m)$, we get $c \exp(-(1/c)k) \leq \frac{1}{2m}$ and thus the $k$ in the exponent adds at most an extra $m \cdot \frac{1}{2m} = 1/2$ in the allowed size of $\Ibad$. Thus, there exists again a simple argument to remove the $k$ in the exponent. We defer these simple arguments and thus the complete proofs of \cref{thm:hoeffding,thm:chernoff} to \cref{sec:hoeffdingchernofffull}.

\begin{lemma}
\label{lem:hoeffding_chernoff}
There exists a constant $c > 0$ such that the following holds. Let $n,m,k \in \mathbb{N}$ be arbitrary and $A \in \mathbb{R}_{\geq 0}^{m \times n}$. Also, let $\p \in \left[\frac{1}{2^k},1\right]^n$ and $\Deltav \in \mathbb{R}_{>0}^n$. For each $i \in [m]$, let

\[\failthree := c \exp(-(1/c) \min(\max(\expone,\exptwo),k))\]

where

\[\expone := \min \left(\frac{\Deltai^2}{\sum_{j\in [n]} \aij^2}, \frac{\Deltai k}{\sum_{j\in [n]} \pj \aij}\right)\]

and

\[\exptwo := \min \left(\frac{\Deltai^2}{\sum_{j\in [n]} \pj \aij (\max_{j \in [n]} \aij)}, \frac{\Deltai}{\max_{j\in [n]} \aij}, \frac{\Deltai k}{\sum_{j\in [n]} \pj \aij} \right).\]

There exists a deterministic parallel algorithm with $\tilde{O}(\max(nnz(A),n,m)\poly(k))$ work and $\poly(\log(nm),k)$ depth that outputs a set $\Ibad \subseteq [m]$ with $|\Ibad| \leq \sum_{i \in [m]} \failthree$ and a vector $\q \in \{0,1\}^n$ such that for every $i \in [m] \setminus \Ibad$ it holds that

\[|\sum_{j \in [n]} \aij(\pj - \qj)| \leq \Deltai.\]
\end{lemma}
\begin{proof}

For a given $\p$, we define $\ell = \lceil - \log(\min_{j \in [n]} \pj) \rceil$. Note that $\ell \leq k$. We prove \cref{lem:hoeffding_chernoff} by proving a slightly different statement by induction on $\ell$. \cref{lem:hoeffding_chernoff} then follows as a simple corollary of that statement. During the induction, we think of $k$ as being fixed.
To state the precise statement, we first introduce two sequences of numbers. Namely, for each $\ell \in [0,k]$, we define $\eps_\ell = \frac{1}{16}\max\left(0.8^\ell,\frac{1}{k}\right)$ and $\alpha_\ell = \frac{1}{2}\left(1 - 0.8^\ell + \frac{\ell}{k} \right)$. We have chosen $\eps_\ell$ and $\alpha_\ell$ such that $\eps_\ell$ is sufficiently large for certain concentrations to hold and such that $\alpha_\ell$ satisfies 

\begin{enumerate}
   \item $\alpha_\ell \in \left[\frac{1}{10},1 \right]$ for every $\ell \in [0,k]$ and
   \item $\alpha_\ell \geq \eps_\ell + (1 + \eps_\ell)\alpha_{\ell -1}$ for every $\ell \in [1,k]$.
\end{enumerate}

The first property follows immediately from the definition of $\alpha_\ell$. For the second one, using that $\alpha_{\ell-1} \leq 1$, we get

\begin{align*}
\eps_\ell + (1 + \eps_\ell)\alpha_{\ell - 1} \leq \alpha_{\ell-1} + 2\eps_{\ell} &\leq \frac{1}{2}(1 - 0.8^{\ell-1} + \frac{\ell-1}{k}) + \frac{1}{8}\max(0.8^\ell,\frac{1}{k}) \\
&\leq \frac{1}{2}\left(1 + \frac{1}{4}0.8^\ell - \frac{5}{4}0.8^\ell + \frac{\ell}{k}\right) = \alpha_\ell.
\end{align*}

The statement we prove by induction on $\ell$ is that for $c$ being sufficiently large, there exists a parallel algorithm with $\tilde{O}(\max(nnz(A),n,m)\poly(k))(\ell + 1)$ work and $\poly(\log(nm),k))(\ell + 1)$ depth that outputs a set $\Ibad \subseteq [m]$ with $|\Ibad| \leq \alpha_\ell \sum_{i\in [m]}\failthree$ and a vector $\q \in \{0,1\}^n$ such that for every $i \in [m] \setminus \Ibad$ it holds that

\[|\sum_{j \in [n]} \aij(\pj - \qj)| \leq \alpha_\ell \Deltai.\]

Note that because $\alpha_\ell \leq 1$ and $\ell \leq k$, \cref{lem:hoeffding_chernoff} follows as a simple corollary once we have proven the statement.

The base case $\ell \leq 1$ is a simple corollary of \cref{lem:weighted_good_for_large_p} for $c$ being sufficiently large. Here, we are using the fact that $\alpha_\ell \geq \frac{1}{10}$.

Now, assume that $\ell > 1$. Note that we can assume without loss of generality that $\failthree \leq 1$ for every $i \in [m]$. Let $\Jsmall := \{j \in [n] \colon \pj < 0.5^{\ell-1}\}$ and let $\Asmall \in \mathbb{R}^{m \times \Jsmall}, \psmallin \in \mathbb{R}^{\Jsmall}$, $\Deltasmall \in \mathbb{R}^m$ and $\ksmall$ be defined as

\begin{enumerate}
    \item $\asmallij = 2\pj\aij$ for every $i \in [m], j \in \Jsmall$,
    \item $\psmallinj = 1/2$ for every $j \in \Jsmall$ and
    \item $\Deltasmalli = \eps_{\ell}\Deltai$ for every $i \in [m]$ and
    \item $\ksmall = k^3.$
\end{enumerate}
By running the algorithm of \cref{lem:weighted_good_for_large_p} with input $\Asmall, \psmallin, \Deltasmall$ and $\ksmall$, we can compute with  $\tilde{O}(\max(nnz(A),n,m)\poly(k))$ work and $\poly(\log(nm),k))$ depth a set $\Ibadsmall \subseteq [m]$ with $|\Ibadsmall| \leq \sum_{i \in [m]} \failtwo(\Asmall,\Deltasmall,\ksmall)$ and a vector $\psmallout \in \{0,1\}^{\Jsmall}$ such that for every $i \in \Igoodsmall := [m] \setminus \Ibadsmall$, it holds that

\[|\sum_{j \in \Jsmall}\asmallij(\psmallinj - \psmalloutj)| \leq \Deltasmalli := \eps_\ell \Deltai.\]

In particular, for $\Jrec := [n] \setminus \{j \in \Jsmall \colon \psmalloutj = 0\}$, we get 

\[|\sum_{j \in \Jsmall} \pj\aij - \sum_{j \in \Jsmall \cap \Jrec} 2\pj\aij| \leq \eps_\ell \Deltai.\]

Let $\Arec \in \mathbb{R}^{\Igoodsmall \times \Jrec}, \precin \in \mathbb{R}^{\Jrec}$ and $\Deltarec \in \mathbb{R}^{\Igoodsmall}$ be defined as 

\begin{enumerate}
    \item $\arecij = \aij$ for every $i \in \Igoodsmall$, $j \in \Jrec$,
    \item $\precinj = \begin{cases}
\pj &, \text{if $j \in [n] \setminus \Jsmall$}\\
2\pj &,\text{if $j \in \Jsmall \cap \Jrec$}
\end{cases}$ and
\item $\Deltareci = \left(1 + \eps_\ell\right)\Deltai$ for every $i \in \Igoodsmall$.
\end{enumerate}
Using the induction hypothesis, we can compute in $\tilde{O}(\max(nnz(A),n,m)\poly(k)) \ell$ work and $\poly(\log(nm),k))\ell$ depth a set $\Ibadrec \subseteq \Igoodsmall$ with $|\Ibadrec| \leq  \alpha_{\ell-1}\sum_{i \in \Igoodsmall} \failthree(\Arec,\Deltarec,\precin,k)$ and a vector $\precout \in \mathbb{R}^{\Jrec}$ such that for every $i \in \Igoodsmall \setminus \Ibadrec$, it holds that

\[|\sum_{j\in \Jrec}\arecij(\precinj - \precoutj)| \leq \alpha_{\ell -1} \Deltareci = (1 + \eps_\ell) \alpha_{\ell-1} \Deltai.\]

Now, the final output is $\Ibad := \Ibadsmall \sqcup \Ibadrec$ and $\q \in \mathbb{R}^n$ with $\qj = \begin{cases}
\precoutj&, \text{if $j \in \Jrec$}\\
0 &,\text{if $j \in [n] \setminus \Jrec$}
\end{cases}$.

The algorithm runs in $\tilde{O}(\max(nnz(A),n,m)\poly(k)) (\ell+1)$ work and $\poly(\log(nm),k))(\ell+1)$ depth. Thus, to finish the induction, it remains to argue about the correctness of the algorithm. 

We start by showing that the deviation $|\sum_{j \in [n]}\aij(\pj-\qj)|$ is upper bounded by $\alpha_\ell \Deltai$ for every $i \in [m] \setminus \Ibad$. The first error comes from the fact that every $p_j \in \Jsmall$ is either increased by a factor of $2$ or set to $0$ by calling \cref{lem:hoeffding_chernoff}. We previously upper bounded the error from this step by

\[|\sum_{j \in \Jsmall}\pj\aij - \sum_{j \in \Jsmall \cap \Jrec} \precinj \aij| = |\sum_{j \in \Jsmall}\pj\aij - \sum_{j \in \Jsmall \cap \Jrec} 2\pj \aij| \leq \eps_\ell \Deltai.\]

We also previously upper-bounded the error introduced by the recursive call by

\[|\sum_{j \in \Jrec} \aij(\precinj - \qj)| = |\sum_{j \in \Jrec} \arecij(\precinj - \precoutj)| \leq (1+\eps_\ell)\alpha_{\ell-1}\Deltai.\]

Hence, using that $q_j = 0$ for every $j \in [n] \setminus \Jrec$, $\precinj = \pj$ for every $j \in [n] \setminus \Jsmall$ and $\alpha_\ell \geq \eps_\ell + (1+\eps_\ell)\alpha_{\ell-1}$, we can upper bound the overall error by

\begin{align*}
|\sum_{j \in [n]} \aij(\pj - \qj) = |\sum_{j \in \Jsmall}\pj\aij - \sum_{j \in \Jsmall \cap \Jrec} \precinj \aij| + |\sum_{j \in \Jrec} \aij(\precinj - \qj)| \leq \alpha_\ell \Deltai.
\end{align*}

Thus, it remains to upper bound the size of $\Ibadsmall$.
Below, we show that for every $i \in [m]$, we have $\failtwo(\Asmall,\Deltasmall,\ksmall) \leq \eps_\ell \failthree(A,\Deltav,\p,k)$ and for every $i \in \Igoodsmall$, we have
$\failthree(\Arec,\Deltarec,\precin,k) \leq \failthree(A,\Deltav,\p,k)$. Thus, by using the fact that $\alpha_\ell \geq \eps_\ell + \alpha_{\ell -1}$, we can conclude that

\begin{align*}
 |\Ibadsmall| &\leq |\Ibadsmall| + |\Ibadrec| \\
              &\leq \sum_{i \in [m]} \failtwo(\Asmall,\Deltasmall,\ksmall) + \alpha_{\ell -1} \sum_{i \in \Igoodsmall}  \failthree(\Arec,\Deltarec,\precin,k)\\
              &\leq \eps_\ell \sum_{i \in [m]}\failthree(A,\Deltav,\p,k) + \alpha_{\ell-1} \sum_{i \in [m]} \failthree(A,\Deltav,\p,k) \\
              &\leq \alpha_{\ell} \sum_{i \in [m]} \failthree(A,\Deltav,\p,k).
\end{align*}

We start by showing that $\failthree(\Arec,\Deltarec,\precin,k) \leq \failthree(A,\Deltav,\p,k)$ for every $i \in \Igoodsmall$. For every $i \in \Igoodsmall$, we have

\begin{align*}
\sum_{j \in \Jrec} \precinj \arecij &= \sum_{j \in [n] \setminus \Jsmall} \pj\aij + \sum_{j \in \Jrec \cap \Jsmall} 2\pj\aij \\
&\leq \sum_{j \in [n]} \pj\aij + |\sum_{j \in \Jsmall} \pj\aij - \sum_{j \in \Jsmall \cap \Jrec} 2\pj\aij| \\
&\leq \sum_{j \in [n]} \pj\aij + \eps_\ell \Deltai \leq (1 + \eps_{\ell})\max\left(\sum_{j \in [n]} \pj\aij, \Deltai \right).
\end{align*}

Therefore,

\[\frac{\Deltareci k}{\sum_{j \in \Jrec} \precinj \arecij} \geq \frac{(1+\eps_\ell)\Deltai k}{(1 + \eps_{\ell})\max\left(\sum_{j \in [n]} \pj\aij, \Deltai \right)} = \min \left( \frac{\Deltai k}{\sum_{j \in [n]} \pj\aij},k\right)\]

and

\[\frac{(\Deltareci)^2}{\sum_{j\in [n]} \precinj \arecij (\max_{j \in [n]} \arecij)}  \geq \min \left( \frac{\Deltai^2}{\sum_{j\in [n]} \pj \aij (\max_{j \in [n]} \aij)}, \frac{\Deltai}{\max_{j\in [n]} \aij}\right).
\]

The former inequality directly implies $\expone(\Arec,\Deltarec,\precin,k) \geq \min(\expone(A,\Deltav,\p,k),k)$ and both inequalities together directly imply  $\exptwo(\Arec,\Deltarec,\precin,k) \geq \min(\exptwo(A,\Deltav,\p,k),k)$.
Therefore,

\[\failthree(\Arec,\Deltarec,\precin,k) \leq \failthree(A,\Deltav,\p,k).\]

Thus, it remains to show that $\failtwo(\Asmall,\Deltasmall,\ksmall) \leq \eps_\ell \failthree(A,\Deltav,\p,k)$ for every $i \in [m]$.
For every $i \in [m]$, using the fact that $\eps_\ell \geq 0.8^\ell$ and $p_j \leq 0.5^{\ell - 1}$ for every $j \in \Jsmall$, we get

\begin{align*}
\frac{(\Deltasmalli)^2}{\sum_{j \in \Jsmall} (\asmallij)^2} =\frac{(\eps_\ell \Delta_i)^2}{\sum_{j \in \Jsmall}(2\pj\aij)^2} 
           &\geq \frac{0.8^{2\ell}}{16 \cdot 0.5^\ell} \frac{\Deltai^2}{\sum_{j \in [n]} \pj \aij^2} \\
           &\geq \frac{1.2^{\ell}}{16}\max\left(\frac{\Deltai^2}{\sum_{j \in [n]} \aij^2},\frac{\Deltai^2}{\sum_{j \in [n]} \pj\aij(\max_{j \in [n]} \aij)} \right).
\end{align*}

Using the fact that $\eps_\ell \geq \frac{1}{k}$ and $\ksmall := k^3$, we get

\[\frac{\Deltasmalli \ksmall}{\sum_{j \in \Jsmall} \asmallij} \geq \frac{\frac{1}{k}\Deltai k^3}{\sum_{j \in \Jsmall} 2\pj\aij} \geq \frac{k}{2}\frac{\Deltai k}{\sum_{j \in [n]} \pj \aij}.\]

Thus, we get

\[\min \left(\frac{(\Deltasmalli)^2}{\sum_{j \in \Jsmall} (\asmallij)^2},\frac{\Deltasmalli \ksmall}{\sum_{j \in \Jsmall} \asmallij}\right) \geq \min\left(\frac{1.2^\ell}{16},\frac{k}{2}\right)\max(\expone,\exptwo).\]

Thus, for $c$ being sufficiently large, and using that $\eps_\ell \gg (1/c)^{\min(1.2^\ell,k)}$ for $c$ being sufficiently large and $\failthree(A,\Deltav,\p,k) \leq 1$, we get

\begin{align*}
\failtwo(\Asmall,\Deltasmall,\ksmall) 
&:= \ctwo\exp\left( - (1/\ctwo)\min \left( \frac{(\Deltasmalli)^2}{\sum_{j \in \Jsmall} (\asmallij)^2},\frac{\Deltasmalli k}{\sum_{j \in \Jsmall} \asmallij}  \right) \right)\\
&\leq \ctwo\exp\bigg( - (1/\ctwo)\min\left(\frac{1.2^\ell}{16},\frac{k}{2}\right)\max(\expone,\exptwo) \bigg) \\
&\leq c\exp\left( - (1/c)\max(\expone,\exptwo)  \right)^{2\min \left(1.2^\ell,k\right)} \\
&= \failthree(A,\Deltav,\p,k)\cdot (\failthree(A,\Deltav,\p,k)/c)^{2\min(1.2^{\ell},k) - 1} \\
&\leq \failthree(A,\Deltav,\p,k)\cdot (1/c)^{\min(1.2^\ell,k)} \\
&\leq \eps_\ell \failthree(A,\Deltav,\p,k).
\end{align*}

This finishes the induction and therefore completes the proof of \cref{lem:hoeffding_chernoff}.
\end{proof}

\subsection{Bernstein-like concentration}

Finally, we prove \cref{thm:variance} by using the algorithm of \cref{thm:chernoff} once. In particular, for each $i \in [m]$ and $\ell \in \mathbb{Z}$, we define bucket $\Bil = \{j \in [n] \colon \aij \in (2^{\ell -1},2^\ell]\}$. We then assign each bucket $\Bil$ an allowed deviation $\Deltail$ such that the following is satisfied: on the one hand $\sum_{\ell} \Deltail \leq \Deltai$ for every $i$ and on the other hand $\sum_{\ell} \failil \leq \fail$, where informally speaking $\failil$ is the failure probability of \cref{thm:chernoff} to ensure that $|\sum_{j \in \Bil}\aij(\pj - \qj)| \leq \Deltail$. 
\variance*
\begin{proof}

Fix some $i \in [m]$. We define

\[\mui = \sum_{j \in [m]} \pj\aij \text{ and } \Vi := \sum_{j \in [m]} \pj \aij^2.\]

Also, for each $\ell \in \mathbb{Z}$, let $\Bil := \{j \in [n] \colon \aij \in (2^{\ell-1},2^\ell]\}$. We also define

\[\muil = \sum_{j \in \Bil} \pj\aij \text{ and } \Vil := \sum_{j \in \Bil} \pj \aij^2.\]

We also define $\lmaxi$ as the largest $\ell \in \mathbb{Z}$ with $\Bil \neq \emptyset$ and $\gammai = \frac{\Vi}{\Deltai2^{\lmaxi}}$.

Moreover, we define

\[\epsone = \sqrt{c \cdot \alphai\min(1,\sqrt{\gammai})\frac{\Vil}{\Vi}}, \epstwo = 0.9^{\lmaxi - \ell}, \epsthree = \frac{\muil}{\mui}, \epsil = \max(\epsone,\epstwo,\epsthree) \text{ and } \Deltail := \frac{\epsil}{100}\Deltai\]

and we also define

 \[\failil := \ccher  \exp \left(-(1/\ccher) \min \left(\frac{\Deltail^2}{\Vil}, \frac{\Deltail}{2^\ell}, \frac{\Deltail c \log(nm) k}{\muil} \right)\right).\]

Now, assume we split the $i$-th constraint into  multiple constraints, one for each nonempty bucket $B_{i,\ell}$. The corresponding row is defined by $a_{(i,\ell),j} = \begin{cases}
\aij &, \text{if $j \in B_{i,\ell}$}\\
0 &,\text{otherwise} 
\end{cases}$ for every $j \in [n]$ and the deviation we allows is $\Deltail$. Then, by using \cref{thm:chernoff} with $k' = c \log(nm) k$, we get a set $\Ibadcher \subseteq [m] \times \mathbb{Z}$ with $|\Ibadcher| \leq \sum_{i \in [m]}\sum_{\ell \colon \Bil \neq \emptyset} \failil$ and a vector $\mathbf{q}$ such that for every $(i,\ell) \notin \Ibadcher$ it holds that

\[|\sum_{j \in \Bil} \aij(\pj - \qj)|\leq \Deltail.\]

We not output $\Ibad = \{i \in [m] \colon \text{there exists an $\ell$ with $(i,\ell) \in \Ibadcher$}\}$ and $\mathbf{q}$. The algorithm runs in the desired work and depth. It thus remains to verify that the output is correct. To that end, we define

\[\Bbigi = \{\ell \leq \lmaxi \colon \Bil \neq \emptyset \text{ and } \epsone \geq 0.8^{\lmaxi - \ell}\} \text{ and } \Bsmalli := \{\ell \leq \lmaxi \colon \Bil \neq \emptyset \text{ and } \epsone < 0.8^{\lmaxi - \ell} \}.\]

We have

\begin{align*}
    \sum_{\ell \leq \lmaxi} \epsil &\leq \sum_{\ell \leq \lmaxi \colon \epsil = \epsone} \epsone  + \sum_{\ell \leq \lmaxi} \epstwo + \sum_{\ell \leq \lmaxi} \epsthree \\
    &\leq \sum_{\ell \in \Bbigi} \epsone + \sum_{\ell \leq \lmaxi} 0.9^{\lmaxi - \ell} + \sum_{\ell \leq \lmaxi} \frac{\muil}{\mui} \\
    &\leq \sum_{\ell \in \Bbigi} \epsone + 10 + 1.
\end{align*}

Using Cauchy-Schwarz, we get

\begin{align*}
    \sum_{\ell \in \Bbigi} \epsone &=  \sqrt{c \cdot \alphai \min(1,\sqrt{\gammai}) } \sum_{\ell \in \Bbigi}\sqrt{\frac{\Vil}{\Vi}} \\
         &\leq \sqrt{c \cdot \alphai \min(1,\sqrt{\gammai}) } \sqrt{|\Bbigi|}\sqrt{\sum_{\ell \in \Bbigi} \frac{\Vil}{\Vi}} \\
         &\leq \sqrt{c \cdot \alphai \min(1,\sqrt{\gammai}) |\Bbigi|}.
\end{align*}

Thus, we have to find a good upper bound on $|\Bbigi|$.

Consider an arbitrary $\ell \in \Bbigi$. From the definition of $\epsone$ and $\Bbigi$, it follows that

\[\sqrt{c \cdot \alphai \min(1,\sqrt{\gammai})\frac{\Vil}{\Vi}} \geq 0.8^{\lmaxi - \ell}\]

and therefore

\[\frac{\Vi}{\Vil} \leq \frac{c \cdot \alphai \min(1,\sqrt{\gammai})}{0.64^{\lmaxi - \ell}} \leq 1.6^{\lmaxi - \ell}.\]

It directly follows from the way $\gammai$ is defined that $\lmaxi = \log\left(\frac{\Vi}{\gammai \Deltai}\right)$. Therefore,

\begin{align*}
 \lmaxi &=\log \left( \frac{\Vi}{\gammai\Deltai} \right) \\
              &\leq \log(1/\gammai) + \log \left(\frac{\Vi}{\muil} \right) + \log \left(\frac{\mui}{\Delta_i}\right) \\
              &\leq \log(1/\gammai) + \log\left(\frac{\Vi 2^\ell} {\Vil} \right) + \frac{1}{c \cdot \alphai}\\
              &\leq \log(1/\gammai) + \log\left(\frac{\Vi}{\Vil} \right) + \ell + \frac{1}{c \cdot \alphai}\\
              &\leq \log(1/\gammai) + \log\left(1.6^{\lmaxi - \ell} \right)+  \ell +  \frac{1}{c \cdot \alphai} \\
              &\leq \log(1/\gammai) + 0.9(\lmaxi - \ell) + \ell + \frac{1}{c \cdot \alphai}.
\end{align*}

Thus, by rearranging terms, we get

\[\lmaxi - \ell \leq  10 \left(\log(1/\gammai)  + \frac{1}{c \cdot \alphai}\right).\]

Therefore, $|\Bbigi| \leq 10 \left(\log(1/\gammai) + \frac{1}{c \cdot \alphai}\right) + 1$ and

\[\sum_{\ell \in \Bbigi} \epsone \leq \sqrt{c \cdot \alphai \min(1,\sqrt{\gammai}) |\Bbigi|} \leq \sqrt{c \cdot \alphai \min(1,\sqrt{\gammai}) \left(10 \left(\log(1/\gammai) + \frac{1}{c \cdot \alphai}\right) + 1\right)} 
 \leq 50.\]

Thus, for $c$ being sufficiently large, we get

\begin{align*}
 \sum_{\ell \leq \lmaxi} \Deltail = \frac{\Deltai}{100}\sum_{\ell \leq \lmaxi} \epsil \leq \frac{\Deltai}{100} \left(11 + \sum_{\ell \in \Bbigi} \epsone \right) \leq \Deltai.
\end{align*}

Thus, for every $i \in [m] \setminus \Ibad$, we get

\[|\sum_{j \in [n]} \aij(\pj - \qj)| \leq \sum_{\ell \leq \lmaxi} |\sum_{j \in \Bil}\aij(\pj  - \qj)| \leq \sum_{\ell \leq \lmaxi} \Deltail \leq \Deltai. \]

Thus, it remains to show that $|\Ibad| \leq \sum_{i \in [m]} \fail$. We have

\begin{align*}
    \frac{\Deltail^2}{\Vil} &= \left(\frac{\epsil}{100\epsone}\right)^2\frac{(\epsone)^2 \cdot \Deltai^2}{\Vil} \\ 
                            &= \left(\frac{\epsil}{100\epsone}\right)^2\frac{c \cdot \alphai \min(1,\sqrt{\gammai})\cdot \frac{\Vil}{\Vi} \Deltai^2}{\Vil} \\
                            &= \left(\frac{\epsil}{100\epsone}\right)^2 \cdot \max(1,1/\sqrt{\gammai}) \cdot \frac{c \cdot \alphai \min(1,\gammai) \Deltai^2}{\Vi} \\ 
                            &= c \cdot \alphai \left(\frac{\epsil}{100\epsone}\right)^2 \cdot \max(1,1/\sqrt{\gammai}) \cdot \frac{ \min(1,\gammai) \Deltai^2}{\Vi} \\
                            &= c \cdot \alphai \left(\frac{\epsil}{100\epsone}\right)^2 \cdot \max(1,1/\sqrt{\gammai}) \cdot \min\left(\frac{\Deltai^2}{\Vi},\frac{\Deltai}{2^{\lmaxi}} \right).
\end{align*}

Also,

\begin{align*}
    \frac{\Deltail}{2^\ell} \geq \frac{1}{100}\frac{\epstwo \Deltai}{2^{\ell}} = \frac{1}{100}\frac{0.9^{\lmaxi - \ell}\Deltai}{2^{\ell}} = \frac{(2 \cdot 0.9)^{\lmaxi - \ell}}{100}\frac{\Deltai}{2^{\lmaxi}} \geq \frac{1.5^{\lmaxi - \ell}}{100} \cdot  \frac{\Deltai}{2^{\lmaxi}}
\end{align*}

and

\begin{align*}
    \frac{\Deltail k \cdot c\log (nm)}{\muil} \geq \frac{c\log(nm)}{100} \cdot \frac{\epsthree \Deltai k}{\muil} \geq \log(nm) \cdot \frac{\Deltai k}{\mui}.
\end{align*}

We now define

\[\betail = \min\left(c \cdot \alphai \left(\frac{\epsil}{100\epsone}\right)^2\max(1,1/\sqrt{\gammai}),\frac{1.5^{\lmaxi - \ell}}{100},\log(nm)\right).\]

Using above, and that we can assume $\failcher(A,\Deltav,\p,k) \leq 1$ without loss of generality, we get

\begin{align*}
\failil &:= \ccher  \exp \left(-(1/\ccher) \min \left(\frac{\Deltail^2}{\Vil}, \frac{\Deltail}{2^\ell}, \frac{\Deltail c \log(nm) k}{\muil} \right)\right) \\
&\leq \ccher  \exp \left(-(\betail/\ccher) \min \left(\frac{\Deltai^2}{\Vi}, \frac{\Deltai}{2^{\lmaxi}}, \frac{\Deltai k}{\mui} \right)\right) \\
        &\leq \ccher \alphai \left(\frac{1}{\alphai}\right) \exp \left(- \alphai \min \left(\frac{\Deltai^2}{\Vi}, \frac{\Deltai}{2^{\lmaxi -1}}, \frac{\Deltai k}{\mui} \right)\right)^{\frac{\betail}{2 \ccher \alphai}} \\
        &\leq \ccher \alphai \cdot \failcher(A,\Deltav,\p,k) \cdot \left( \alphai \failcher(A,\Deltav,\p,k)\right)^{\frac{\betail}{2 \ccher \alphai} - 1} \\
        &\leq \ccher \alphai \cdot \failcher(A,\Deltav,\p,k) \cdot \left(\frac{1}{c}\right)^{\frac{\betail}{4 \ccher \alphai}}.
\end{align*}

Consider an arbitrary $\ell \in \Bsmalli$. It holds that $\frac{\epsil}{\epsone} \geq \frac{0.9^{\lmaxi - l}}{0.8^{\lmaxi - l}} \geq 1.1^{\lmaxi - \ell}$ and therefore,

\[\betail \geq \min \left(\frac{c \cdot \alphai}{100^2}1.2^{\lmaxi - \ell}, \frac{1.5^{\lmaxi - \ell}}{100},\log(nm)\right) \geq \min \left(\frac{c \cdot \alphai}{10000}1.2^{\lmaxi - \ell},\log(nm) \right)\]

and thus for sufficiently large $c$,

\[\ccher \alphai \left(\frac{1}{c}\right)^{\frac{\betail}{4 \ccher \alphai}} \leq \frac{1}{100}\max\left(0.5^{\lmaxi - \ell}, \frac{1}{n}\right).\]

Thus, we get

\begin{align*}
    \sum_{\ell \in \Bsmalli} \failil  \leq \sum_{\ell \Bsmalli} \frac{1}{100}\max\left(0.5^{\lmaxi - \ell},\frac{1}{n}\right) \failcher(A,\Deltav,\p,k)  \leq \frac{1}{2}\failcher(A,\Deltav,\p,k).
\end{align*}

It always holds that $\beta_{i,\ell} \geq \min \left(\frac{c}{100^2} \dot \alphai \max(1,1/\sqrt{\gammai}),\frac{1.5^{\lmaxi - \ell}}{100},\log(nm) \right)$.
Thus, it holds that

\[\ccher \alphai \left(\frac{1}{c}\right)^{\frac{\betail}{4 \ccher \alphai}} \leq \frac{\alphai}{c}\max\left(\min(1,\gammai) ,0.5^{\lmaxi - \ell},\frac{1}{n}\right).\]

Thus, 

\begin{align*}
\sum_{\ell \in \Bbigi} \failil &\leq \frac{1}{3}\failcher(A,\Deltav,\p,k) +  |\Bbigi|\frac{\alphai \min(1,\gammai)}{c}\failcher(A,\Deltav,\p,k) \\
&\leq \frac{1}{2}\failcher(A,\Delta,\p,k).
\end{align*}

Therefore,

\begin{align*}
 |\Ibad| \leq |\Ibadcher| \leq \sum_{i \in [m]} \left(\sum_{\ell \colon \Bil \neq \emptyset} \failil \right) &\leq \sum_{i \in [m]} \left( \sum_{\ell \in \Bsmalli} \failil + \sum_{\ell \in \Bbigi} \failil \right) \\
 &\leq \sum_{i \in [m]} \failcher(A,\Deltav,\p,k).
\end{align*}
\end{proof}

\subsection*{Acknowledgment}
We thank Danil Koževnikov for help in proving \cref{cl:danil_genius}.

\bibliographystyle{alpha}
\bibliography{ref}

\appendix
\section{A proof sketch of \Cref{lem:pairwise}}
\label{app:pairwise}

\begin{proof}[Proof Sketch of \Cref{lem:pairwise}]
We provide only an informal sketch. The result is implicit in the work of ~\cite{luby1988removing} and has been used under different phrasings. See \cite[Section 3.2]{berger1994efficient} or Harris~\cite[Section 2]{harris2019deterministic}. 

Let each random variable $x_i$, for $i\in [n]$, be defined by $x_i=-1+2(\vec{i}\bullet \vec{z})$. Here, $\vec{z}$ is a random vector in $\{0, 1\}^{\log n + 1}$, and $\vec{i}$ is the $L=(\log n+1)$ dimensional vector indicating the binary representation of number $i$ where we append a $1$ in the least significant place to ensure that the vector is not all zeros. The sign $\bullet$ is used for the dot product of the two vectors mod $2$. It is easy to see that $\Pr[x_i=+1]=\Pr[x_i=-1]=1/2$ and that these variables are pairwise independent, i.e., any two $x_i, x_j$---for $i, j\in [n]$ such that $i\neq j$---are independent of each other. The task is to find a fixing of $\vec{z}\in \{0, 1\}^{\log n + 1}$ such that the function $f()$ is upper bounded by its expectation. 

For that, we fix the bits of $\vec{z}$ one by one. The core challenge is how to do one step: in such a step, some number $r \in [0, L-1]$ of the bits of $\vec{z}$ have already been fixed, and the task is to compute the conditional expectation of $f()$ for the two possibilities of setting the next bit of $\vec{z}$ equal to $0$ or $1$.  Thus, each of these two problems is to compute the conditional expectations for $f()$ over the random space of $\vec{z}$, in the conditional space where the first $r+1$ bits of $\vec{z}$ are set in one particular way, and the rest remain random. Doing this computation for the constant term and the linear terms in $f()$ is easy, in $||Q_k||\cdot \poly(\log n)$ work and $\poly\log(n)$ depth. It is the quadratic terms that require some ingenuity. Notice that, once some $r+1$ bits of $\vec{z}$ have been fixed and the rest remain random, the random variables $x_i, x_j$ are no longer necessarily pairwise independent. For instance, two variables $x_i$ and $x_j$ such that the binary representations of $\vec{i}$ and $\vec{j}$ have the same last $L-(r+1)$ bits are deterministic functions of each other. Luby's idea is that we can group $i\in [1,n]$ based on the binary representation of $\vec{i}$ in the remaining $L-(r+1)$ bits (e.g., with sorting based on only these bits). Variables that are in different groups are still pairwise independent and thus their products cancel out from the computation. Variables inside one group can be broken into two subgroups, depending on the mod-$2$ dot product of their binary representation in the first $(r+1)$ bits with the chosen $r+1$ bits of vector $\vec{z}$. Then the computation boils down to calculating the summation inside each subgroup of coefficients $\alpha_i$ and $\beta_j$ respectively for $A$ and $B$. This can be performed using $||Q_k||\cdot \poly(\log n)$ work and $\poly\log(n)$ depth. Please see ~\cite[Section 2]{luby1993removing},~\cite[Section 3.2]{berger1994efficient}, or~\cite[Section 2]{harris2019deterministic} for details.
\end{proof}

\section{Proof of \cref{thm:hoeffding} and \cref{thm:chernoff}}
\label{sec:hoeffdingchernofffull}
% \todo{add more explanation}
\begin{proof}
We now use \cref{lem:hoeffding_chernoff} to prove \cref{thm:hoeffding,thm:chernoff}. Note that it suffices to show that one can remove the assumption in \cref{lem:hoeffding_chernoff} that $\pj \geq \frac{1}{2^k}$ for every $j \in [n]$ and to remove the $k$ in the exponent of the failure probability. 
Let $A,\p,\Deltav,k$ be some general input to \cref{thm:hoeffding} or \cref{thm:chernoff}. We can assume without loss of generality that \[\min(\failhoef(A,\Deltav,\p,k),\failcher(A,\Deltav,\p,k)) \leq 1.\]

Let $\pres \in \mathbb{R}^n,\Deltares \in \mathbb{R}^{m},\kres \in \mathbb{N}$ be defined as

\begin{enumerate}
\item $\presj = \max\left(\pj,\frac{1}{n}\right)$ for every $j \in [n]$,
\item $\Deltaresi = \Deltai/2$ for every $i \in [m]$ and
\item $\kres = \lceil c^2\log(2nm)\rceil k$.
\end{enumerate}

Note that $\presj \geq \frac{1}{n} \geq \frac{1}{2^{\kres}}$ for every $j \in [n]$. Thus, we can use \cref{lem:hoeffding_chernoff} to compute using computational work $\tilde{O}(\max(nnz(A),n,m)\poly(\ksmall)) = \tilde{O}(\max(nnz(A),n,m)\poly(k))$ and depth bounded by $\poly(\log(nm),\ksmall) = \poly(\log(nm),k)$ a set $\Ibad \subseteq [m]$ with $|\Ibad|\leq \sum_{i \in [m]} \failthree(A,\Deltares,\pres,\kres)$ and a vector $\q$ such that for every $i \in [m] \setminus \Ibad$, it holds that

\[|\sum_{j \in [n]} \aij(\presj - \qj)| \leq \Deltaresi.\]

Note that

\[|\sum_{j \in [n]} \aij(\pj - \presj)| \leq \frac{1}{n}\sum_{j \in [n]} \aij \leq \max_{j \in [n]} \aij.\]

Moreover, for $c$ being sufficiently large, it holds that $\max_{j \in [n]} \aij \leq \Deltai/2$ because 

\[1 \geq \min(\failhoef,\failcher) \geq c\exp\left(-\left(\frac{1}{c}\right)\max\left(\frac{\Deltai^2}{(\max_{j \in [n]} \aij)^2},\frac{\Deltai}{\max_{j\in [n]} \aij}\right)\right).\]

Therefore,

\[|\sum_{j \in [n]} \aij(\pj - \qj)| \leq |\sum_{j \in [n]} \aij(\pj - \presj)| + |\sum_{j \in [n]} \aij(\presj - \qj)| \leq \Deltai/2 + \Deltaresi = \Deltai.\]

Thus, it remains to verify that $|\Ibad| \leq \sum_{i \in [m]} \failhoef(A,\Deltav,\p,k)$ in order to prove \cref{thm:hoeffding} and $|\Ibad| \leq \sum_{i \in [m]} \failcher(A,\Deltav,\p,k)$ in order to prove \cref{thm:chernoff}. To that end, we first show that for every $i \in [m]$ and $c$ being sufficiently large, it holds that

\[\failthree(A,\Deltares,\pres,\kres) \leq \frac{1}{2}\min(\failhoef(A,\Deltav,\p,k),\failcher(A,\Deltav,\p,k)) + \frac{1}{2m}.\]

We have argued above that for $c$ being sufficiently large, it holds that

\[\sum_{j \in [n]} \presj \aij \leq \sum_{j \in [n]} \pj \aij + \Delta_i \leq 2\max \left(\sum_{j \in [n]} \pj \aij, \Deltai \right).\]

Therefore,

\[\frac{\Deltaresi \kres}{\sum_{j\in [n]} \presj \aij} \geq \frac{\Deltai c\log(2m)k}{\max(\sum_{j \in [n]} \pj \aij,\Deltai)} \geq \min\left( \frac{\Deltai k}{\sum_{i \in [n]} \pj \aij},c\log(2m) \right).\]

Together with 

\[\frac{(\Deltaresi)^2}{\sum_{j \in [n]} \aij^2} \geq \frac{1}{4}\frac{\Deltai^2}{\sum_{j \in [n]} \aij^2},\]

this implies that

\[\expone(A,\Deltares,\pres,\kres) \geq \min\left(\frac{1}{4} \expone(A,\Deltav,\p,k), c\log(2m)\right).\]

Using again that $\sum_{j\in [n]} \presj \aij \leq 2\max(\sum_{j \in [n]} \pj \aij, \Deltai)$, we get

\begin{align*}
\frac{(\Deltaresi)^2}{\sum_{j \in [n]} \presj \aij(\max_{j \in [n]} \aij)} &\geq \frac{1}{8}\frac{\Deltai^2}{\max \left( \sum_{j \in [n]} \pj\aij (\max_{j \in [n]} \aij), \Deltai (\max_{j \in [n]} \aij\right)} \\
&= \frac{1}{8} \min \left(\frac{\Deltai^2}{\sum_{j\in [n]} \pj \aij (\max_{j \in [n]} \aij)}, \frac{\Deltai}{\max_{j\in [n]} \aij} \right) \\
&\geq \frac{1}{8}\exptwo(A,\Deltav,\p,k).
\end{align*}

Therefore,

\begin{align*}
\exptwo(A,\Deltares,\pres,\kres) &:= \min \left(\frac{(\Deltaresi)^2}{\sum_{j\in [n]} \presj \aij (\max_{j \in [n]} \aij)}, \frac{\Deltaresi}{\max_{j\in [n]} \aij}, \frac{\Deltaresi k}{\sum_{j\in [n]} \presj \aij} \right) \\
&\geq \min \left(\frac{1}{8} \exptwo(A,\Deltav,\p,k),\frac{1}{2}\frac{\Deltai}{\max_{j\in [n]} \aij},\frac{\Deltai k}{\sum_{j\in [n]} \pj \aij}, c \log(2m) \right) \\
&\geq  \min \left(\frac{1}{8}\exptwo(A,\Deltav,\p,k), c \log(2m) \right).
\end{align*}

Thus, for sufficiently large $c$, we indeed get

\begin{align*}
&\failthree(A,\Deltares,\pres,\kres) \\
&= \cthree \exp \left(- \left(\frac{1}{\cthree}\right) \min(\max \left( \expone \left(A,\Deltares,\pres,\kres \right),\exptwo \left(A,\Deltares,\pres,\kres \right)\right),\kres)\right) \\
&\leq \cthree \exp \left(- \left(\frac{1}{\cthree}\right)\min \left(\max \left(\frac{1}{4} \expone(A,\Deltav,\p,k),\frac{1}{8}\exptwo(A,\Deltav,\p,k) \right),c\log(2m)\right)\right) \\
&\leq 0.5 c\exp \left(- \left(\frac{1}{c}\right)\max \left(\expone(A,\Deltav,\p,k),\exptwo(A,\Deltav,\p,k)\right)\right) + \cthree \exp\left(- \frac{c}{\cthree} \log(2m)\right) \\
&\leq 0.5\min(\failhoef(A,\Deltav,\p,k),\failcher(A,\Deltav,\p,k)) + \frac{1}{2m}.
\end{align*}

Therefore,

\begin{align*}
|\Ibad| &\leq \bigg\lfloor \sum_{i \in [m]} \failthree(A,\Deltares,\pres,\kres) \bigg\rfloor \\
        &\leq \bigg\lfloor \sum_{i \in [m]} \left(0.5\min(\failhoef(A,\Deltav,\p,k),\failcher(A,\Deltav,\p,k))+ \frac{1}{2m} \right)\bigg\rfloor \\ &\leq \bigg\lfloor 0.5 + 0.5\sum_{i \in [m]} 0.5\min(\failhoef(A,\Deltav,\p,k),\failcher(A,\Deltav,\p,k))\bigg\rfloor \\
        &\leq \min \left(\sum_{i \in [m]} \failhoef(A,\Deltav,\p,k),\sum_{i \in [m]} \failcher(A,\Deltav,\p,k)\right).
\end{align*}

Thus, we have proven both \cref{thm:hoeffding} and \cref{thm:chernoff}.
\end{proof}

\end{document}